\documentclass[letter]{aa}
\usepackage[varg]{txfonts}
\usepackage[breaklinks=true]{hyperref}
\hypersetup{
    colorlinks,
    linkcolor={blue},
    citecolor={blue},
    urlcolor={blue}
}

\usepackage{natbib,twoopt}
\bibpunct{(}{)}{;}{a}{}{,} 
\makeatletter
\newcommandtwoopt{\citeads}[3][][]{\href{http://adsabs.harvard.edu/abs/#3}%
{\def\hyper@linkstart##1##2{}%
\let\hyper@linkend\@empty\citealp[#1][#2]{#3}}}
\newcommandtwoopt{\citepads}[3][][]{\href{http://adsabs.harvard.edu/abs/#3}%
{\def\hyper@linkstart##1##2{}%
\let\hyper@linkend\@empty\citep[#1][#2]{#3}}}
\newcommandtwoopt{\citetads}[3][][]{\href{http://adsabs.harvard.edu/abs/#3}%
{\def\hyper@linkstart##1##2{}%
\let\hyper@linkend\@empty\citet[#1][#2]{#3}}}
\newcommandtwoopt{\citeyearads}[3][][]%
{\href{http://adsabs.harvard.edu/abs/#3}
{\def\hyper@linkstart##1##2{}%
\let\hyper@linkend\@empty\citeyear[#1][#2]{#3}}}
\makeatother

\newcommand\jwst{JWST\xspace}
\newcommand\hst{HST\xspace}
\newcommand\smacs{SMACS~J0723\xspace}

\graphicspath{{./figures/}}

\begin{document}

\title{First JWST observations of a gravitational lens}
\subtitle{Mass model from new multiple images with near-infrared observations of SMACS~J0723.3$-$7327}

\author{G.~B.~Caminha       \inst{\ref{mpa}}                        
                            \thanks{e-mail address: \href{mailto:caminha@mpa-garching.mpg.de}{caminha@mpa-garching.mpg.de}. The MUSE redshift catalogue (Table \ref{tab:muse_cat}) and lens model files are available at the CDS via anonymous ftp to cdsarc.u-strasbg.fr (130.79.128.5) or via \url{http://cdsarc.u-strasbg.fr/viz-bin/cat/J/A+A/666/L9}.} \and
        S.~H.~Suyu           \inst{\ref{mpa},\ref{tum},\ref{sinica}}      \and      
        A.~Mercurio \inst{\ref{inafcapo}} \and
        G.~Brammer \inst{\ref{dawn},\ref{niels_bohr}} \and
        P.~Bergamini \inst{\ref{unimilano}, \ref{inafbologna}} \and
        A.~Acebron \inst{\ref{unimilano}, \ref{inafmilano}} \and
        E.~Vanzella  \inst{\ref{inafbologna}}
        }
\institute{
Max-Planck-Institut f\"ur Astrophysik, Karl-Schwarzschild-Str. 1, D-85748 Garching, Germany \label{mpa} \and
Technische Universit\"at M\"unchen, Physik-Department, James-Franck Str. 1, 85748 Garching, Germany \label{tum} \and
Academia Sinica Institute of Astronomy and Astrophysics (ASIAA), 11F of ASMAB, No.1, Section 4, Roosevelt Road, Taipei 10617, Taiwan \label{sinica} \and
INAF - Osservatorio Astronomico di Capodimonte, Via Moiariello 16, I-80131 Napoli, Italy\label{inafcapo} \and
Cosmic Dawn Center (DAWN), Denmark \label{dawn} \and
Niels Bohr Institute, University of Copenhagen, Jagtvej 128, DK-2200 Copenhagen N, Denmark \label{niels_bohr} \and
Dipartimento di Fisica, Università degli Studi di Milano, Via Celoria 16, I-20133 Milano, Italy \label{unimilano} \and
INAF -- OAS, Osservatorio di Astrofisica e Scienza dello Spazio di Bologna, via Gobetti 93/3, I-40129 Bologna, Italy \label{inafbologna} \and
INAF - IASF Milano, via A. Corti 12, I-20133 Milano, Italy \label{inafmilano}
}

\abstract{

We present our lens mass model of SMACS J0723.3$-$7327, the first strong gravitational lens observed by the \textit{James Webb} Space Telescope (\jwst).
We use data from the {\textit{Hubble} Space Telescope} and the Multi Unit Spectroscopic Explorer (MUSE) to build our `pre-\jwst' lens model and then refine it with newly available \jwst near-infrared imaging in our \jwst model.
To reproduce the positions of all multiple lensed images with good accuracy, the adopted mass parameterisation consists of one cluster-scale component, accounting mainly for the dark matter distribution, the galaxy cluster members, and an external shear component.
The pre-\jwst model has, as constraints, 19 multiple images from six background sources, of which four have secure spectroscopic redshift measurements from this work.
The \jwst model has more than twice the number of constraints: 30 additional multiple images from another 11 lensed sources.
Both models can reproduce the multiple image positions very well,  with a $\delta_{\rm rms}$ of $0\farcs39$ and $0\farcs51$ for the pre-\jwst and \jwst models, respectively.
The total mass estimates within a radius of 128~kpc  (roughly the Einstein radius) are $7.9_{-0.2}^{+0.3} \times 10^{13} \rm M_{\odot}$ and $ 8.7_{-0.2}^{+0.2} \times 10^{13} \rm M_{\odot}$ for the pre-\jwst and \jwst models, respectively.
We predict with our mass models the redshifts of the newly detected \jwst sources, which is crucial information, especially for systems without spectroscopic measurements, for further studies and follow-up observations.
Interestingly, one family detected with \jwst is found to be at a very high redshift, $z>7.5$ (68\% confidence level), and with one image that has a lensing magnification of $|\mu|= 9.5_{-0.8}^{+0.9}$, making it an interesting case for future studies.
The lens models, including magnification maps and redshifts estimated from the model, are made publicly available, along with the full spectroscopic redshift catalogue from MUSE.
}


\maketitle

\section{Introduction}
\label{sec:introduction}

\begin{figure*}
  \centering
  \includegraphics[width = 1.\textwidth]{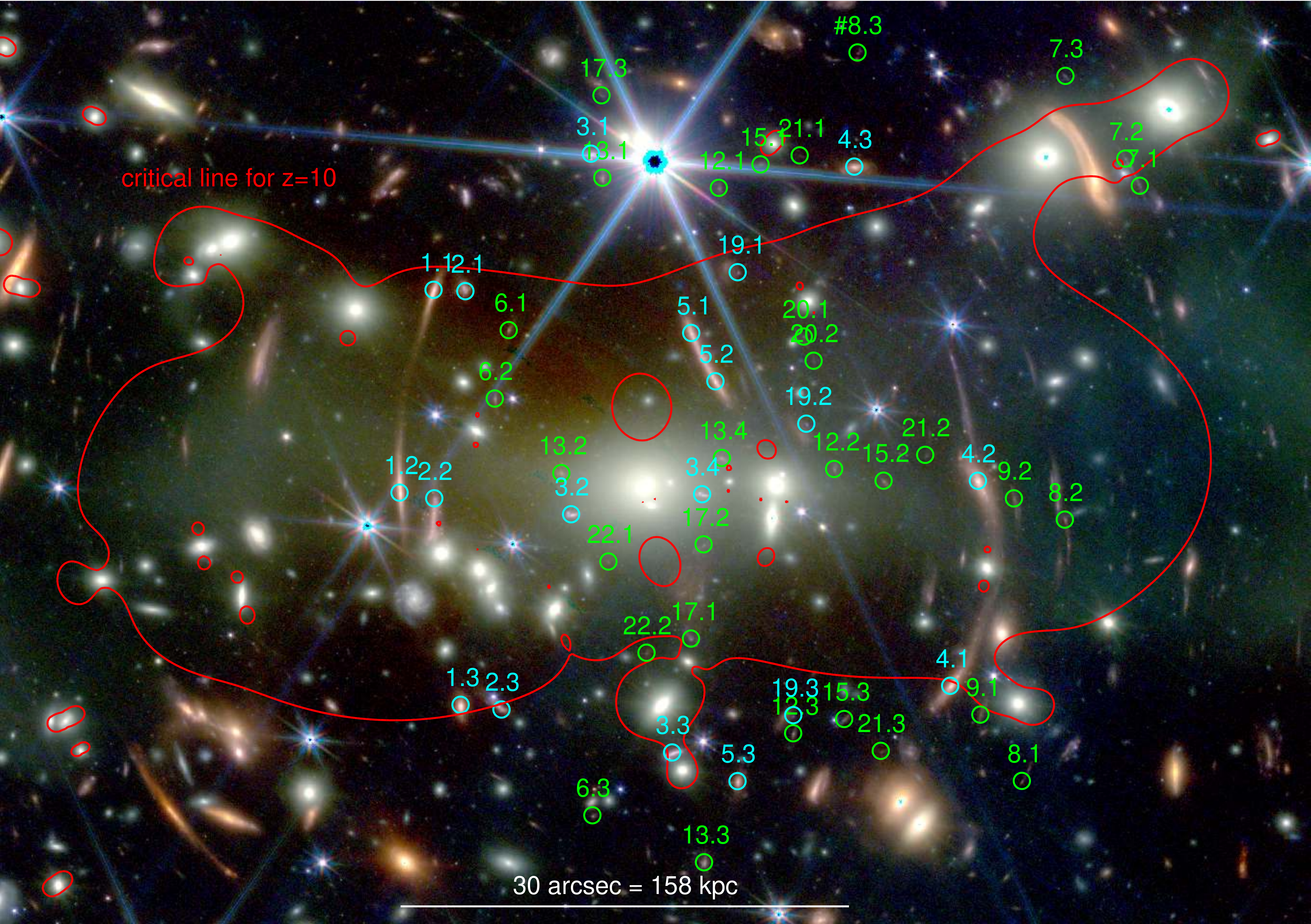}
  \caption{The galaxy cluster SMACS~J0723. Colour composed image using \jwst/NIRCam imaging with the filters F090W, F150W in blue; F200W, F277W in green; and F356W and F444W in red. Cyan circles show the positions of multiple images identified in the pre-\jwst lens model and the green circles are the newly identified multiple images with \jwst. The red line is the critical line for a source at redshift $z=10$.}
  \label{fig:JWST_image}
\end{figure*}

Galaxy clusters are the most massive structures in the Universe and are therefore powerful cosmic telescopes for observing faint and distant sources via the gravitational lensing effect \citep[e.g.][]{2015ApJ...800...18A}.  When a background source galaxy lies behind a galaxy cluster, the light rays that we see from the source are deflected by the gravitational potential of the foreground galaxy cluster, that is, the source is gravitationally lensed by the cluster. In the regime of strong lensing, we see multiple magnified and distorted images of the background source. The lensing magnifications produced by galaxy clusters can reach factors of $\sim 100-1000$, allowing us to detect the faintest sources, which would otherwise be  impossible \citep[see e.g.][]{2016A&A...595A.100C, 2020MNRAS.494L..81V, 2021A&A...646A..57V}.  The positions and morphologies of the multiple lensed images also allow us to probe the distribution of matter, particularly dark matter, in the galaxy cluster, which is crucial for understanding the nature of dark matter \citep[e.g.][]{2006ApJ...652..937B, 2006ApJ...648L.109C}.  Furthermore, an individual galaxy cluster often lenses several background sources into several corresponding `families' of multiple images that straddle different locations around the galaxy cluster.  Families of multiple images formed by sources of different redshifts provide measurements of angular diameter distance ratios between the cluster and sources, allowing us to probe the geometry of the Universe \citep{2010Sci...329..924J, 2017MNRAS.470.1809A,2016A&A...587A..80C,2022A&A...657A..83C}.  In cases where a background source is time varying, such as a quasar or supernova, the time delay(s) between the multiple images provide measurements of the `time-delay distance' and the Hubble constant that sets the expansion rate of the Universe \citep{1964MNRAS.128..307R, 2010ApJ...711..201S, 2015Sci...347.1123K, 2018ApJ...860...94G, 2020ApJ...898...87G}.  Strong lensing galaxy clusters are therefore excellent laboratories for astrophysical and cosmological studies.  

The {\textit{James Webb} Space Telescope} \citep[\jwst;][]{2022arXiv220705632R} operating in the infrared will provide unprecedented observations of high-redshift galaxies, in terms of both sensitivity and angular resolution.  Studying these distant galaxies is crucial for understanding how the first galaxies formed and evolved into the structures that we see today.  These first galaxies are inherently faint, and the combination of \jwst and galaxy cluster lensing is therefore the best way to detect and study these faintest galaxies.  As part of the Early Release Observations (ERO), the \jwst team publicly released the observations of its first cosmic targets, including the galaxy cluster SMACS~J0723.3$-$7327 (hereafter \smacs, discovered by \citealt{2001ApJ...553..668E} and \citealt{2018MNRAS.479..844R}). In this Letter we demonstrate the power of strong gravitational lensing and \jwst to unveil and study faint distant galaxies.  In particular, we identify new sources in the \jwst images that were previously undetected in \textit{Hubble} Space Telescope (\hst) imaging, more than doubling the number of families of multiple images.  Even though these new families of images do not yet have spectroscopic redshifts, we demonstrate the utility of our cluster lens mass model to constrain the redshifts of these new source galaxies, including one at $z>6$.  

Our paper is organised as follows.  In Sect. \ref{sec:pre_data} we present the pre-\jwst archival data of the galaxy cluster \smacs.
The \jwst observations used in this work are presented in Sect. \ref{sec:jwst_data}.
In Sect. \ref{sec:lens_model} we present our lens models based on archival data only (pre-\jwst), the updated version that includes new multiple images identified with \jwst, and a comparison to other models in the literature.
Finally, we provide a summary in Sect. \ref{sec:conclusions}.

\begin{table}[!]
\centering
\tiny
\caption{Multiple image list.}
\begin{tabular}{l c c l l} \hline \hline
ID & RA & Dec & $z$ & $|\mu|$\\
\hline
1.1  &  110.8405365164  &  $-$73.4509957549  &  1.4503  &  $9.8_{-1.0}^{+1.3}$  \\
1.2  &  110.8427758496  &  $-$73.4547637941  &  1.4503  &  $21.9_{-3.7}^{+4.2}$  \\
1.3  &  110.8388129336  &  $-$73.4587087766  &  1.4503  &  $6.7_{-0.6}^{+0.7}$  \\
\hline        
2.1  &  110.8384630963  &  $-$73.4510225059  &  1.3782  &  $9.2_{-0.9}^{+1.2}$  \\
2.2  &  110.8405113048  &  $-$73.4548709934  &  1.3782  &  $113_{-38}^{+111}$  \\
2.3  &  110.8361429009  &  $-$73.4587973008  &  1.3782  &  $6.0_{-0.6}^{+0.6}$  \\
\hline                  
3.1  &  110.8302653775  &  $-$73.4484703860  &  $1.87_{-0.09}^{+0.10}$  &  $4.4_{-0.3}^{+0.4}$  \\
3.2  &  110.8315839500  &  $-$73.4551704511  &  $''$  &  $3.6_{-0.5}^{+0.6}$  \\
3.3  &  110.8250107058  &  $-$73.4595997391  &  $''$  &  $16.1_{-3.5}^{+6.6}$  \\
3.4  &  110.8230386287  &  $-$73.4548011166  &  $''$  &  $1.4_{-0.2}^{+0.3}$  \\
\hline                  
4.1  &  110.8068620127  &  $-$73.4583655711  &  $''$  &  $9.3_{-1.1}^{+1.3}$  \\
4.2  &  110.8050431160  &  $-$73.4545561509  &  $''$  &  $13.5_{-2.0}^{+2.2}$   \\
4.3  &  110.8130876147  &  $-$73.4487156625  &  $2.14_{-0.08}^{+0.09}$  &  $5.7_{-0.5}^{+0.7}$  \\
\hline                  
5.1  &  110.8237341404  &  $-$73.4518040104  &  1.4254  &  $28.1_{-4.9}^{+6.2}$  \\
5.2  &  110.8221590678  &  $-$73.4527025651  &  1.4254  &  $49.2_{-5.2}^{+7.1}$  \\
5.3  &  110.8207468206  &  $-$73.4601341978  &  1.4254  &  $3.9_{-0.3}^{+0.3}$  \\
\hline     
19.1  &  110.8207007748  &  $-$73.4506760494  &  1.3825  &  $9.0_{-1.0}^{+1.2}$  \\
19.2  &  110.8162335526  &  $-$73.4534926339  &  1.3825  &  $12.7_{-1.5}^{+1.6}$  \\
19.3  &  110.8171039789  &  $-$73.4589088990  &  1.3825  &  $5.2_{-0.4}^{+0.5}$  \\
\hline                  
\hline
6.1$\dagger$  &  110.8356387152  &  $-$73.4517465412  &  $1.69_{-0.03}^{+0.04}$  &  $45.3_{-9.5}^{+13.1}$  \\
6.2$\dagger$  &  110.8365547868  &  $-$73.4530182644  &  $''$  &  $18.3_{-1.7}^{+1.9}$  \\
6.3$\dagger$  &  110.8302221440  &  $-$73.4607653275  &  $''$  &  $3.9_{-0.3}^{+0.3}$  \\
\hline  
7.1$\dagger$  &  110.7944672364  &  $-$73.4490759821  &  5.1727  &  $7.9_{-0.8}^{+1.0}$  \\
7.2$\dagger$  &  110.7954106149  &  $-$73.4485708909  &  $''$  &  $20.1_{-4.5}^{+6.9}$  \\
7.3$\dagger$  &  110.7993143225  &  $-$73.4470330844  &  $''$  &  $4.7_{-0.4}^{+0.4}$  \\
\hline
8.1$\dagger$  &  110.8021918024  &  $-$73.4601348271  &  $>7.5$  &  $6.1_{-0.6}^{+0.7}$  \\
8.2$\dagger$  &  110.7993733038  &  $-$73.4552793576  &  $''$  &  $9.5_{-0.8}^{+0.9}$  \\
\#8.3$\dagger$  &  110.8128699466  &  $-$73.4465944181  &  ---  &  $4.5_{-0.4}^{+0.5}$  \\
\hline                  
9.2$\dagger$  &  110.8026934208  &  $-$73.4548842598  &  $3.00_{-0.22}^{+0.27}$  &  $12.9_{-1.8}^{+2.4}$  \\
9.1$\dagger$  &  110.8048945912  &  $-$73.4589021013  &  $''$  &  $8.8_{-0.9}^{+1.2}$  \\
\hline  
12.1$\dagger$  &  110.8219242869  &  $-$73.4491076284  &  $1.69_{-0.05}^{+0.06}$  &  $5.3_{-0.5}^{+0.5}$  \\
12.2$\dagger$  &  110.8144209400  &  $-$73.4543348989  &  $''$  &  $4.3_{-0.5}^{+0.6}$  \\
12.3$\dagger$  &  110.8171117496  &  $-$73.4592453685  &  $''$  &  $5.5_{-0.5}^{+0.5}$  \\
\hline    
13.1$\dagger$  &  110.8295177750  &  $-$73.4489165365  &  $2.75_{-0.21}^{+0.28}$  &  $6.2_{-0.6}^{+0.7}$  \\
13.2$\dagger$  &  110.8217443301  &  $-$73.4541282431  &  $''$  &  $6.0_{-0.9}^{+1.1}$  \\
13.3$\dagger$  &  110.8229393279  &  $-$73.4616409371  &  $''$  &  $3.9_{-0.3}^{+0.3}$  \\
13.4$\dagger$  &  110.8322177002  &  $-$73.4544051992  &  $''$  &  $3.6_{-0.5}^{+0.7}$  \\
\hline  
15.1$\dagger$  &  110.8192030143  &  $-$73.4486729318  &  $1.87_{-0.06}^{+0.07}$  &  $5.5_{-0.5}^{+0.5}$  \\
15.2$\dagger$  &  110.8111947441  &  $-$73.4545527988  &  $''$  &  $6.7_{-0.8}^{+0.9}$  \\
15.3$\dagger$  &  110.8137839019  &  $-$73.4589815364  &  $''$  &  $6.4_{-0.6}^{+0.7}$  \\
\hline   
17.1$\dagger$  &  110.8237732631  &  $-$73.4574832535  &  $''$  &  $21.8_{-3.2}^{+3.7}$  \\
17.2$\dagger$  &  110.8229423834  &  $-$73.4557325725  &  $''$  &  $10.6_{-1.3}^{+1.5}$  \\
17.3$\dagger$  &  110.8295676112  &  $-$73.4473859288  &  $1.92_{-0.07}^{+0.08}$  &  $3.5_{-0.2}^{+0.3}$  \\
\hline                  
20.1$\dagger$  &  110.8163947858  &  $-$73.4518737634  &  $1.47_{-0.21}^{+0.56}$  &  $46.8_{-25.5}^{+36.8}$  \\
20.2$\dagger$  &  110.8157525822  &  $-$73.4523225418  &  $''$  &  $44.2_{-28.9}^{+43.4}$  \\
\hline                  
21.1$\dagger$  &  110.8166488501  &  $-$73.4485063451  &  $2.29_{-0.10}^{+0.12}$  &  $5.2_{-0.5}^{+0.5}$  \\
21.2$\dagger$  &  110.8084789438  &  $-$73.4540734888  &  $''$  &  $7.7_{-0.9}^{+1.1}$  \\
21.3$\dagger$  &  110.8113960548  &  $-$73.4595739183  &  $''$  &  $5.6_{-0.5}^{+0.5}$  \\
\hline                  
22.1$\dagger$  &  110.8291533467  &  $-$73.4560496998  &  $2.15_{-0.30}^{+0.43}$  &  $9.5_{-1.4}^{+2.1}$  \\
22.2$\dagger$  &  110.8266763430  &  $-$73.4577453680  &  $''$  &  $19.7_{-3.8}^{+4.9}$  \\
\hline
\hline
\end{tabular}
\label{tab:multiple_images}
\tablefoot{IDs, coordinates (RA and Dec based on the RELICS world coordinate system) used as inputs to our lens model, redshifts ($z$) and magnifications ($|\mu|$) of all multiple images. Exact redshift values are spectroscopic confirmations, and values with 68\% confidence level uncertainties are obtained from the \jwst lens model. Lensing magnifications are also from the \jwst model. IDs marked with $\dagger$ are objects detected with the release of \jwst data. Image \#8.3 is only a candidate (indicated by the \#) and was not used in the model, although its morphology and colours strongly indicate it is correctly identified as a image of family 8. The IDs reported here are the same as in the last version of \citet{2022arXiv220707102P} and \citet{2022arXiv220707101M}.}
\end{table}

\section{Pre-\jwst archival data}
\label{sec:pre_data}

In this section we describe the data used in this work, consisting of \hst photometry and Multi Unit Spectroscopic Explorer (MUSE) spectroscopy, to build the pre-\jwst lens model.

\subsection{\hst}
\label{sec:HST_data}

Observations from \hst of SMACS~J0723 were obtained under the treasury programme Reionization Lensing Cluster Survey \citep[RELICS;][]{2019ApJ...884...85C}.
Imagings were obtained in the filters F435W, F606W, and F814W using the Advanced Camera for Surveys, and in F105W, F125W, F140W, and F160W using Wide Field Camera 3.
All reduced images and catalogues are publicly available at the Mikulski Archive for Space Telescope website\footnote{\url{https://archive.stsci.edu/prepds/relics/}}.
The imaging and colour information were used to identify multiple images and cluster members with no spectroscopic information.
More details are presented in Sect. \ref{sec:pre_model}.

\subsection{VLT/MUSE}
\label{sec:MUSE_data}

MUSE data of this field are available at the ESO Science Archive Facility website\footnote{\url{http://archive.eso.org/cms.html}}.
The field was observed as part of programme ID 0102.A-0718 (P.I.: A. Edge) under fair conditions, with seeing $\approx 0.7\arcsec$ and a moon illumination of 68\%, and with an exposure time of 2910 seconds on target.
We processed the raw data following all steps in the standard reduction pipeline \citep[version 2.8.3;][]{2020A&A...641A..28W}.
To improve the sky-subtraction, we used the auto-calibration mode and applied the {\tt Zurich Atmosphere Purge} \citep{2016MNRAS.458.3210S} to the data.
The final datacube has a 1~arcmin$^2$ field of view and covers the wavelength range  $4750\AA$ to $9350\AA$.

We extracted the 1D spectra from all \hst-detected sources to measure their redshifts.
In a second step, we performed a blind search to detect faint emission lines with no clear photometric counterpart.
We refer the reader to our previous works \citep{2019A&A...632A..36C,2017A&A...607A..93C,2017A&A...600A..90C} for more details on this procedure.
Our final redshift catalogue contains 78 secure and precise redshift measurements; it is presented in Table \ref{tab:muse_cat} and is available in electronic form at the CDS.

\section{\jwst Early Release Observations}
\label{sec:jwst_data}

SMACS~J0723 is the first gravitational lens observed by \jwst.
It was one of the targets of the ERO carried out during the telescope commissioning.
The observations obtained photometric data from the Near Infrared Camera (NIRCam; e.g. \citealt{2005SPIE.5904....1R}) and the Mid-Infrared Instrument (MIRI; \citealt{2015PASP..127..584R}, \citealt{2015PASP..127..612B}) and spectroscopy from the Near Infrared Spectrograph (NIRSpec; \citealt{2022A&A...661A..80J}) and the Near Infrared Imager and Slitless Spectrograph (NIRISS; \citealt{2012SPIE.8442E..2RD}).
Since MIRI imaging has a very large point spread function ($>1\arcsec$ for wavelengths $\rm >7.5~\mu m$ ) compared to NIRCam, we do not consider mid-infrared data in this work.
NIRSpec targeted multiple images: 4.1, 4.2, and 7.1 (see Table \ref{tab:multiple_images}).
No secure spectroscopic redshift has been obtained for images 4.1 or 4.2.
For image 7.1, a secure redshift of $z=5.1727$ is reported in \citet{2022arXiv220707101M} from \ion{O}{III} and \ion{H}{$\alpha$} detections.
We incorporated this measurement into our model.  We did not consider NIRISS data since slitless spectroscopy is subject to strong spectral contamination, especially in crowded fields such as the core of \smacs.
In the next sections we focus on NIRCam imaging to identify new multiple image families not detected with \hst.

NIRCam ERO observations of \smacs were carried out in the filters  F090W, F150W, F200W, F277W, F356W, and F444W, covering the wavelength range from 0.8~$\rm \mu m$ to $\rm \approx5.0~\mu m$, under programme ID 2736 (P.I.: K. Pontoppidan).
The total exposure time in each filter was 7537 seconds divided over nine dither patterns, ensuring enough depth to detect and resolve faint sources.
We used the standard NIRCam reduction pipeline with the calibration files available with the first data release.
We produced mosaics with resolutions of $\rm 20mas/pixel$ for the F090W, F150W, and F200W filters, and $\rm 40mas/pixel$ for F277W, F356W, and F444W.

The final mosaics cover the core of \smacs and the entire region of multiple image formation.
In Fig. \ref{fig:JWST_image} we show the colour composite image created using the software \texttt{trilogy} \citep{2012ApJ...757...22C}.
We aligned the final images to the RELICS \hst data to maintain the same coordinate reference as previous observations.
This introduces a small rigid offset with respect to the nominal coordinates of the ERO release.
The astrometric precision between the \jwst and \hst data we obtained has an r.m.s. of $0\farcs015$.
Moreover, because of the nature of these early \jwst observations, the calibration files are not ideal and the absolute flux calibration might have some issues.
Nonetheless, these caveats do not influence the colours or morphology of multiple images in the imaging, with no significant effect on our model.

\section{Strong lens model}
\label{sec:lens_model}
We used the software \texttt{lenstool} \citep{1996ApJ...471..643K, 2007NJPh....9..447J,2009MNRAS.395.1319J} to model the mass distribution of SMACS~J0723. The fiducial mass model consists of an elliptical cluster-scale dark matter halo (six free parameters), a truncated spherical isothermal mass profile for each cluster galaxy member (two free parameters with a constant M/L scaling relation for the members), and an external shear (two free parameters).
The model constraints are the positions of multiple images and (when available) spectroscopic redshifts.
For families of multiple images with no spectroscopic measurements, we leave the redshift values free to vary in the model.
We built two versions of the lens model: The first uses information from \hst and MUSE only (i.e. data available prior to the \jwst ERO release).
The second model was built upon \jwst NIRCam imaging, which allowed us to confirm 30 additional, secure multiple images from 11 individual sources.
Naturally, these new identifications are extremely faint or not detected in the \hst optical and near-infrared imaging.
We carefully checked the MUSE datacube around these positions, but as expected, no optical emission was detected.
In the following sections we describe the two models we present in this work.

\begin{figure}
      \includegraphics[width = 1.0\columnwidth]{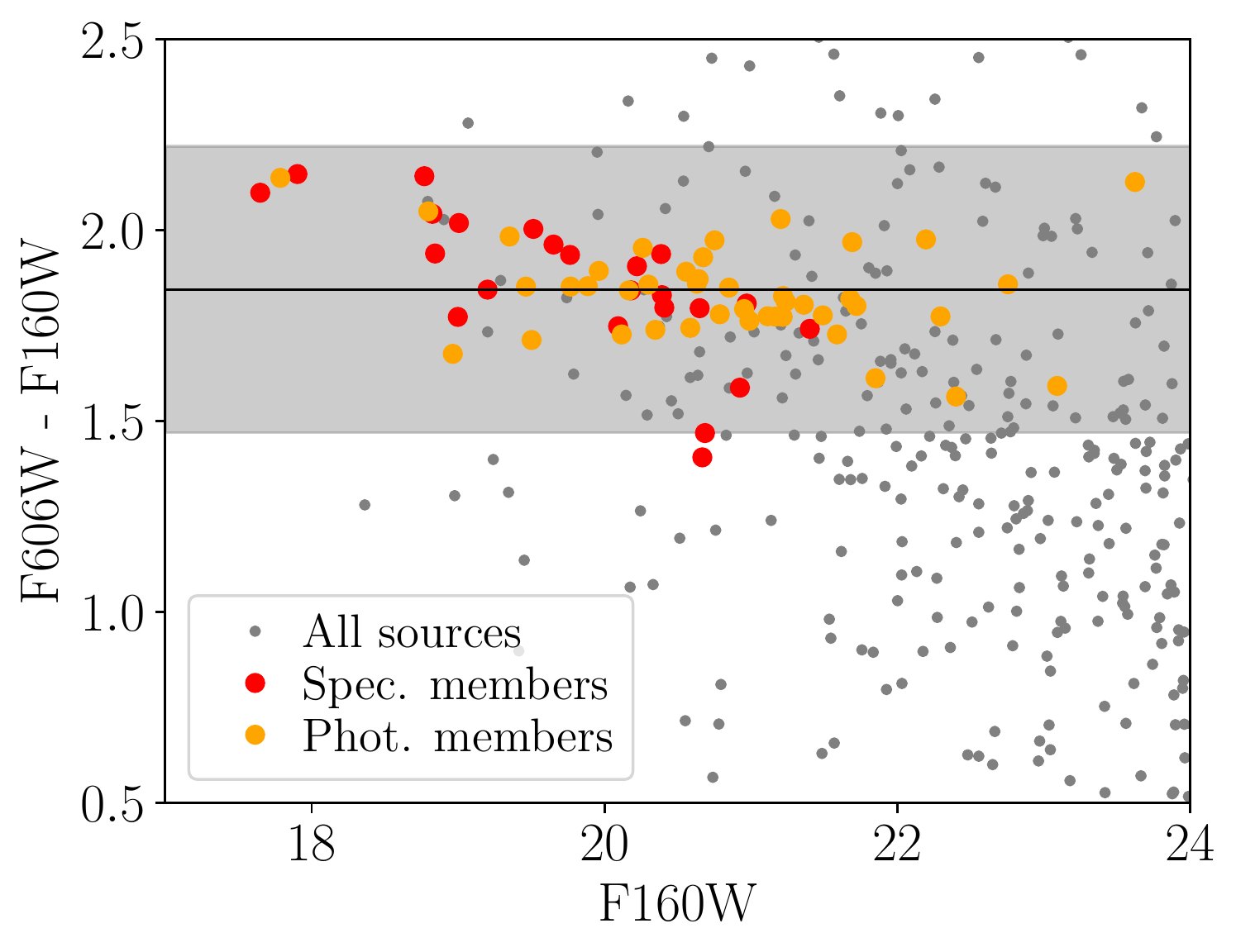}
  \caption{Colour-magnitude diagram of galaxies in the field of view of \smacs.  The spectroscopically confirmed cluster members (red points) define the red sequence. The colour and photometric redshift distribution of spectroscopically confirmed members are used to select 38 additional cluster members with no spectroscopic information (orange points) down to a magnitude of 24 in F160W.}
  \label{fig:color_mag}
\end{figure}

\subsection{Pre-\jwst lens model}
\label{sec:pre_model}
Before the release of \jwst/NIRCam imaging, the best dataset available for SMACS~J0723 was photometry from the RELICS program and spectroscopy from MUSE (see Sects. \ref{sec:HST_data} and \ref{sec:MUSE_data}).
From the spectroscopic data, we confirm a total of 23 cluster members in the redshift range of $0.367 - 0.408$, corresponding to a rest-frame velocity of $\pm 3000$~km/s from the cluster mean redshift of z=0.387.
However, the spectroscopically confirmed members are limited to the MUSE field of view of 1 arcmin$^2$.
We computed the 68\% confidence intervals from the colour and photometric redshift distributions of spectroscopically confirmed members and used these intervals to select additional cluster members with no spectroscopic information.
In Fig. \ref{fig:color_mag} we show the colour-magnitude diagram for the \hst filters F606W and F160W, where the red sequence is clearly defined by the spectroscopically confirmed members.
In total, we selected 38 photometric members down to a magnitude of 24 in F160W.

From the MUSE spectroscopy, we confirmed the redshifts of four families of multiple images.
In Table \ref{tab:multiple_images} we list all the positions and redshifts of multiple images used as constraints in the lens model.
In this first version of the model, two families have no secure spectroscopic confirmations, but are clearly detected in \hst and have colours and parities as expected in strong lensing.
Our pre-\jwst model thus has 19 multiple images as input from six families, and only two of the redshifts are free to vary in the model.

The mass profiles we adopted are the pseudo-isothermal elliptical parameterisation \citep[PIEMD;][]{1993ApJ...417..450K}, to describe the extended cluster mass (dark matter), and the dual pseudo-isothermal mass profile with axial symmetry \citep{2007arXiv0710.5636E, 2010A&A...524A..94S}, to describe the 61 cluster members.
As commonly done in the literature, we assumed a constant mass-to-light ratio for the cluster members.
With this, the number of free parameters describing these components is reduced to two normalisations, namely $\sigma_{\rm v,gals}^{\rm norm.}$ and $r_{\rm cut, gals}^{\rm norm.}$.
Finally, we also considered an external shear component to account for external massive perturbers that can affect the light deflection of background sources.

Our fiducial cluster mass model was obtained following the procedure in \citet{2019A&A...632A..36C}, in order to find the combination of mass profiles that can reproduce the position of multiple images well while not overfitting the data.
We tested a range of combinations of mass profiles, from the simplest one, composed of a single PIEMD halo profile plus cluster members, to models with additional cluster halo components, external shear, and individual cluster members with mass parameters varying freely outside the mass-to-light scaling relation.
We find that the parameterisation with the smallest values of the Bayesian information criterion \citep{schwarz1978} and Akaike information criterion \citep{1974ITAC...19..716A} is the one with ten parameters for the mass distribution (as described at the beginning of Sect. \ref{sec:lens_model}) and the two unknown redshifts of families 1 and 4 (see Table \ref{tab:multiple_images}).
In this model, the best-fit offset between the multiple observed images and the model predicted positions is $\delta_{\rm rms} = 0\farcs39$.
We note that our best-fit $\delta_{\rm rms}$ is significantly lower than that previously obtained by \citet{2022arXiv220705007G}, who quoted a value of $\delta_{\rm rms}=2\farcs3$.

\subsection{\jwst lens model}
\label{sec:pos_model}

With the new \jwst/NIRCam imaging, at wavelengths not reachable with \hst and the high spatial resolution, we were able to significantly improve our lens model constraints.
We carefully inspected the near-infrared imaging in order to find new multiple images that were not previously identified.
Thanks to the colour information and resolution, we obtained a sample of 30 new secure multiple images from 11 background galaxies, increasing by a factor of two the number of model constraints compared to the pre-\jwst (\hst and MUSE only) lens model.

Given that these sources are very faint in observed optical wavelengths, none have spectroscopic confirmation from MUSE and, in most cases, they have no clear \hst counterpart.
In Table \ref{tab:multiple_images} we list all new multiple images as well as their model redshifts.
Moreover, in Fig. \ref{fig:cut_outs} we show the \jwst NIRCam cutouts around all multiple images used in our lens model.

The lens model with the new multiple images has a best-fit $\delta_{\rm rms}$ of $0\farcs51$.
Considering that the number of model constraints is around twice that of the pre-\jwst model, this small increase in $\delta_{\rm rms}$ is expected.
Nonetheless, it is significantly lower than what was reported in previous published works on SMACS~J0723.
In Fig. \ref{fig:JWST_image} we show the critical line for a source at $z=10$ overlaid on the \jwst/NIRCam imaging.
The lens model, including magnification maps and \texttt{lenstool} configuration files, are available in the electronic version of this manuscript and upon request to the authors.
We encourage the community to further explore the lens model presented in this work.

In Table \ref{tab:model_params} we list the constrained total mass parameters, and in Fig. \ref{fig:mass_profiles} we show the total mass and density profiles.
We find a mass of $8.7_{-0.2}^{+0.2} \times 10^{13} \rm M_{\odot}$ at a radius of 128~kpc, approximately the Einstein radius for a source at $z \rightarrow \infty$.
At this radius, this mass estimate is $51\%$ more accurate when compared to our pre-\jwst lens model.

One special case is family 8.
Images 8.1 and 8.2, shown in Fig. \ref{fig:cut_outs}, are secure multiple images thanks to their colours and morphology.
They are composed of two main clumps and some smaller possible substructures.
Image \#8.3 is very likely a counter-image, showing similar colour and the double-clump morphology.
In order to avoid any bias in our lens model, we did not include \#8.3 as an observational constraint.
The total lensing magnifications are $6.1_{-0.6}^{+0.7}$, $9.5_{-0.8}^{+0.9}$, and $4.5_{-0.4}^{+0.5}$ for 8.1, 8.2, and \#8.3, respectively.
The multiple images are located in the easternmost region of the cluster (see Fig. \ref{fig:JWST_image}).
This makes it difficult to precisely constrain its redshift from the model.
However, the lens model indicates it might be at $z>7.5$ from the posterior probability distribution.

\subsection{Comparison to models in the literature}
\label{sec:comparison}

\citet{2022arXiv220707101M} and \citet{2022arXiv220707102P} also used the new \jwst observations, particularly the imaging, to model \smacs.  The model by \citet{2022arXiv220707102P} was built upon the model of \citet{2022arXiv220705007G}, which was based on \hst observations and created before the \jwst ERO. We briefly compare our model ingredients and results to these studies.

All studies identified new families of multiple images in the NIRCam imaging: 16 in \citet{2022arXiv220707101M}, 14 in \citet{2022arXiv220707102P}, and 11 in this work. Since no families have spectroscopic redshift confirmations (except one identified by \citealt{2022arXiv220707101M} with NIRSpec), we deliberately used only highly secure families of multiple images (in Table \ref{tab:multiple_images}) to constrain our mass model in order to avoid misidentifications of families that could bias our mass model parameters. This explains our lower number of reported families. 

All three models are parametric in having a cluster halo component with cluster members, and we have a further external shear component that is not present in the other two studies.  Our robust selection of 61 cluster members is based on both MUSE spectroscopy and \hst photometry, whereas \citet{2022arXiv220707101M} and \citet{2022arXiv220707102P} relied on \hst photometry to identify $\approx 130$ members (which will be refined in future work by including spectroscopy).

\citet{2022arXiv220707102P} have posted three versions of their model with  measured photometric redshifts based on the NIRCam photometry and also predicted the redshifts based on their mass model. The first version from \citet{2022arXiv220707102P}\footnote{\url{https://arxiv.org/abs/2207.07102v1}} produced ten redshift estimates for the new \jwst lensed sources, as did our model, although only six of the ten are the same.  While the model redshifts from both models for five of these sources are all in the range $\approx 1-3$, only one of the five agrees within the 1$\sigma$ uncertainty. The sixth source is our family 7 at $z>7.5$, whereas its redshift is unconstrained by \citet{2022arXiv220707102P}. The model redshifts are therefore highly model dependent and rely on the goodness of the model fits.  Future spectroscopic redshift measurements of some of these sources would be valuable.
The third version of \citet{2022arXiv220707102P} was posted after the release of our model and now has nine redshift estimates in common with ours, all agreeing within the $1\sigma$ uncertainty.

Even though a direct comparison of the different models is not straightforward given the different families of multiple images used, the $\delta_{\rm rms}$ of the models provides an indication of the goodness of the model fit.  Our $\delta_{\rm rms}$ of $0\farcs51$ is a factor of $\approx 2$ lower than the $1\farcs08$ from \citet{2022arXiv220707101M} and the  $0\farcs93$ from the first version of \citet{2022arXiv220707102P}.
We attribute our better fitting model to (1) our use of only the secure families of multiple images and (2) our robust cluster member selection based on both spectroscopy and photometry throughout our work since the beginning. 
The third version of \citet{2022arXiv220707102P} quotes $\delta_{\rm rms}=0\farcs48$, similar to the value obtained by our model.

\section{Conclusions}
\label{sec:conclusions}

We present our strong lens model of \smacs, the first gravitational lens ever observed by \jwst, using newly available data from NIRCam.
With the new data, we have doubled the number of model constraints, increasing the number of multiple images from the 19 identified with \hst and MUSE data to 49.
This demonstrates the unique capabilities of \jwst in discovering faint and/or high-redshift galaxies that were previously undetected in \hst observations.
Our lens model is capable of reproducing the positions of all multiple images with a high accuracy of $\delta_{\rm rms} = 0\farcs51$.
We report the lens model redshifts for 12 lensed galaxies, most of which are very faint; it remains challenging to obtain spectroscopic information for them.
While most of the lensed galaxies have modelled redshifts $\lesssim$3, one object stands out as a high-redshift galaxy at $z>7.5$, with image 8.2 having a magnification of $|\mu| \approx 9.5_{-0.8}^{+0.9}$.
This might be the first galaxy with spatially resolved substructures at the epoch of re-ionisation of the Universe (see Fig. \ref{fig:cut_outs}) and an interesting target for follow-up deep spectroscopy.
We report spectroscopic redshifts from MUSE archival data and their lensing magnifications.
Finally, our lens models, including magnification maps and configuration files for the software \texttt{lenstool}, are made publicly available.

These first observations by \jwst of \smacs have proven to be a treasure trove, allowing us to reduce the uncertainties in our cluster mass measurements by $51\%$ even without incorporating any spectroscopic information on the newly identified lensed galaxies. We anticipate future \jwst observations of other strong lensing galaxy clusters to be transformative in both constraining cluster mass distributions and studying high-redshift galaxies.

\begin{acknowledgements}
The authors thank the anonymous referee for the useful comments on the manuscript.
We thank C.~Grillo, P.~Rosati and D.~Sluse for constructive discussions during the preparation of this work. GBC and SHS acknowledge the Max Planck Society for financial support through the Max Planck Research Group for SHS and the academic support from the German Centre for Cosmological Lensing.
This work is partially funded by the Deutsche Forschungsgemeinschaft (DFG, German Research Foundation) under Germany's Excellence Strategy -- EXC 2094 -- 390783311.
AM, PB, AA and EV acknowledge financial contributions by PRIN-MIUR 2017WSCC32 ``Zooming into dark matter and proto-galaxies with massive lensing clusters'' (P.I.: P.Rosati), INAF ``main-stream'' 1.05.01.86.20: ``Deep and wide view of galaxy clusters'' (P.I.: M. Nonino) and INAF ``main-stream'' 1.05.01.86.31 ``The deepest view of high-redshift galaxies and globular cluster precursors in the early Universe'' (P.I.: E. Vanzella).
AA has received funding from the European Union’s Horizon 2020 research and innovation programme under the Marie Skłodowska-Curie grant agreement No 101024195 - ROSEAU.

\end{acknowledgements}

\bibliographystyle{aa}
\bibliography{references}

\begin{appendix}
\section{Supporting material}
\begin{figure}
      \includegraphics[width = 0.4975\columnwidth]{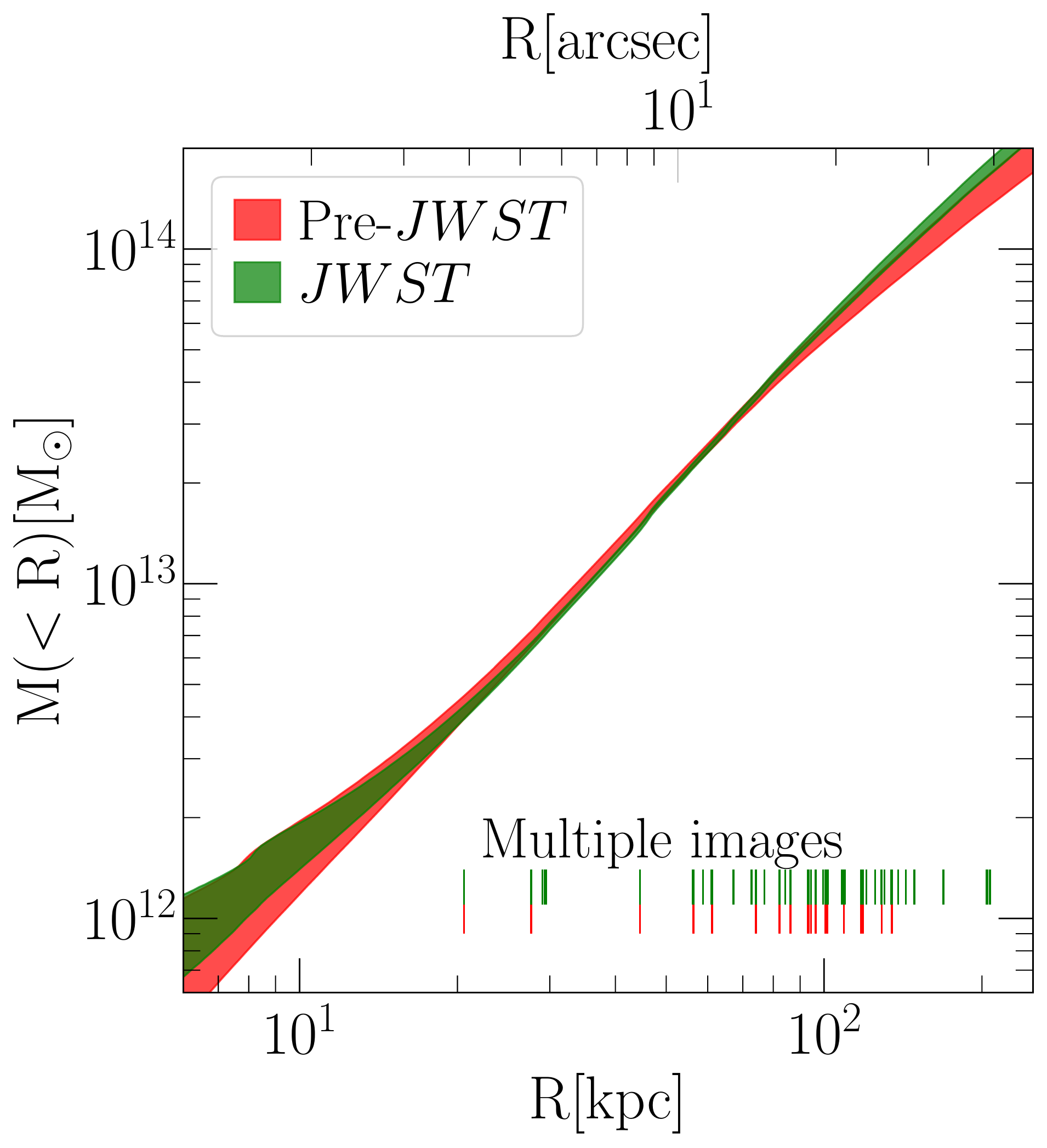}
      \includegraphics[width = 0.4825\columnwidth]{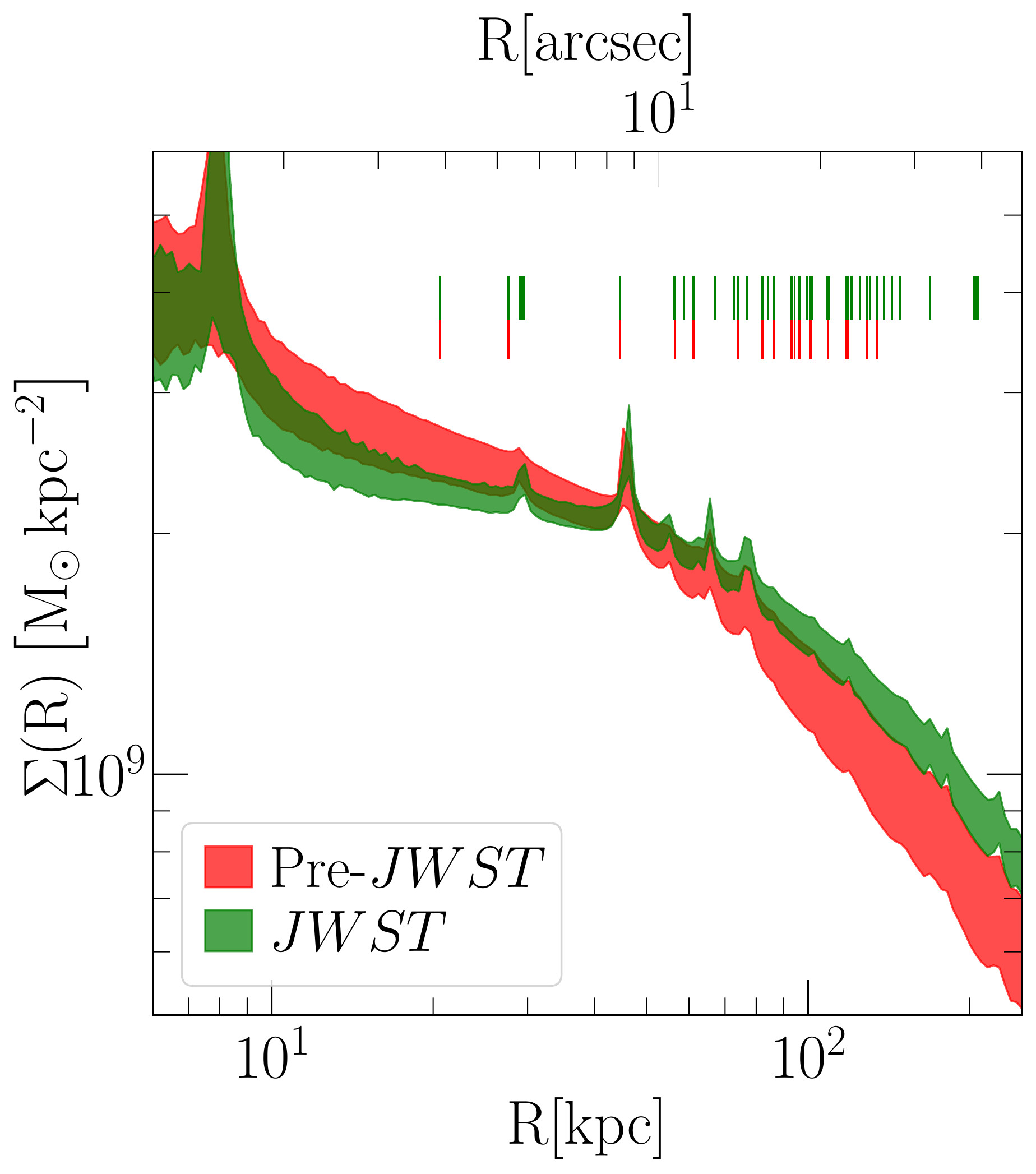}
  \caption{Total mass and density profiles of \smacs from our pre-\jwst (red) and \jwst (green) lens models. The regions correspond to the 95\% confidence, and the positions of multiple images are indicated by vertical lines.}
  \label{fig:mass_profiles}
\end{figure}

\begin{figure}
      \includegraphics[width = 1\columnwidth]{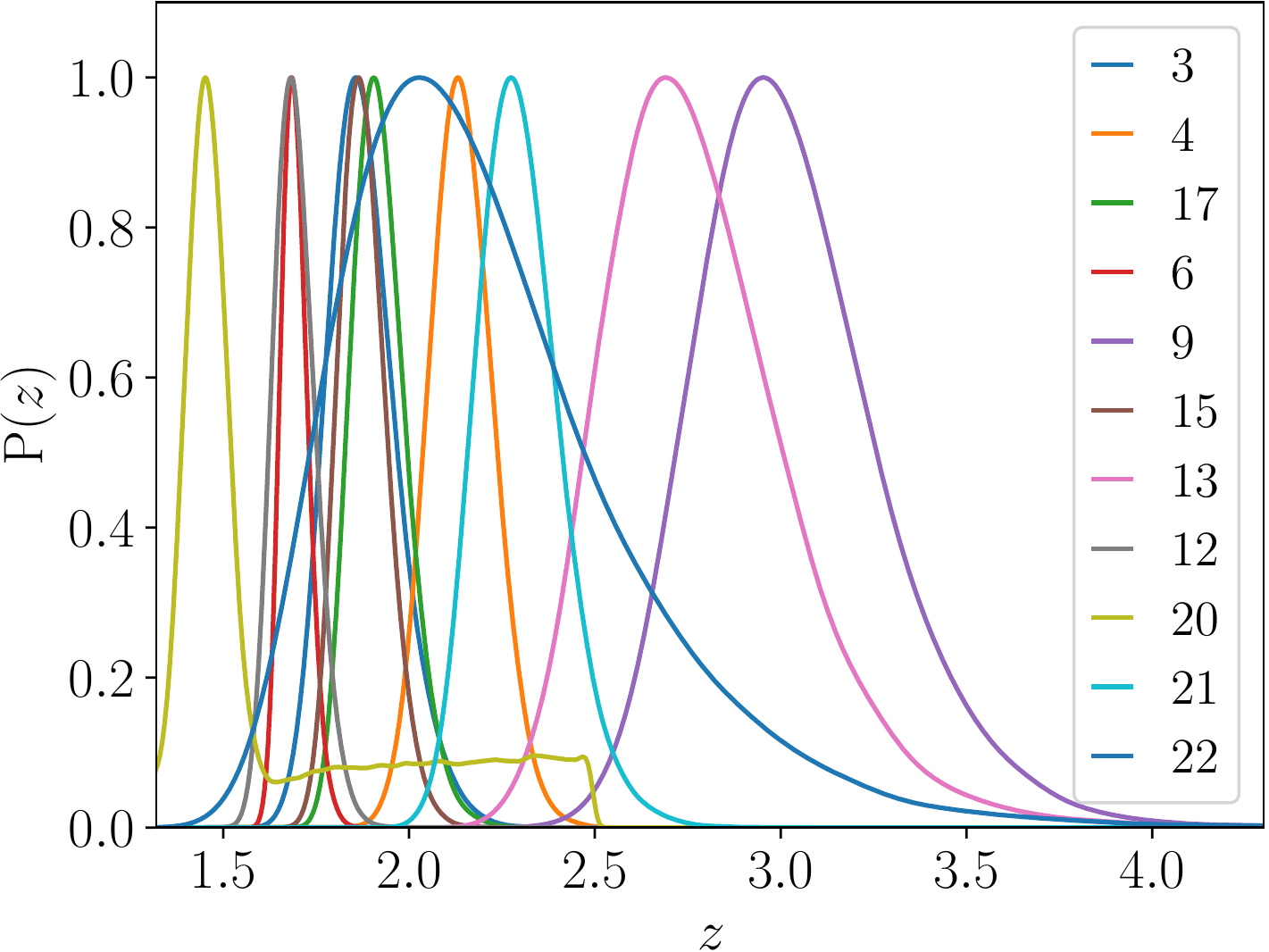}
  \caption{Probability distribution functions (PDF) for lens model redshifts for multiple image families. Median and 68\% confidence levels are listed in Table \ref{tab:multiple_images}. The PDF for family 8 is not shown in this figure for better visualisation of the other families' PDFs.}
  \label{fig:model_z}
\end{figure}

\begin{figure}
      \includegraphics[width = 0.4975\columnwidth]{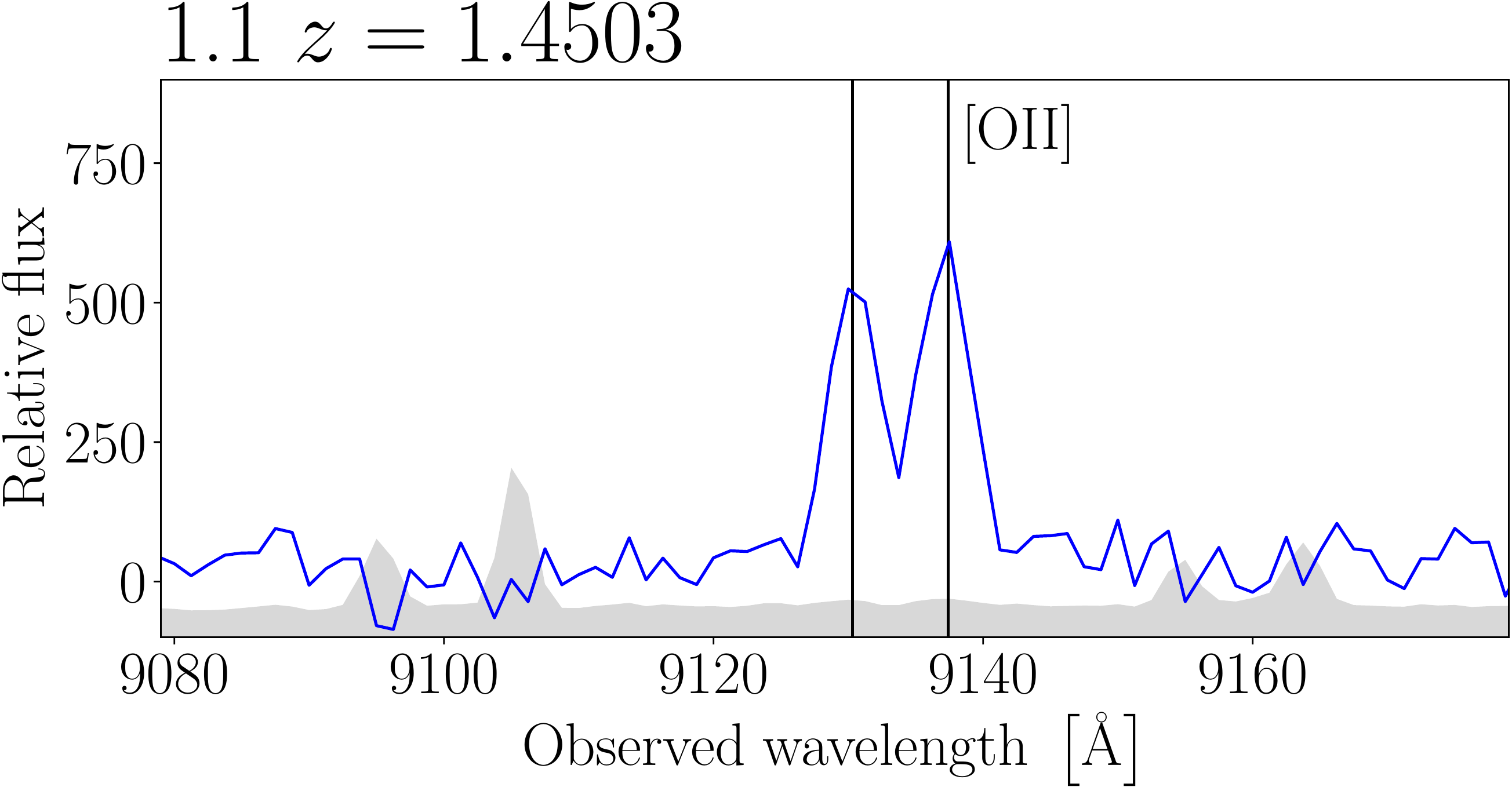}
      \includegraphics[width = 0.4825\columnwidth]{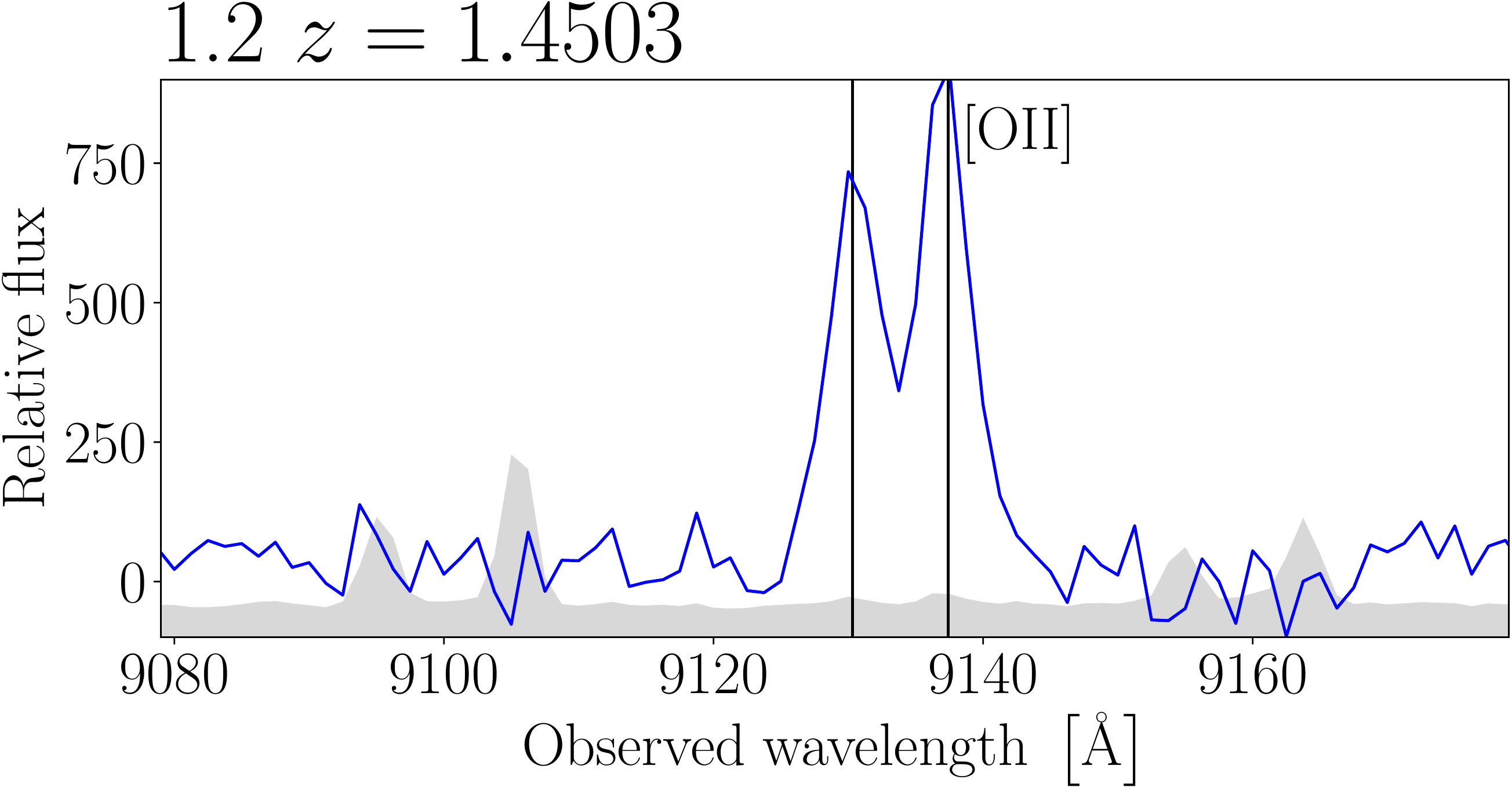}
      
      \includegraphics[width = 0.4975\columnwidth]{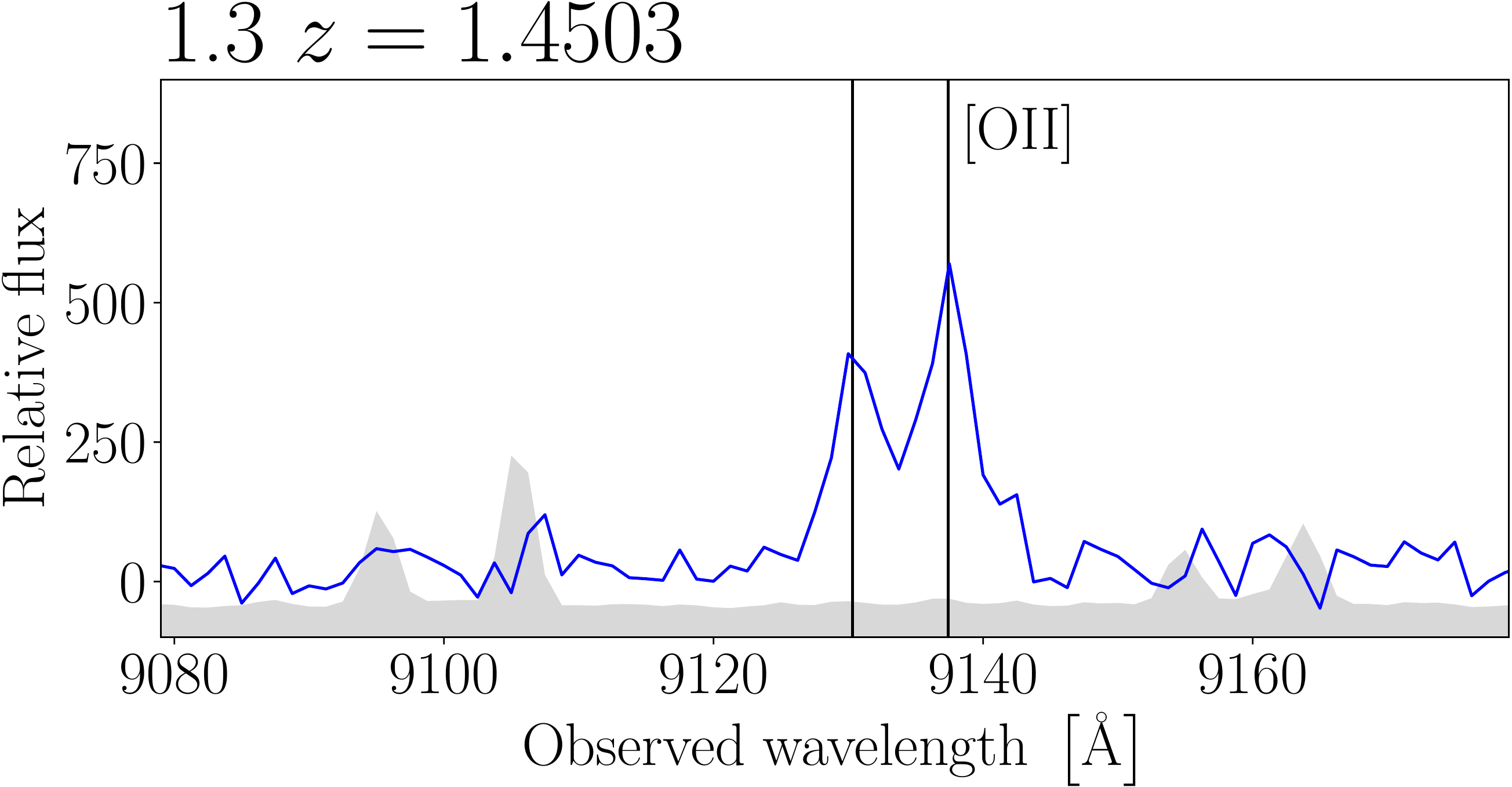}
      \includegraphics[width = 0.4825\columnwidth]{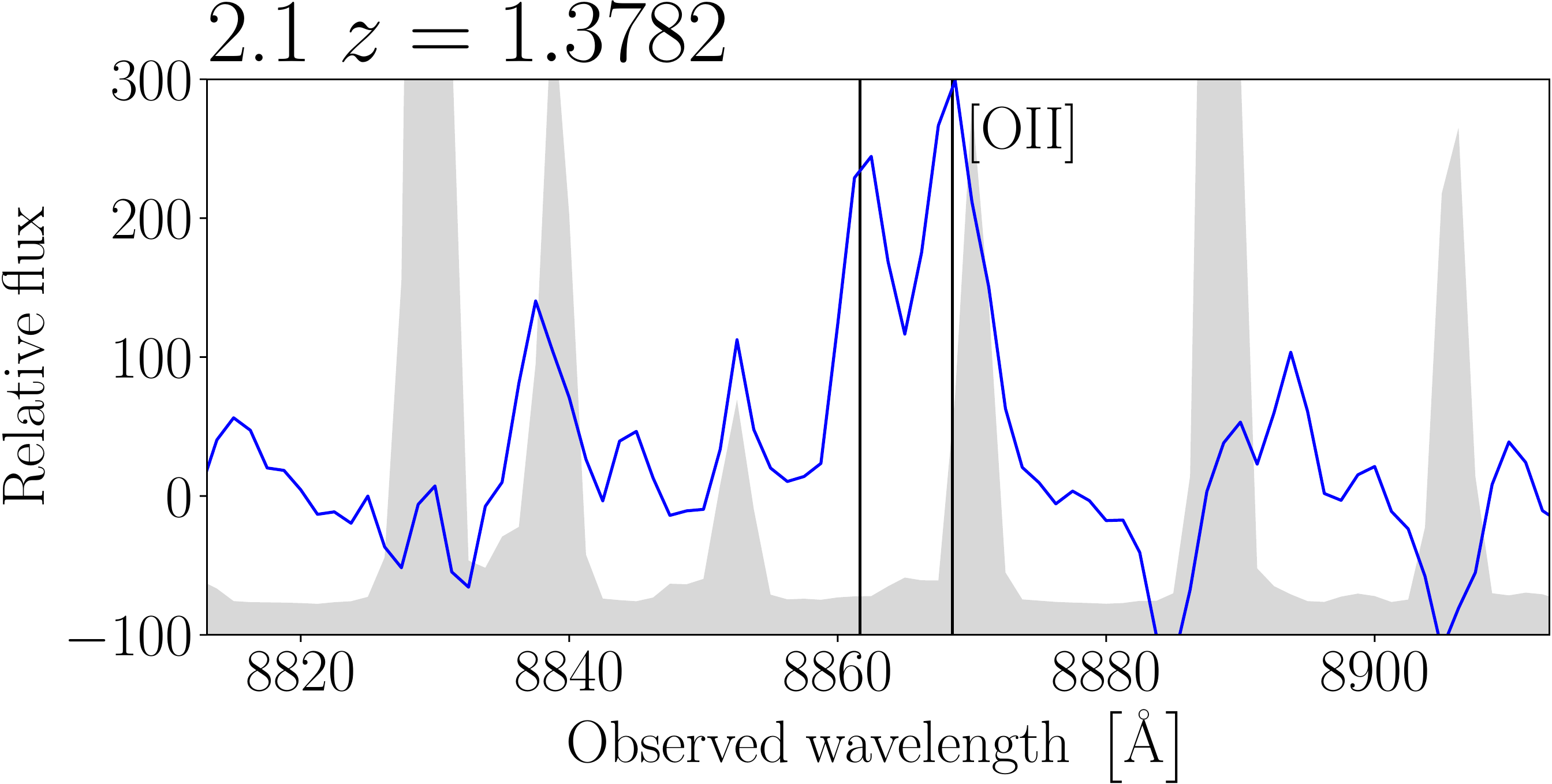}
      
      \includegraphics[width = 0.4975\columnwidth]{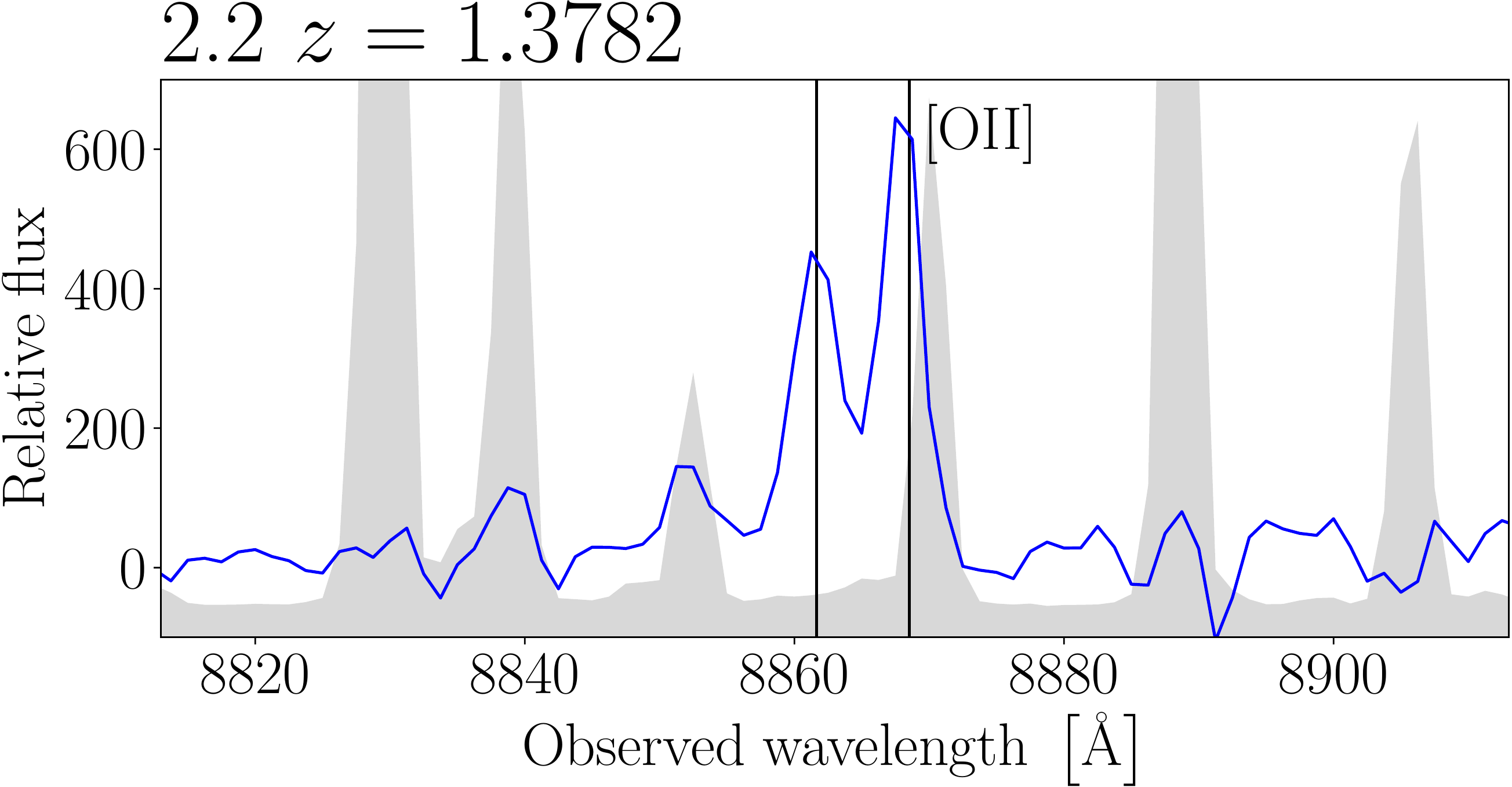}
      \includegraphics[width = 0.4825\columnwidth]{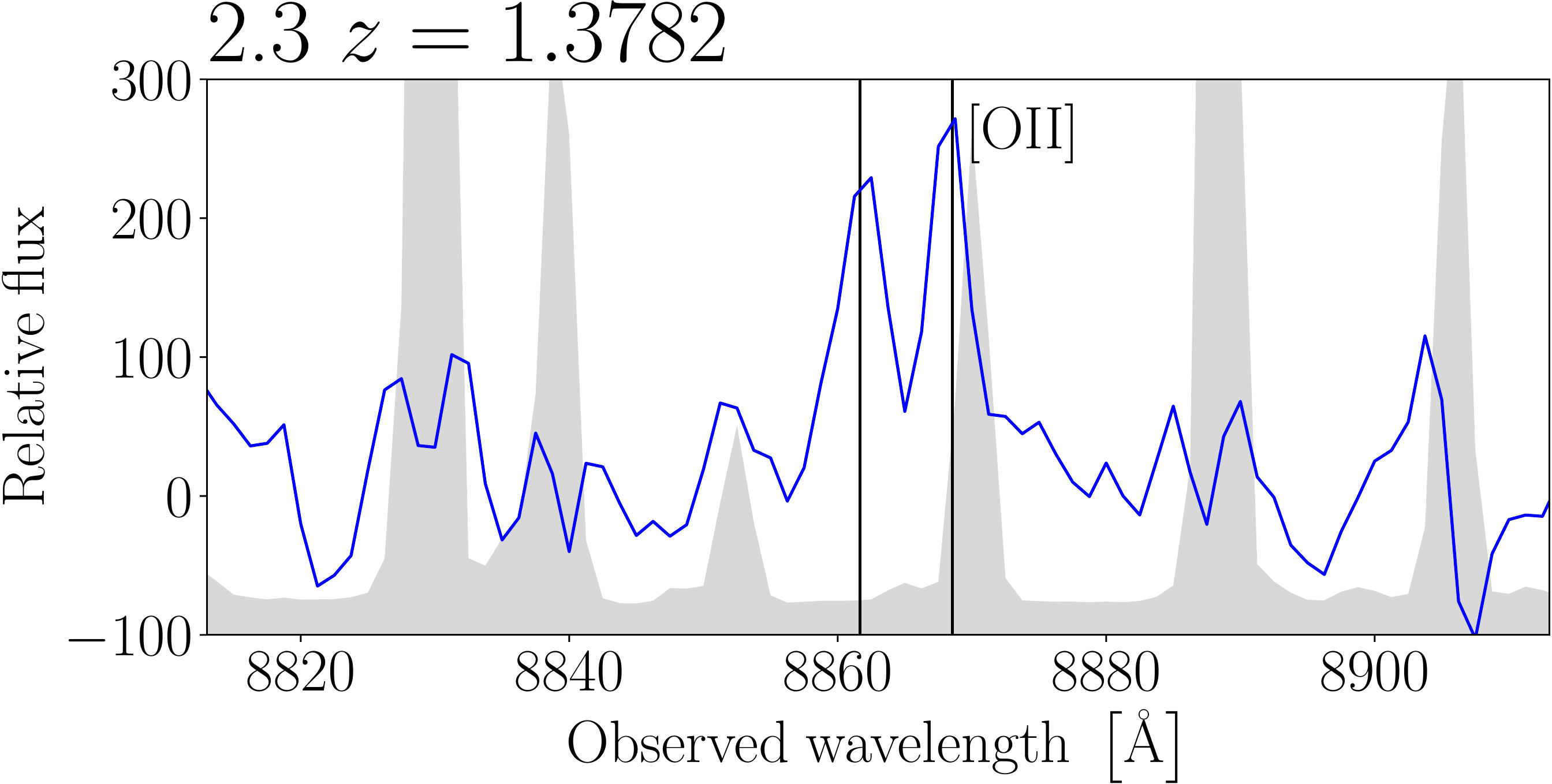}

      \includegraphics[width = 0.4825\columnwidth]{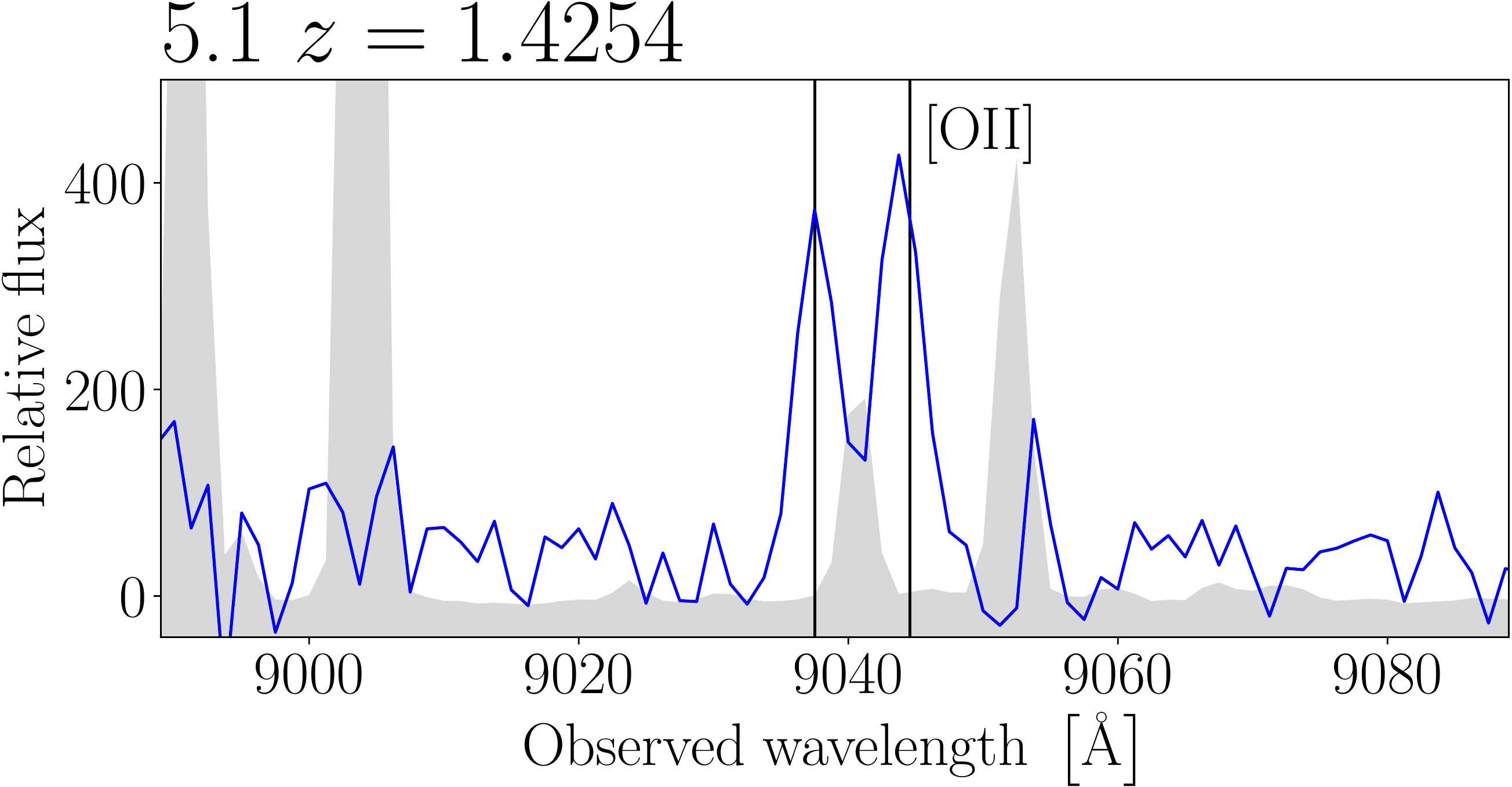}
      \includegraphics[width = 0.4825\columnwidth]{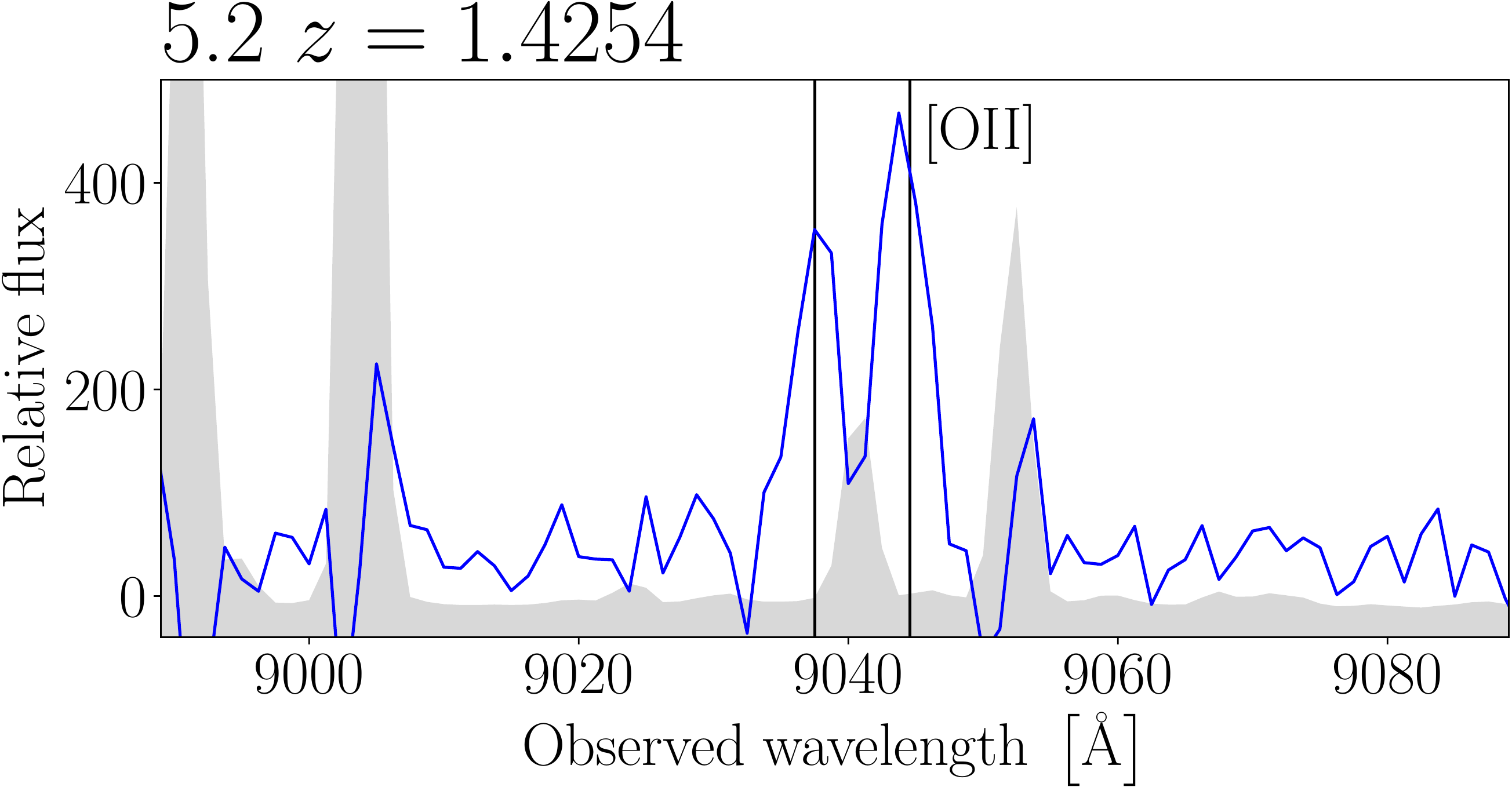}
      
      \includegraphics[width = 0.4975\columnwidth]{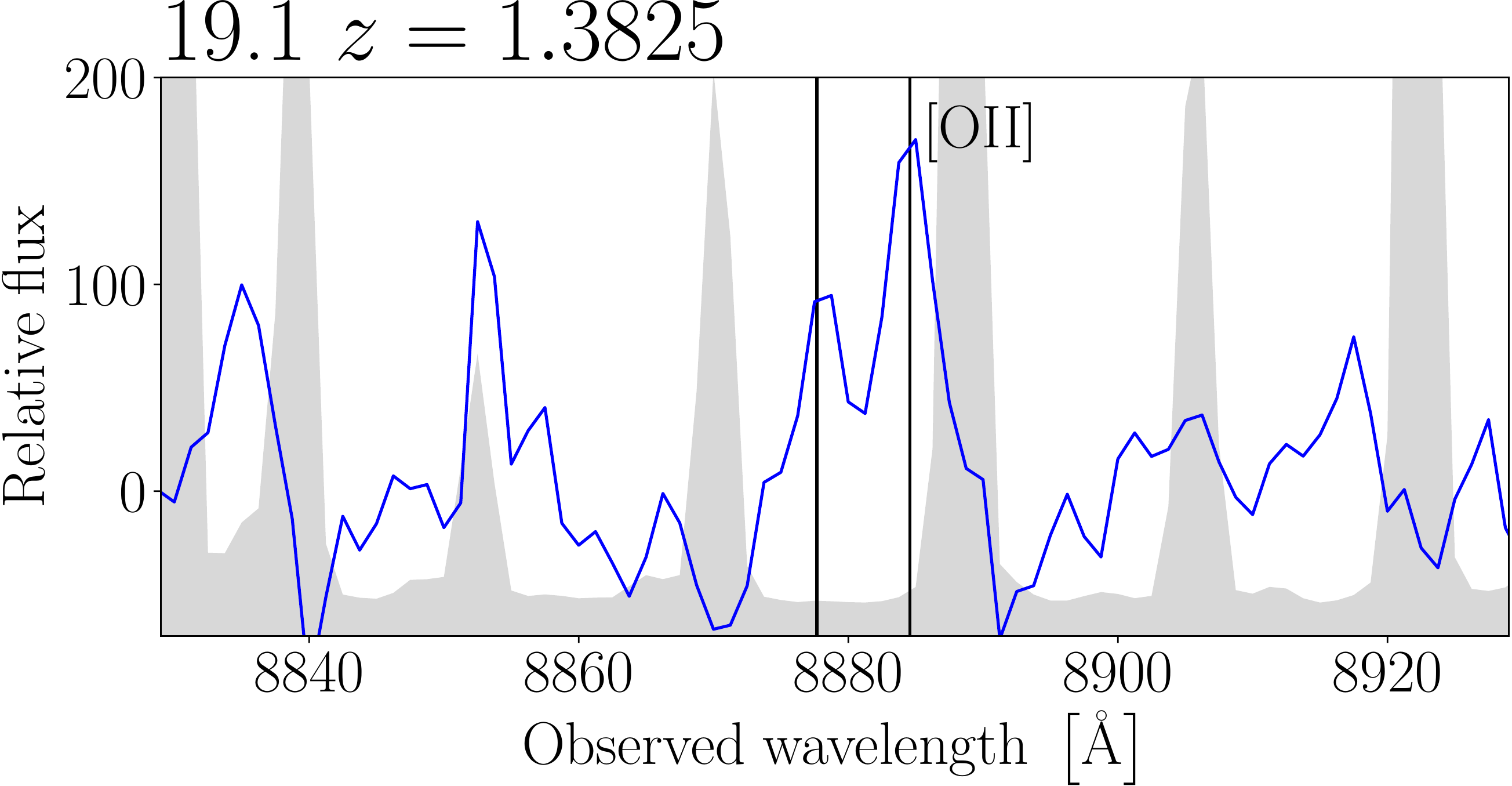}
      \includegraphics[width = 0.4825\columnwidth]{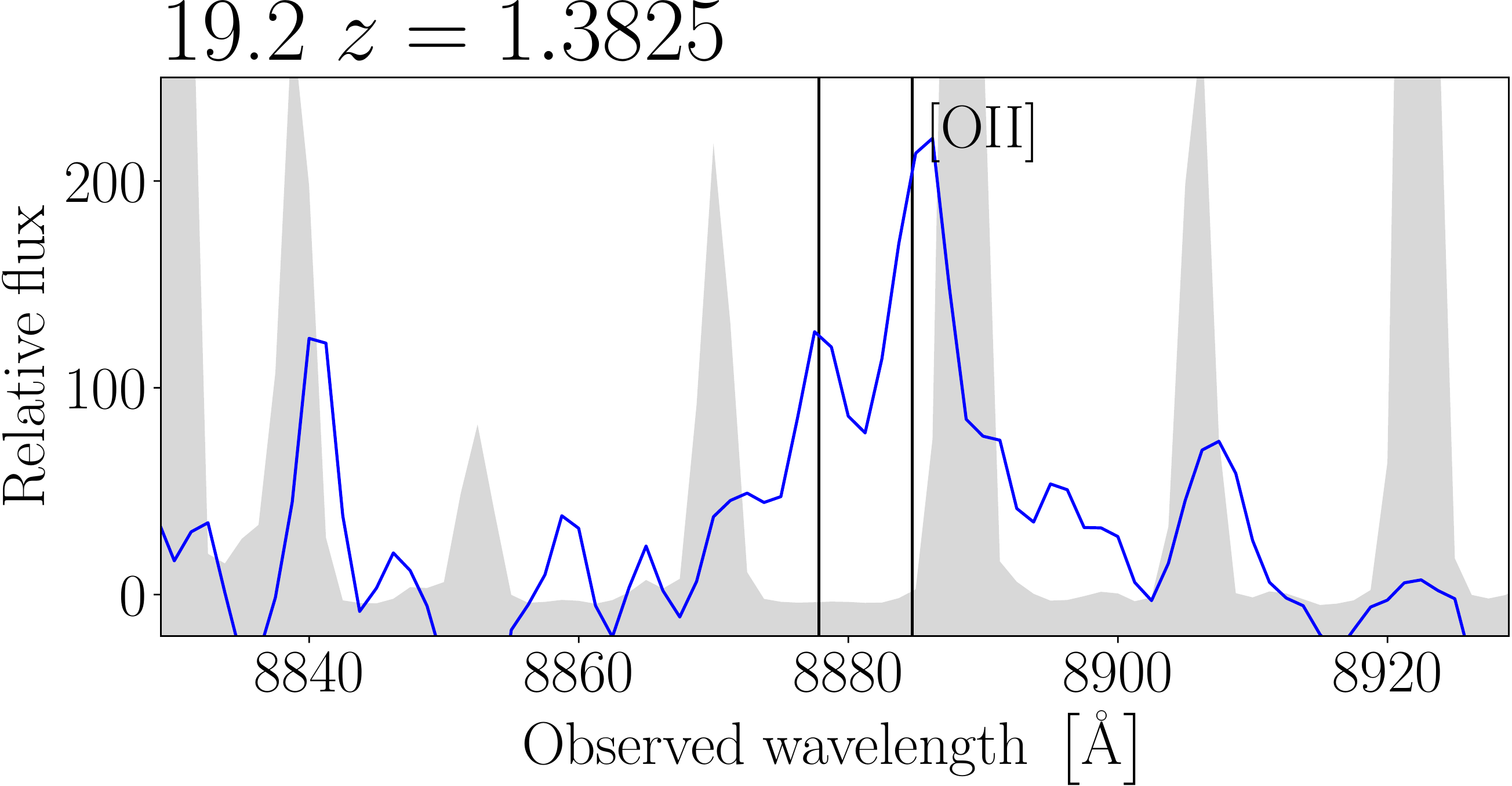}
      
      \includegraphics[width = 0.4975\columnwidth]{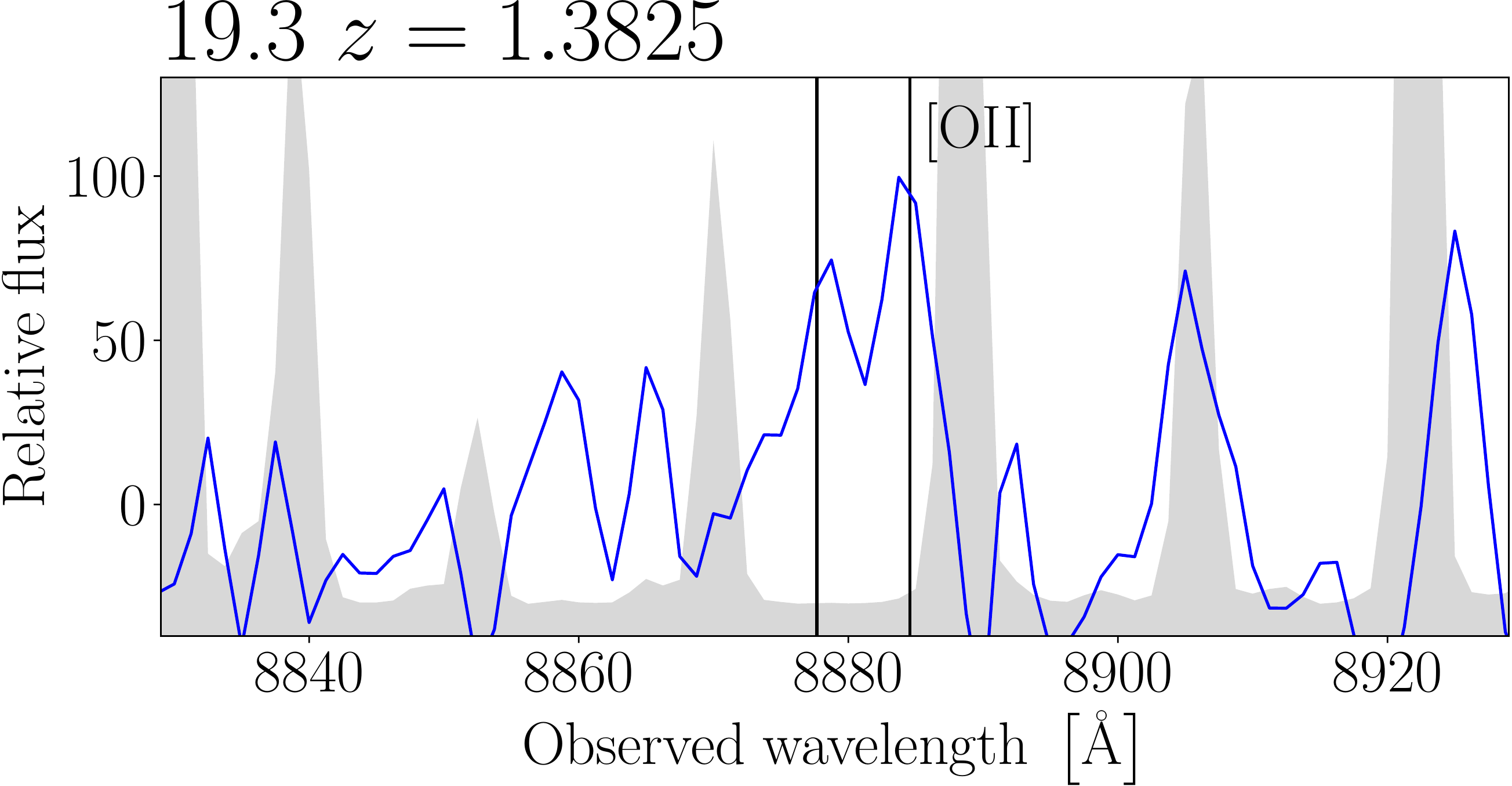}
  \caption{MUSE spectrum of confirmed multiple images. Blues lines are the spectra extracted from the data cube, and the shaded regions are the re-scaled variance. Vertical lines indicate the \ion{O}{II} emission at the source redshift.}
  \label{fig:spectra}
\end{figure}

In this appendix we present the supporting material of this work.
The full redshift catalogue from the MUSE data and the lensing magnification estimates from the \jwst model are presented in Table \ref{tab:muse_cat}.
In Table \ref{tab:model_params} we show the recovered mass model parameters for the \jwst lens model, and the total mass and density profiles in Fig. \ref{fig:mass_profiles}.
In Fig. \ref{fig:model_z} we present the normalised probability density function of the lens model-z for all multiple image families, except for family 8 because its redshift value is above 7.5 and otherwise unconstrained by our model.
Figure \ref{fig:cut_outs} shows the cutouts of all multiple images used in our lens models, and Fig. \ref{fig:spectra} the spectra of confirmed multiple images.

\begin{figure}
    \includegraphics[width = 0.328\columnwidth]{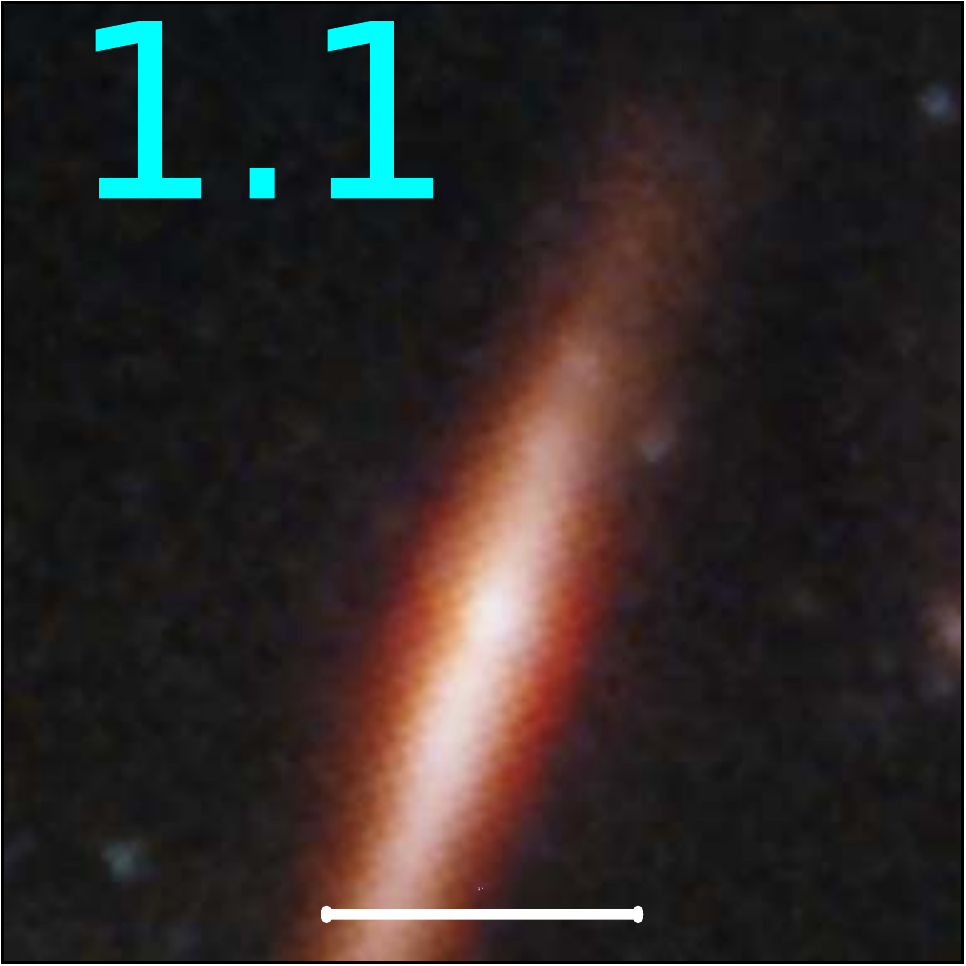}
    \includegraphics[width = 0.328\columnwidth]{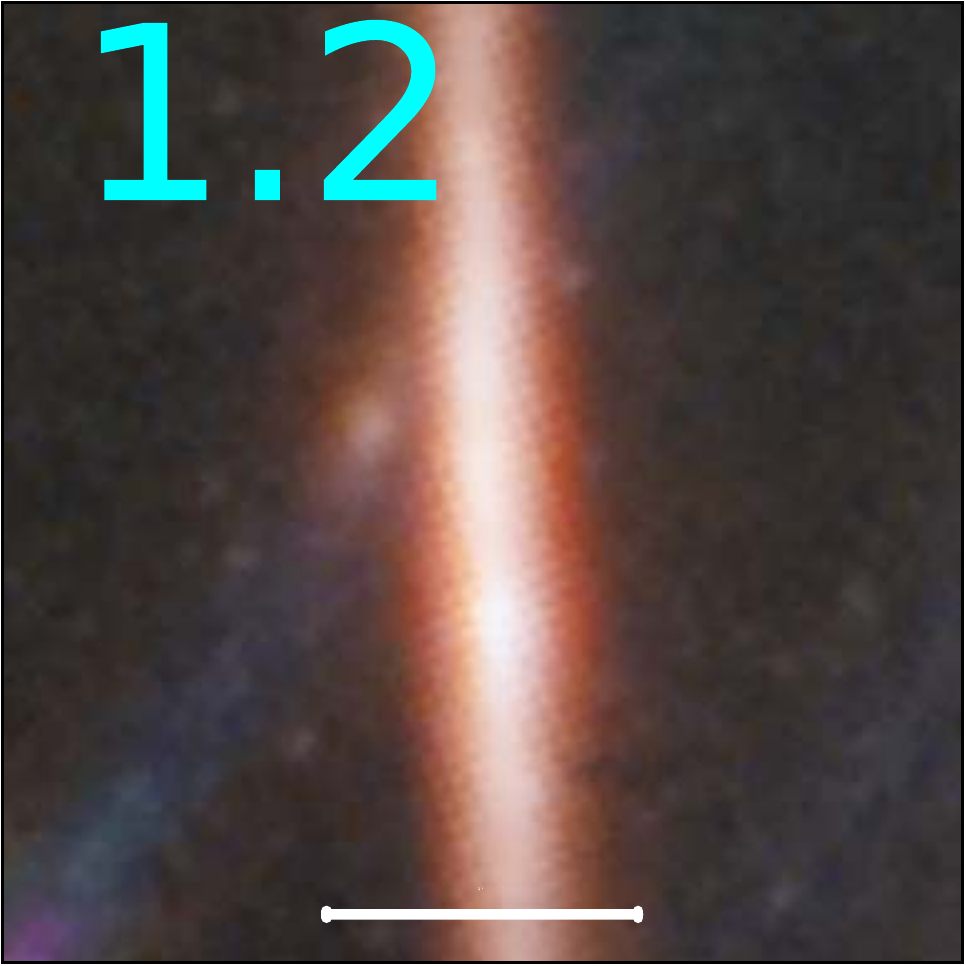}
    \includegraphics[width = 0.328\columnwidth]{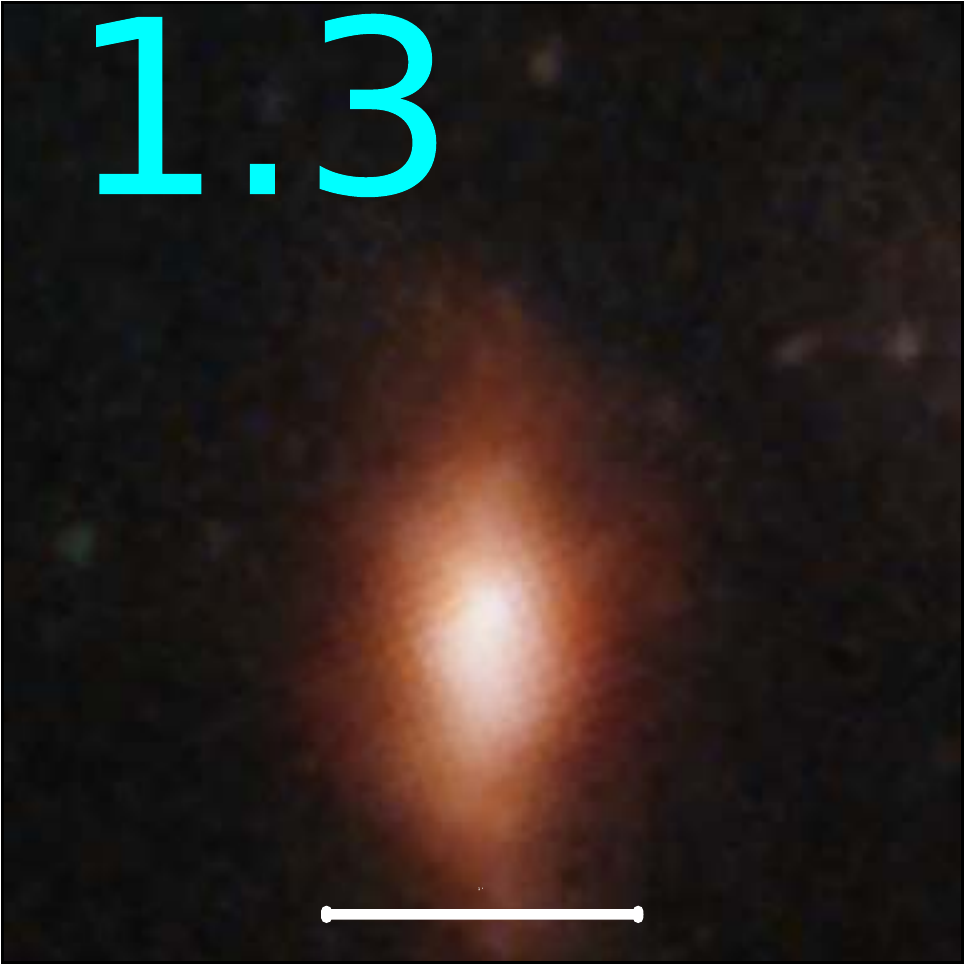}

    \includegraphics[width = 0.328\columnwidth]{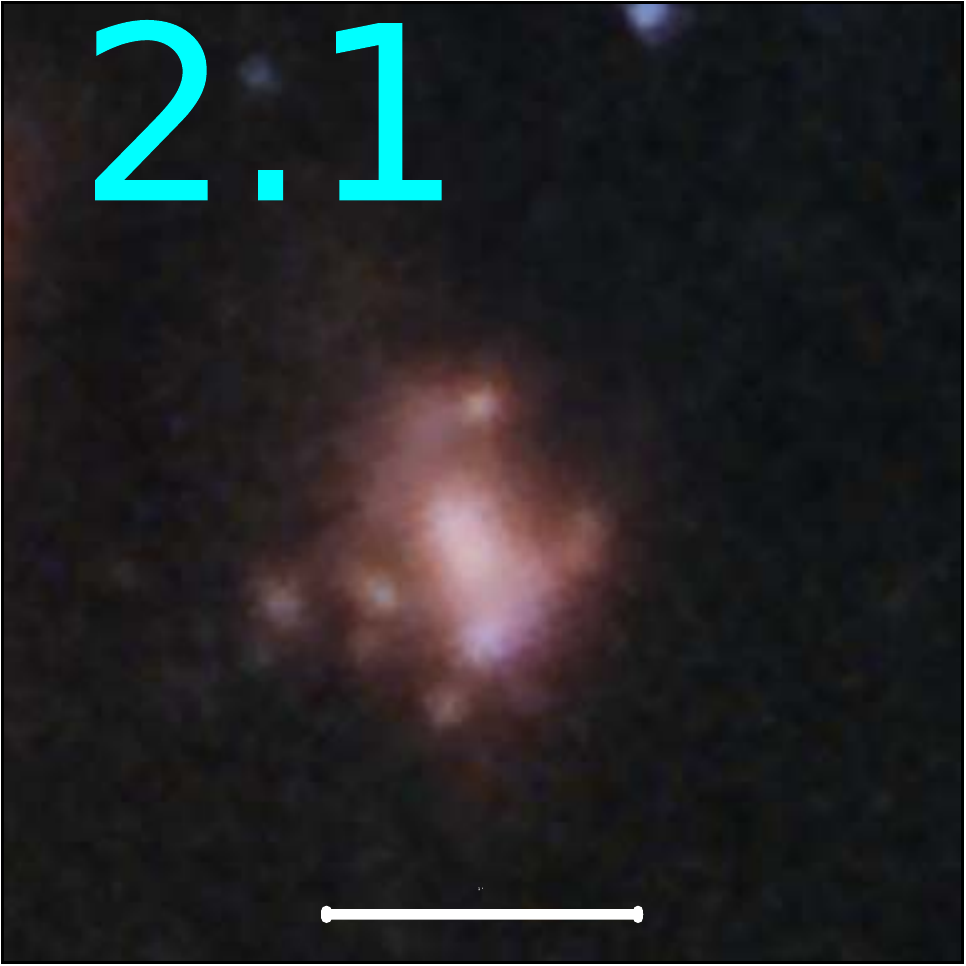}
    \includegraphics[width = 0.328\columnwidth]{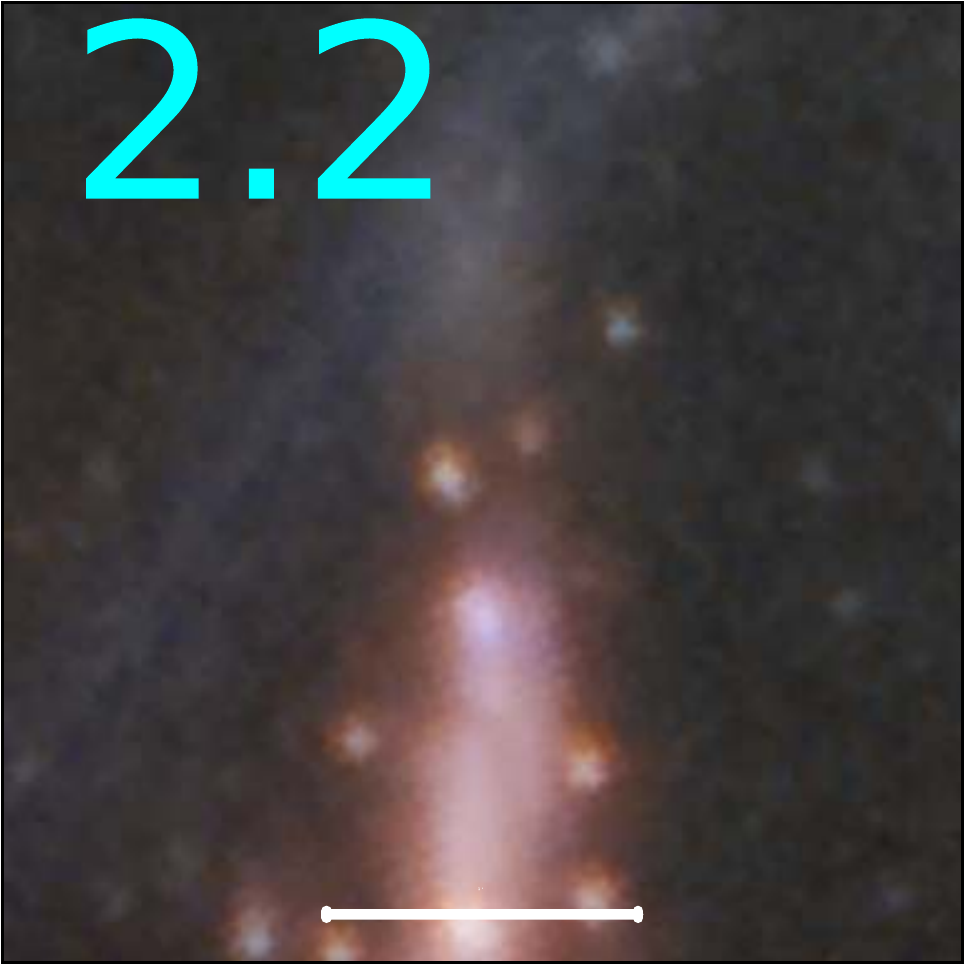}
    \includegraphics[width = 0.328\columnwidth]{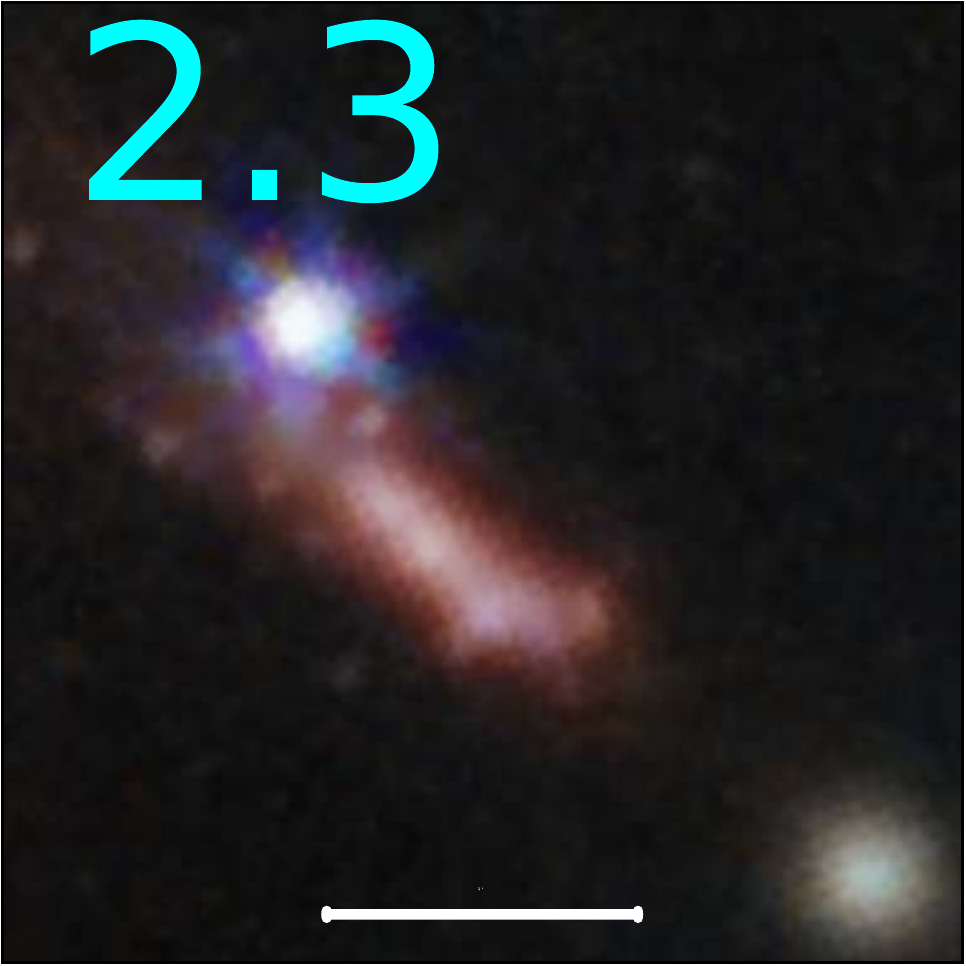}

    \includegraphics[width = 0.328\columnwidth]{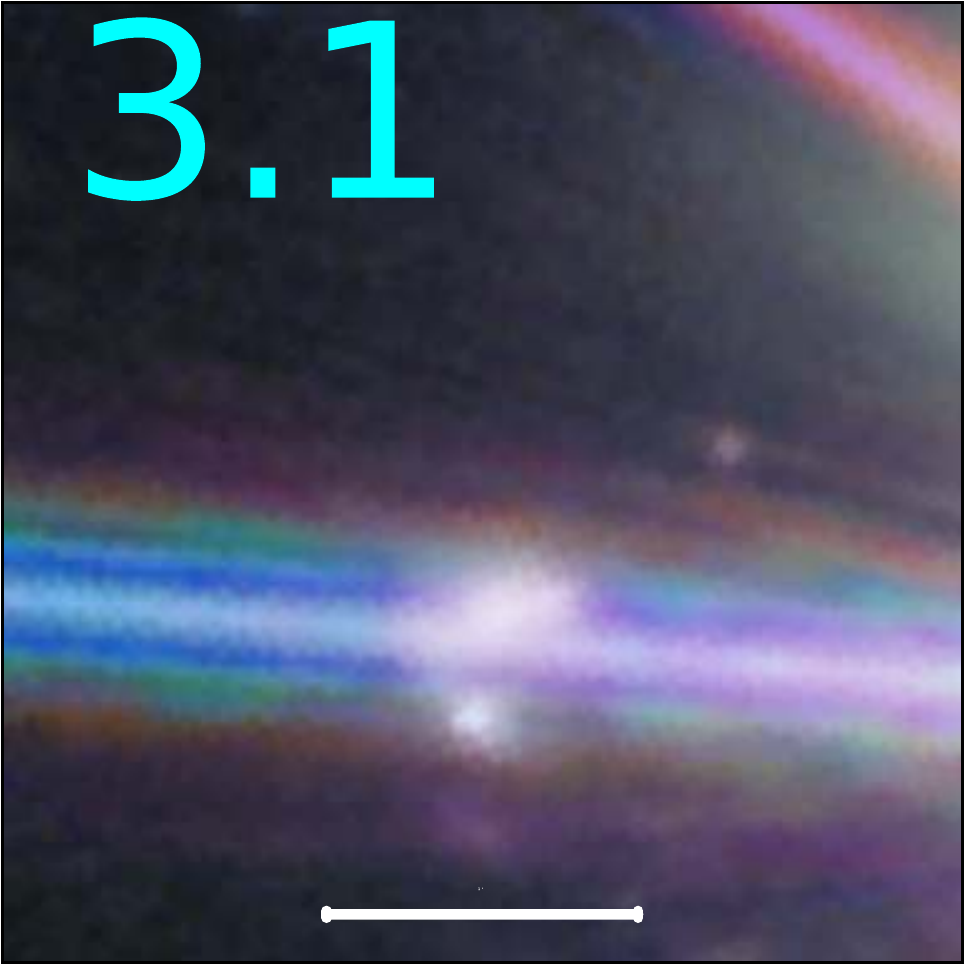}
    \includegraphics[width = 0.328\columnwidth]{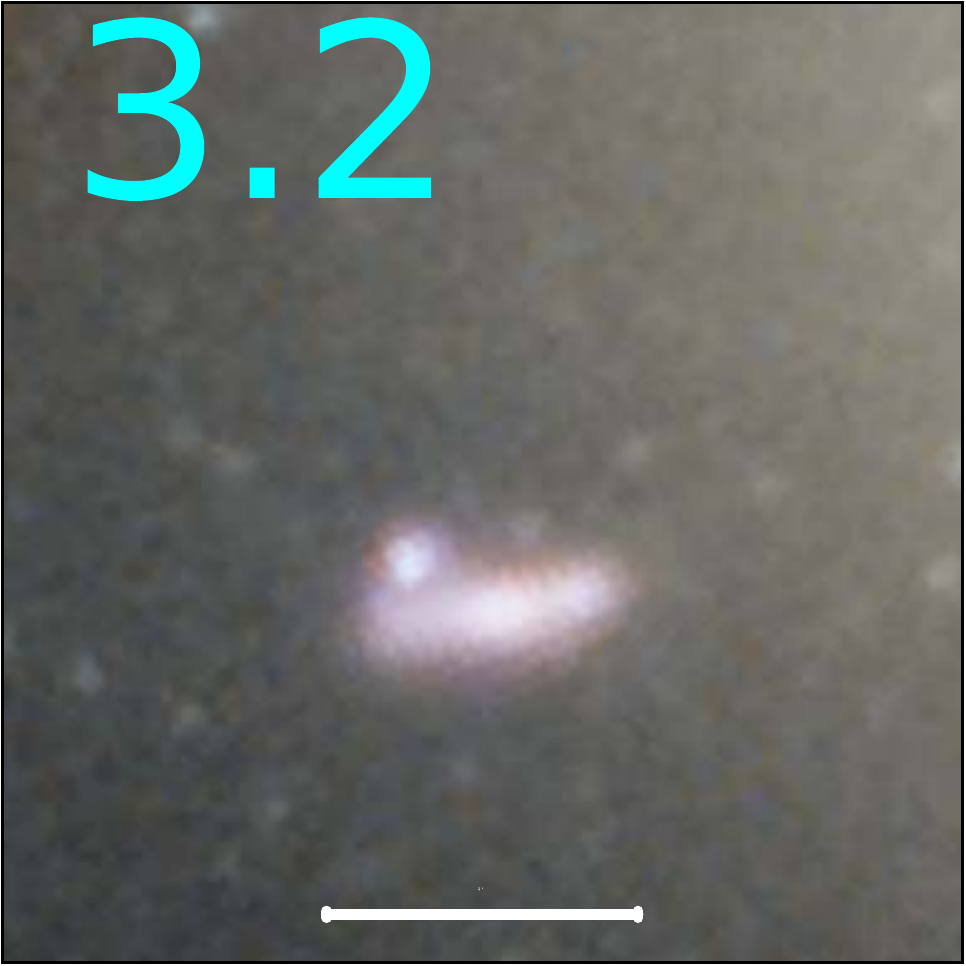}
    \includegraphics[width = 0.328\columnwidth]{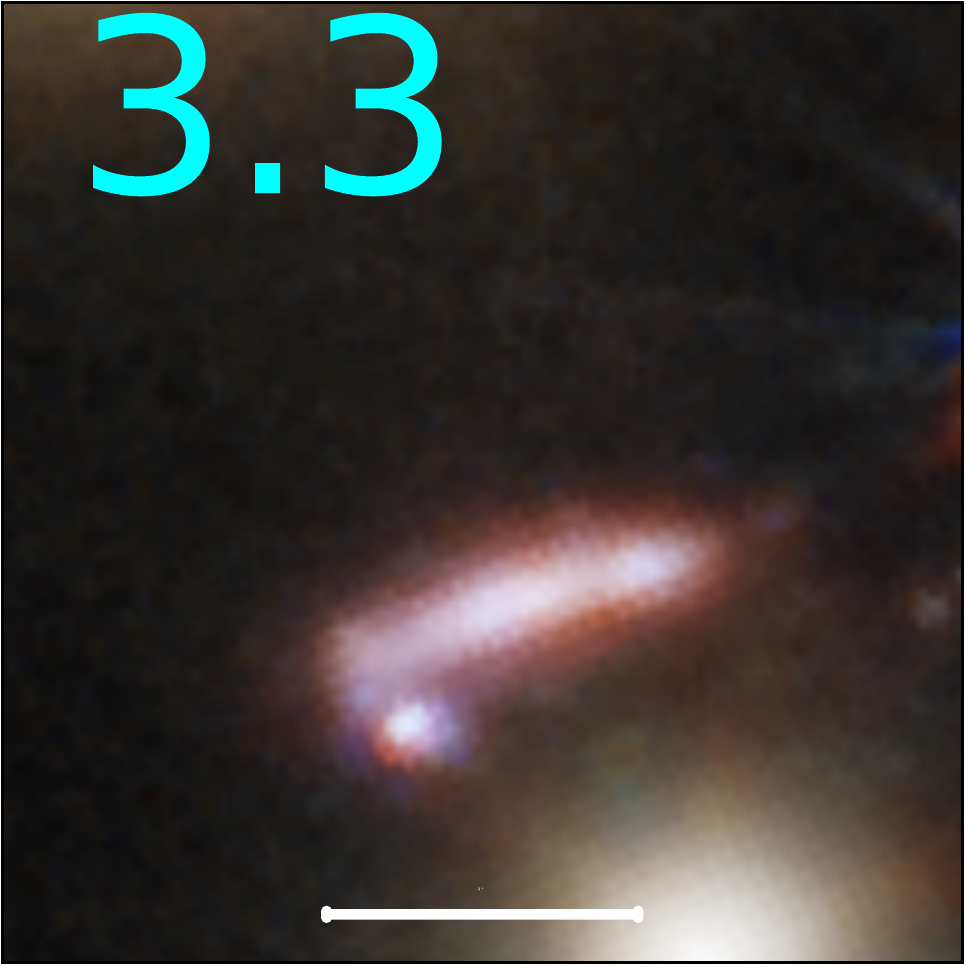}
    
    \includegraphics[width = 0.328\columnwidth]{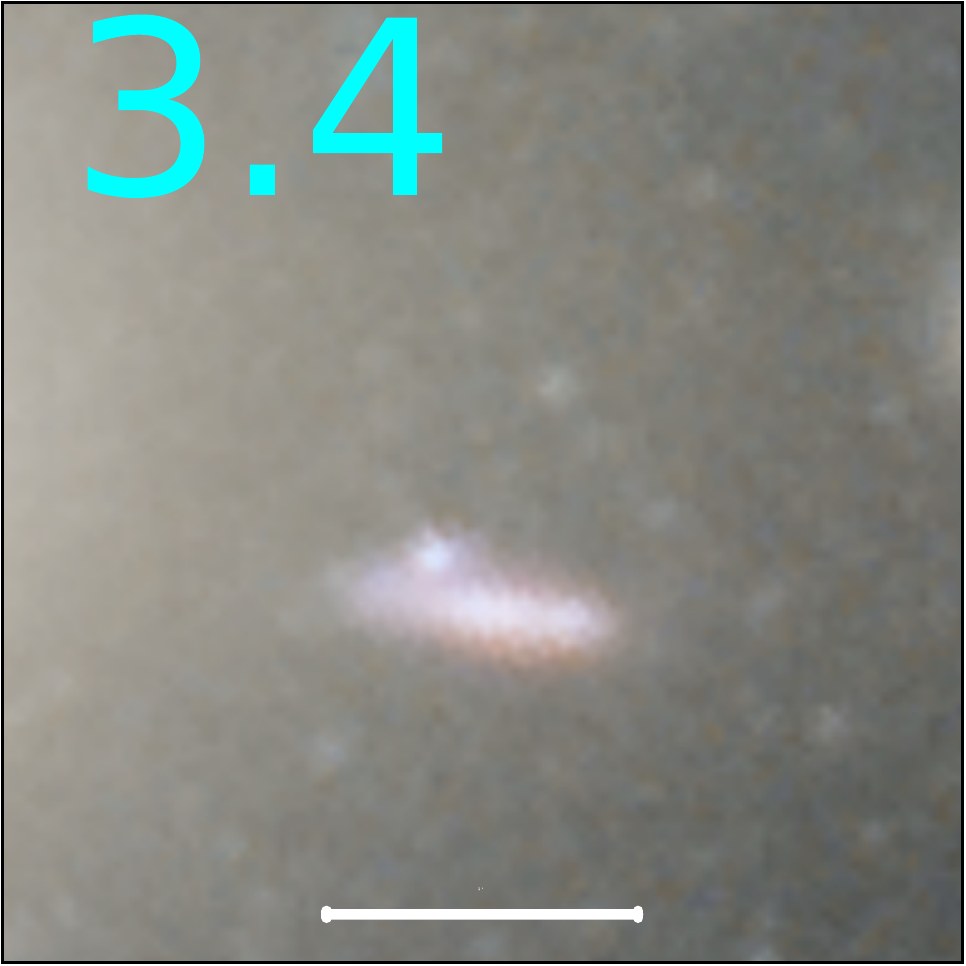}

    \includegraphics[width = 0.328\columnwidth]{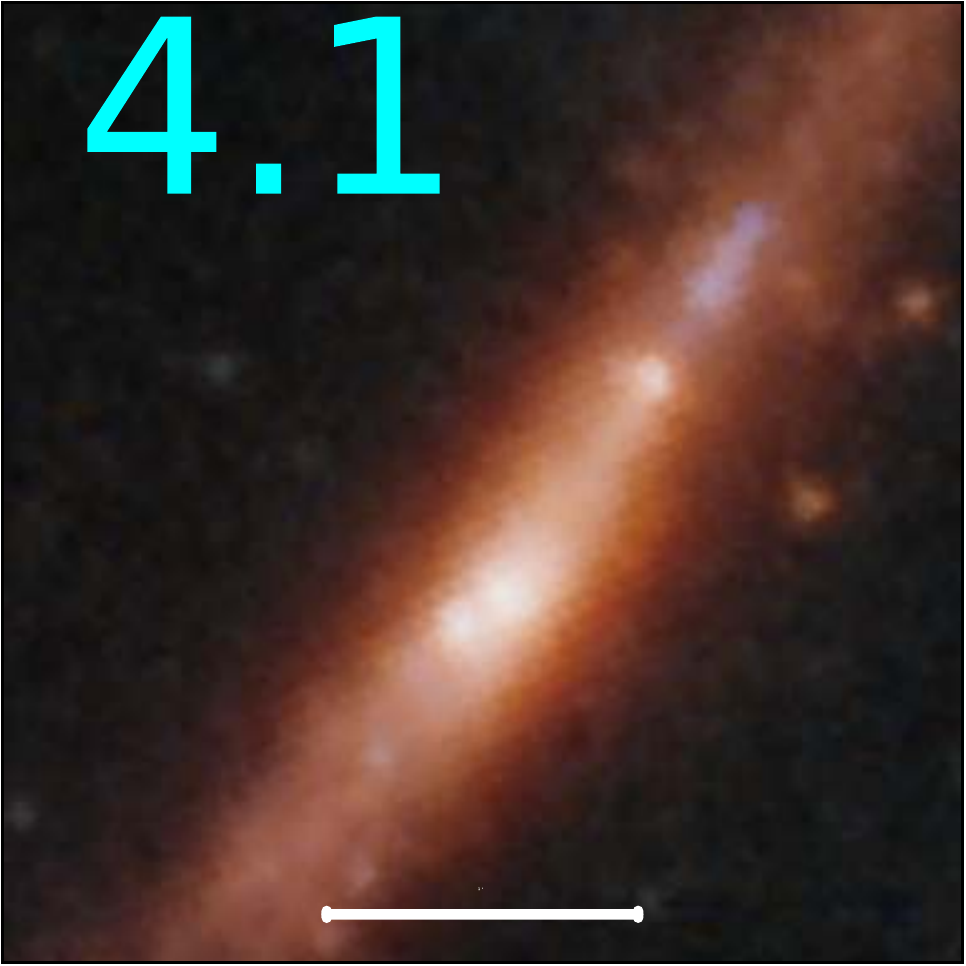}
    \includegraphics[width = 0.328\columnwidth]{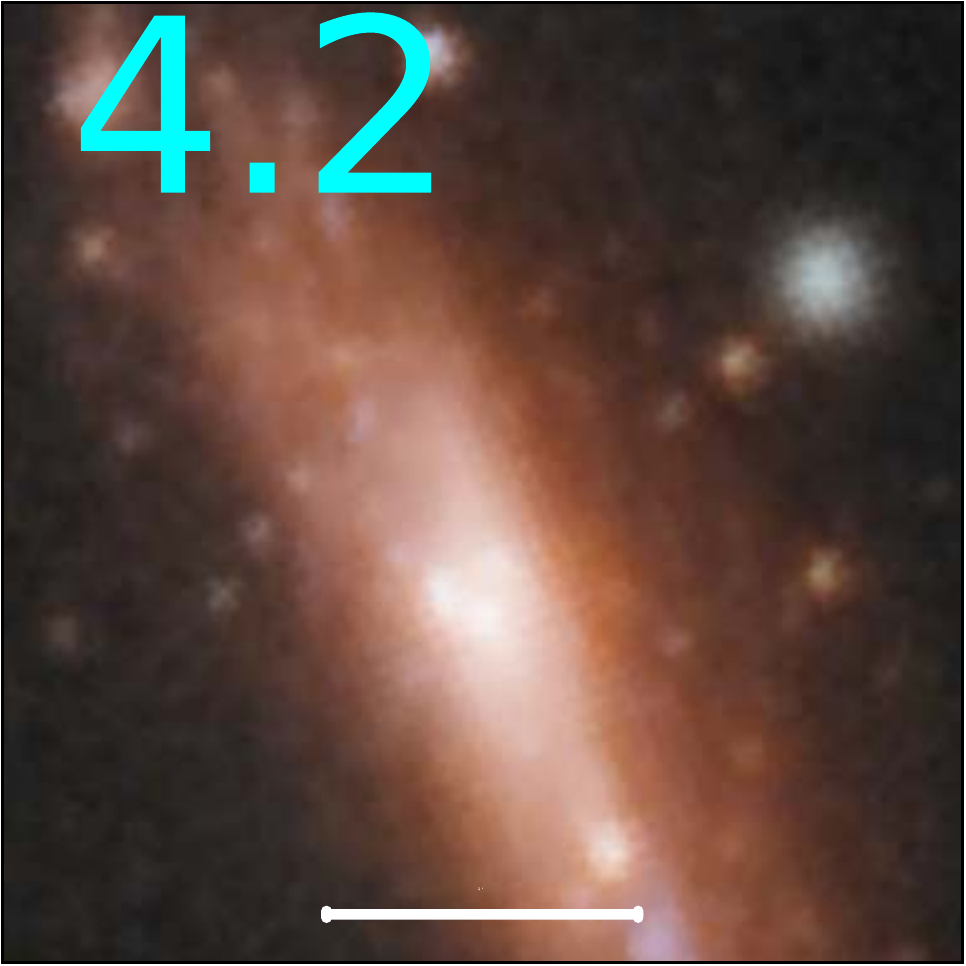}
    \includegraphics[width = 0.328\columnwidth]{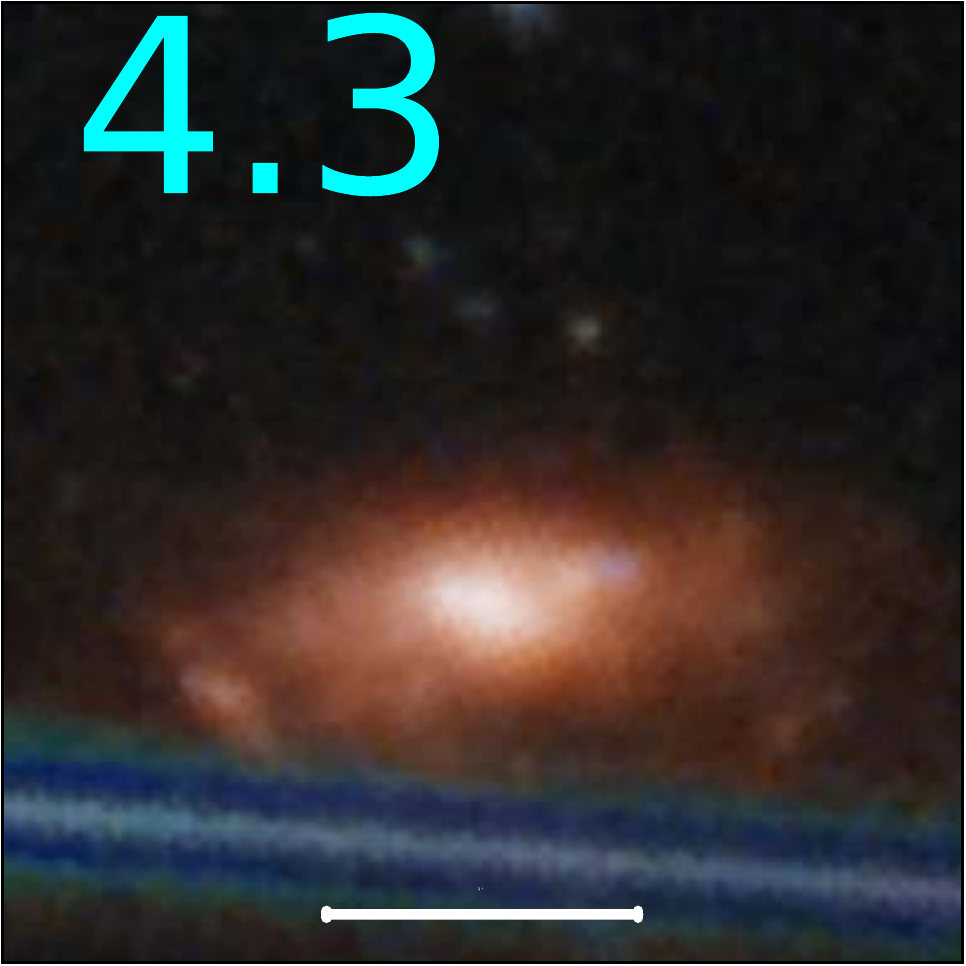}

    \includegraphics[width = 0.328\columnwidth]{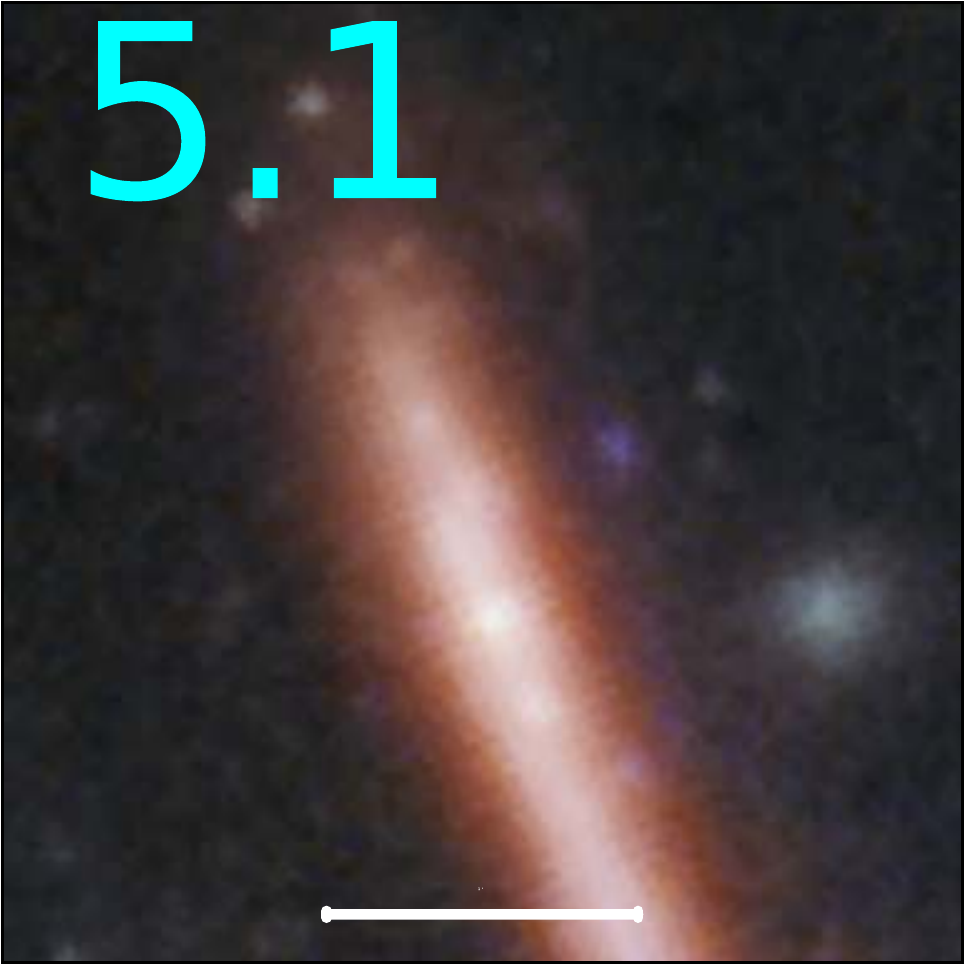}
    \includegraphics[width = 0.328\columnwidth]{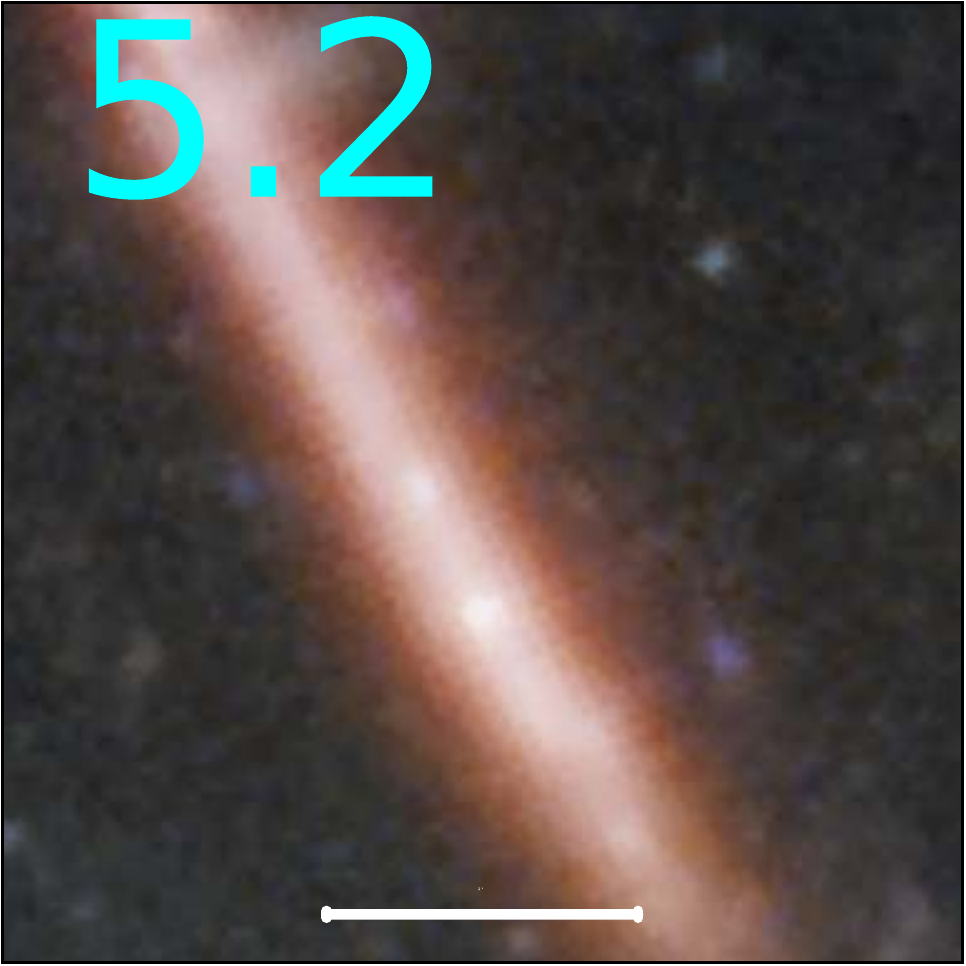}
    \includegraphics[width = 0.328\columnwidth]{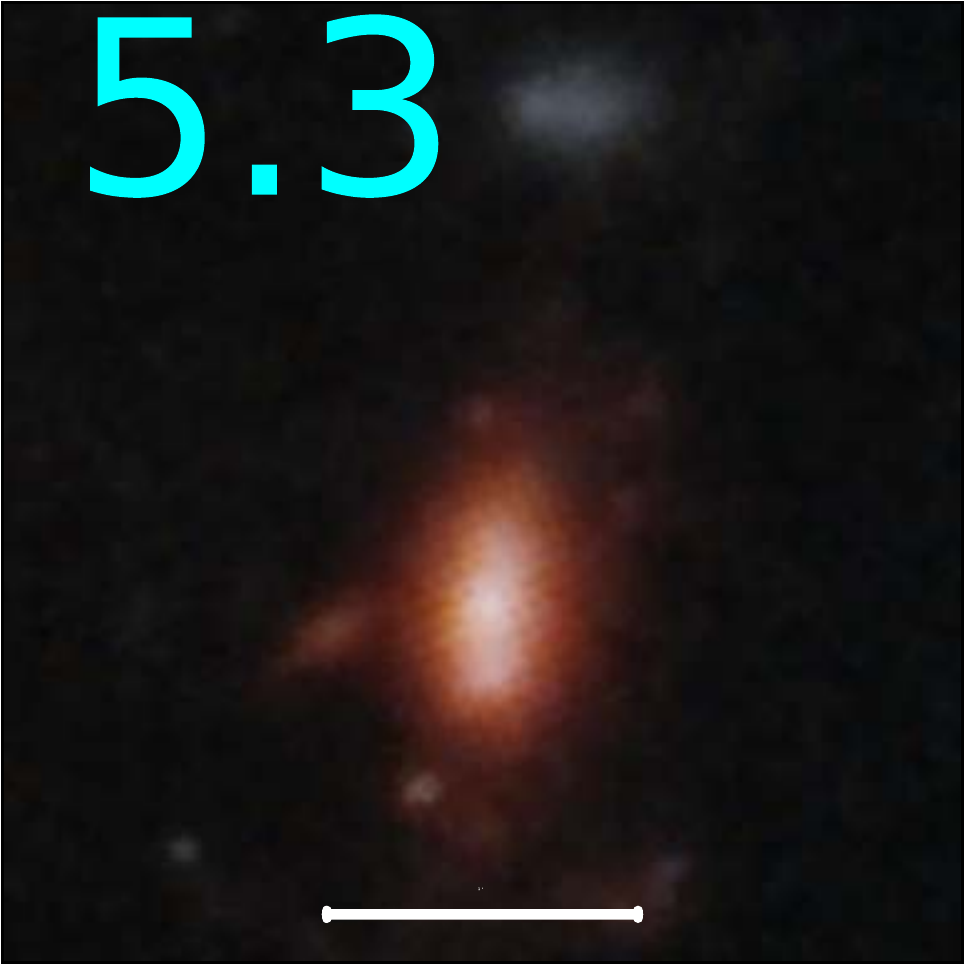}
    
    \includegraphics[width = 0.328\columnwidth]{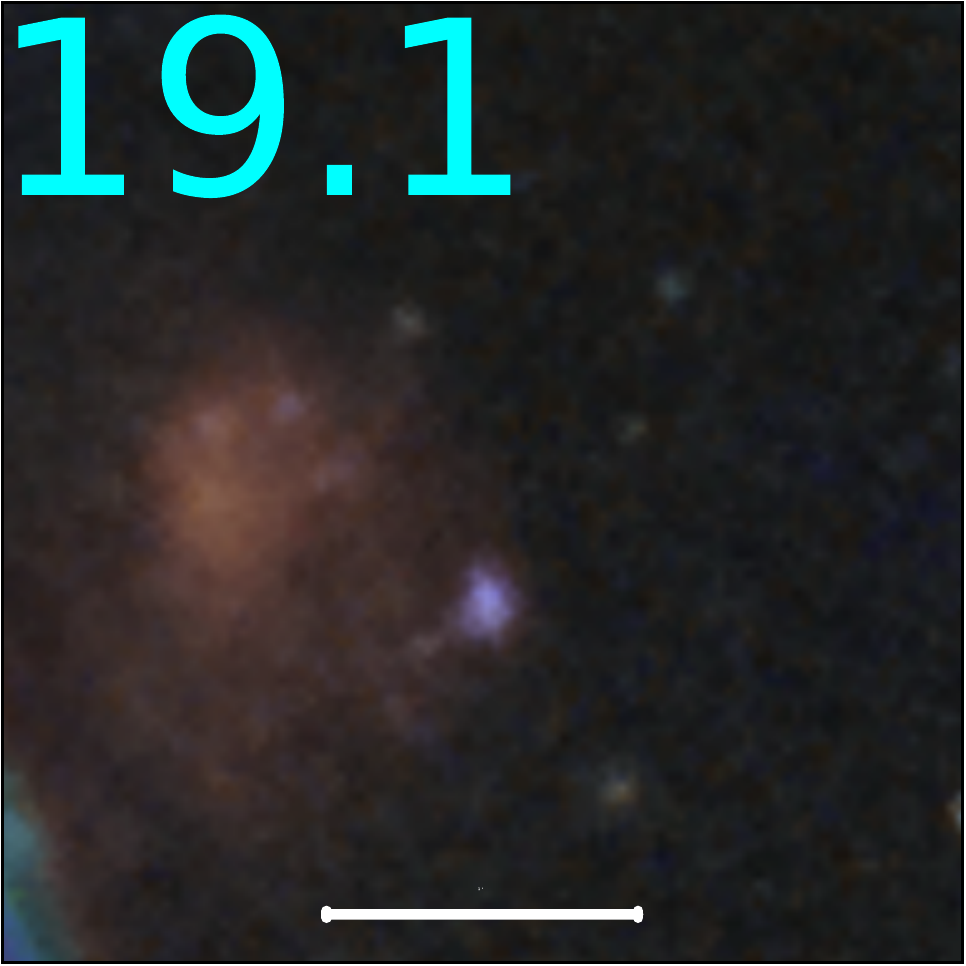}
    \includegraphics[width = 0.328\columnwidth]{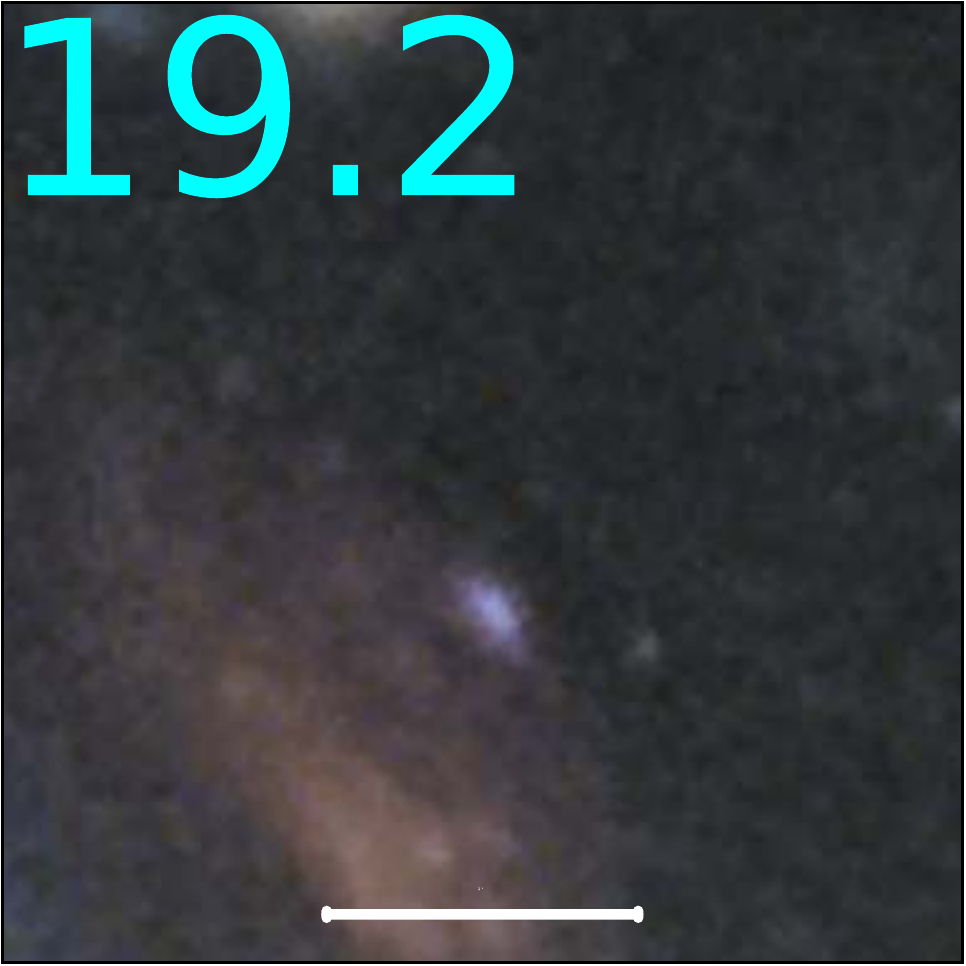}
    \includegraphics[width = 0.328\columnwidth]{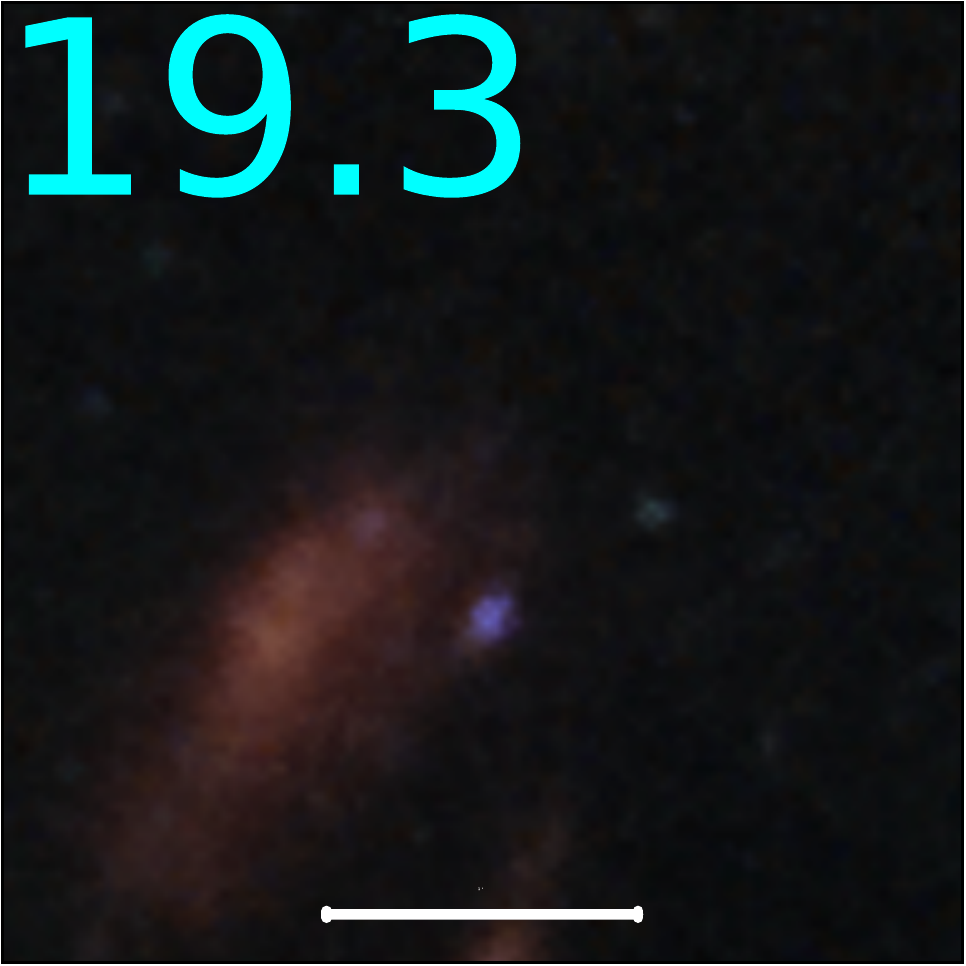}

    \includegraphics[width = 0.328\columnwidth]{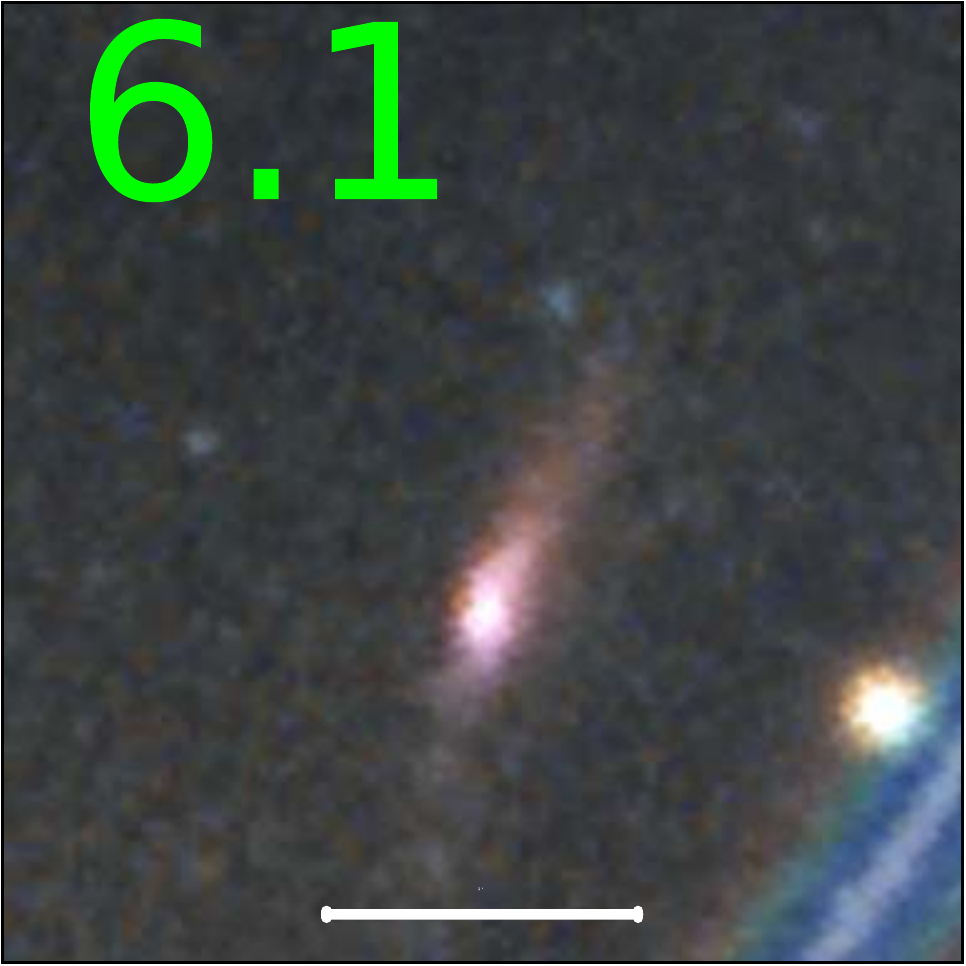}
    \includegraphics[width = 0.328\columnwidth]{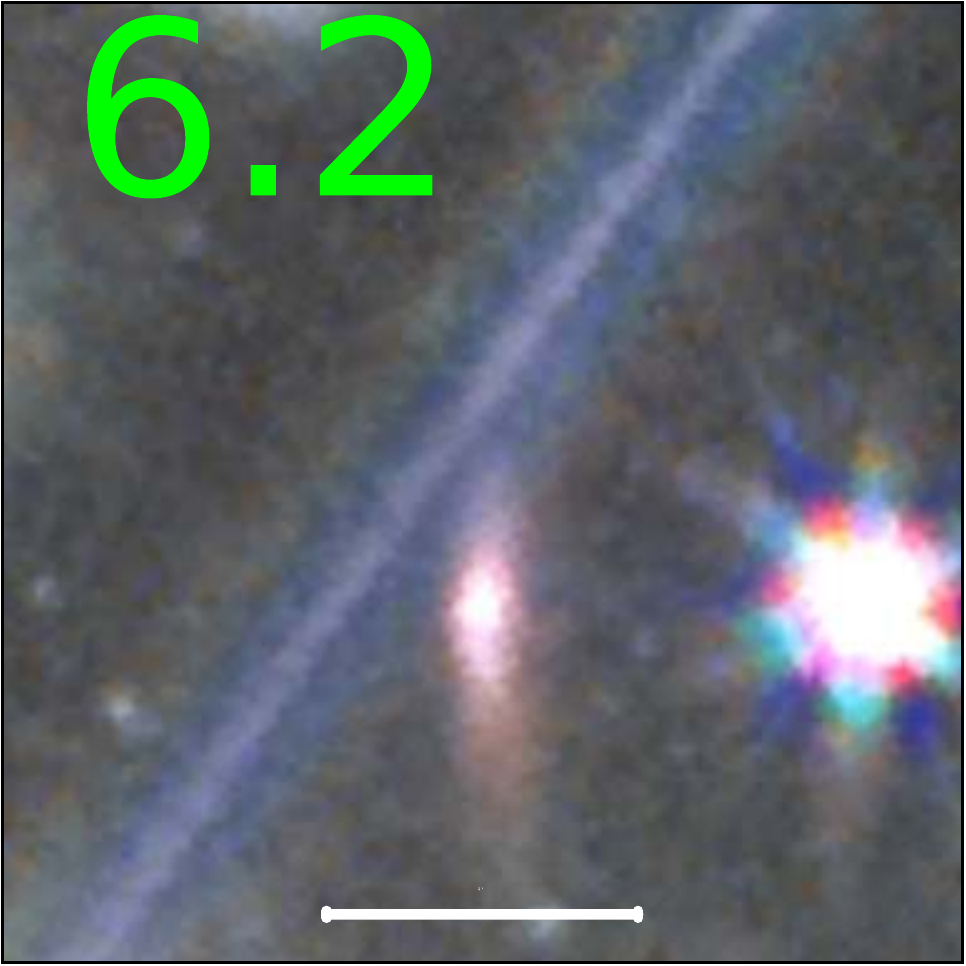}
    \includegraphics[width = 0.328\columnwidth]{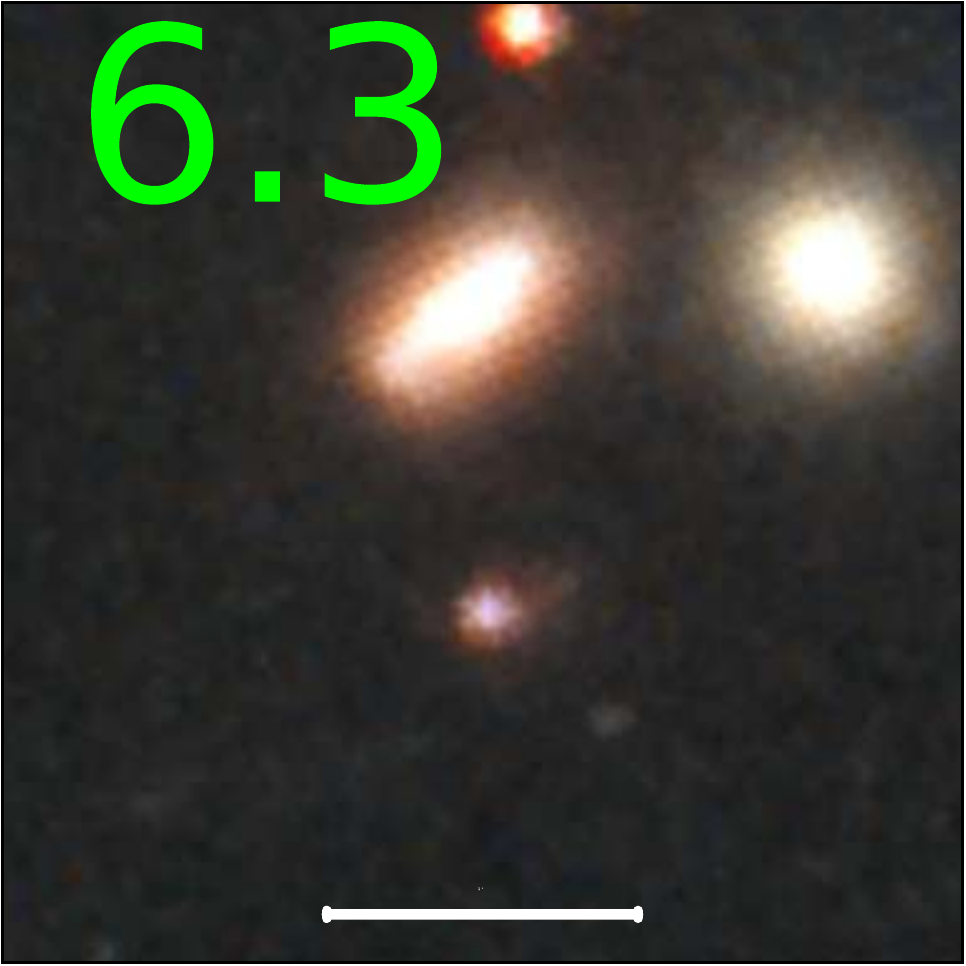}
    
  \caption{Multiple image cutouts. The white scale bar has a length of $1\arcsec$.}
  \label{fig:cut_outs}
\end{figure}    

\begin{figure}
    \setcounter{figure}{\value{figure}-1}

    \includegraphics[width = 0.328\columnwidth]{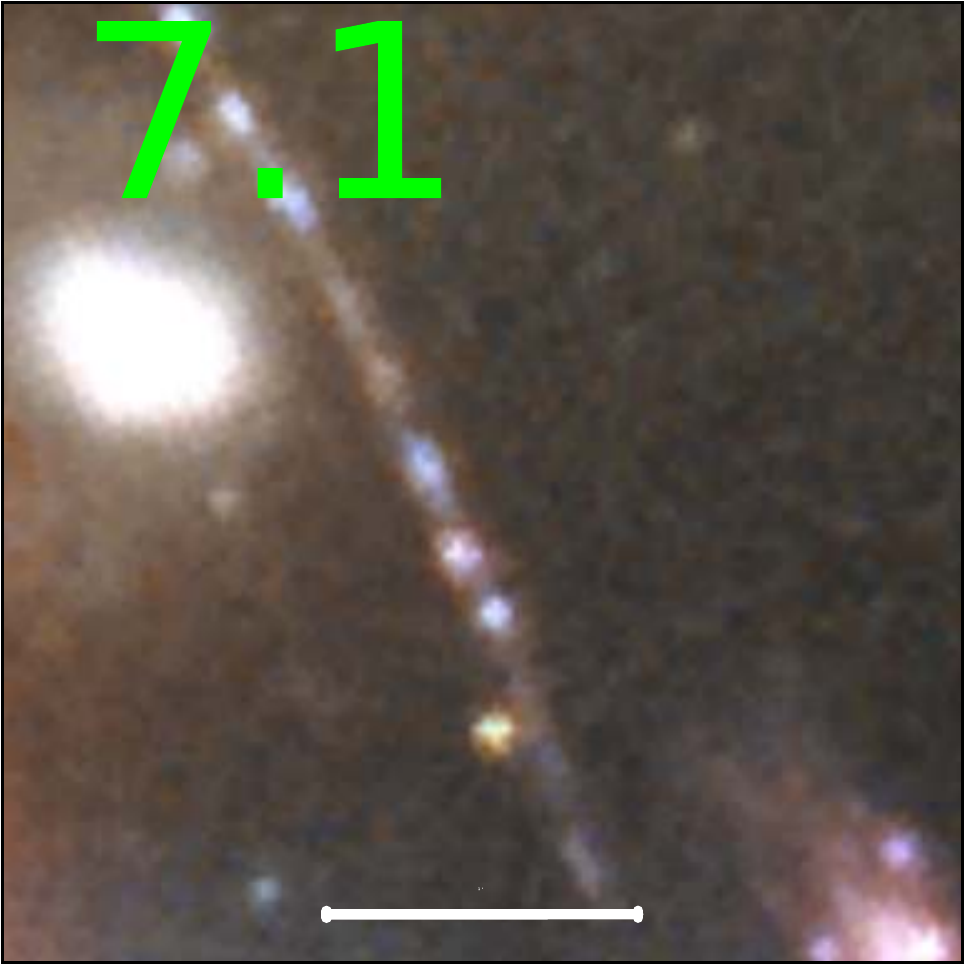}
    \includegraphics[width = 0.328\columnwidth]{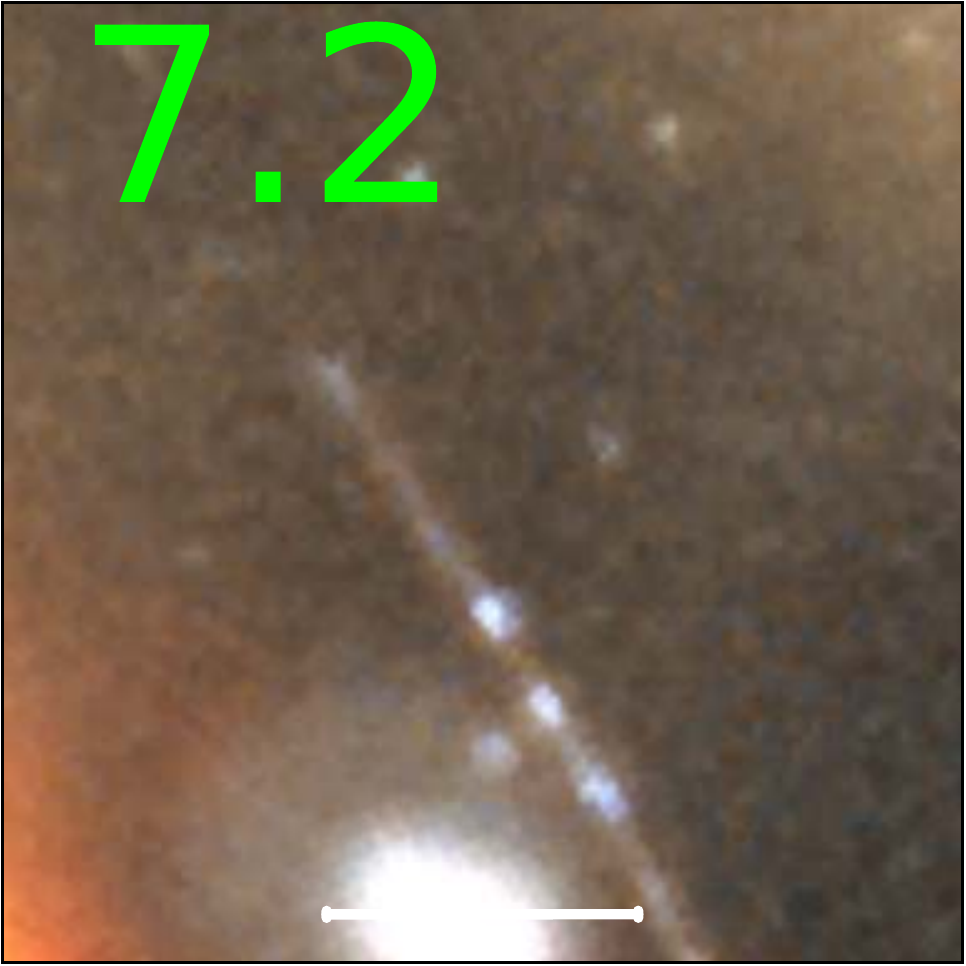}
    \includegraphics[width = 0.328\columnwidth]{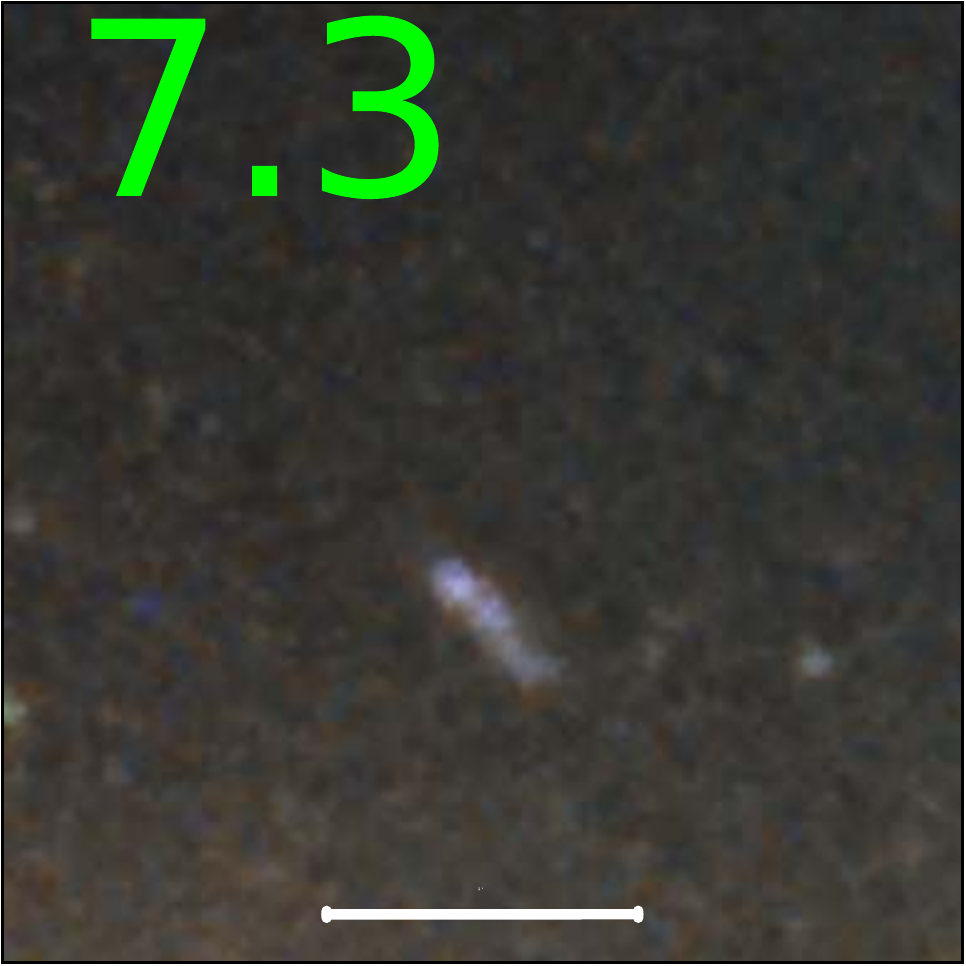}

    \includegraphics[width = 0.328\columnwidth]{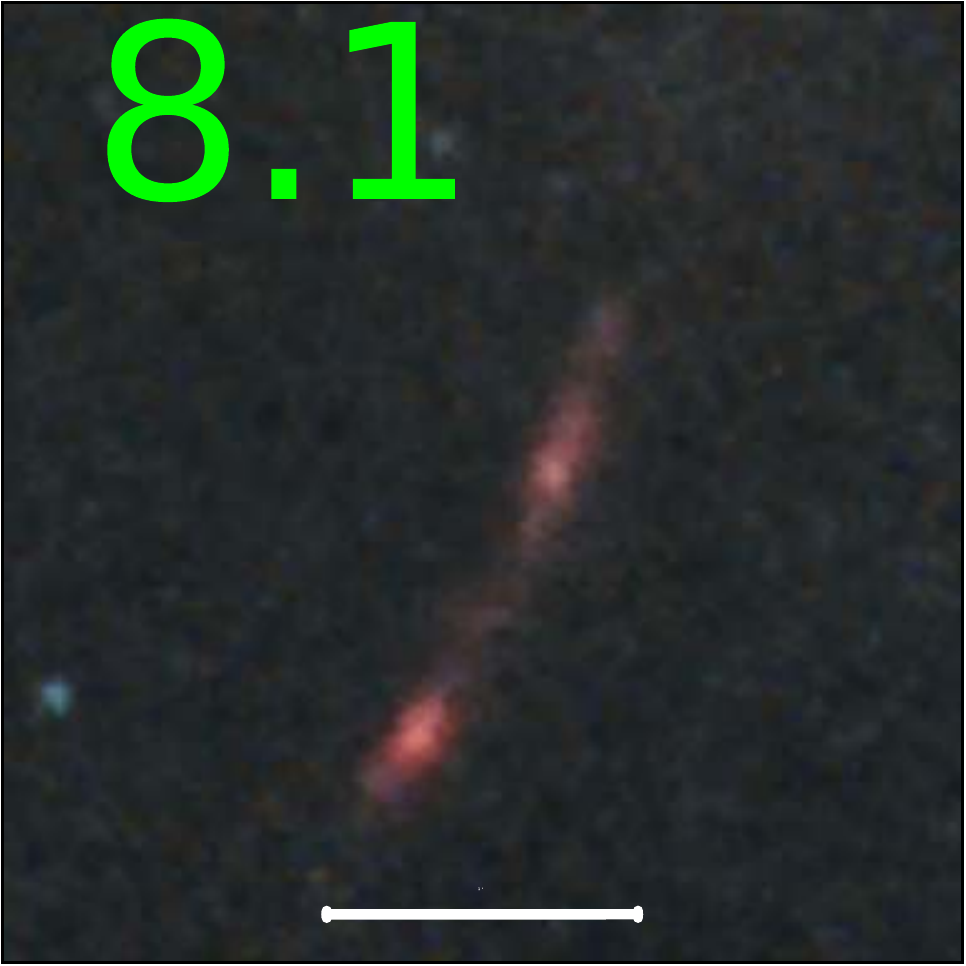}
    \includegraphics[width = 0.328\columnwidth]{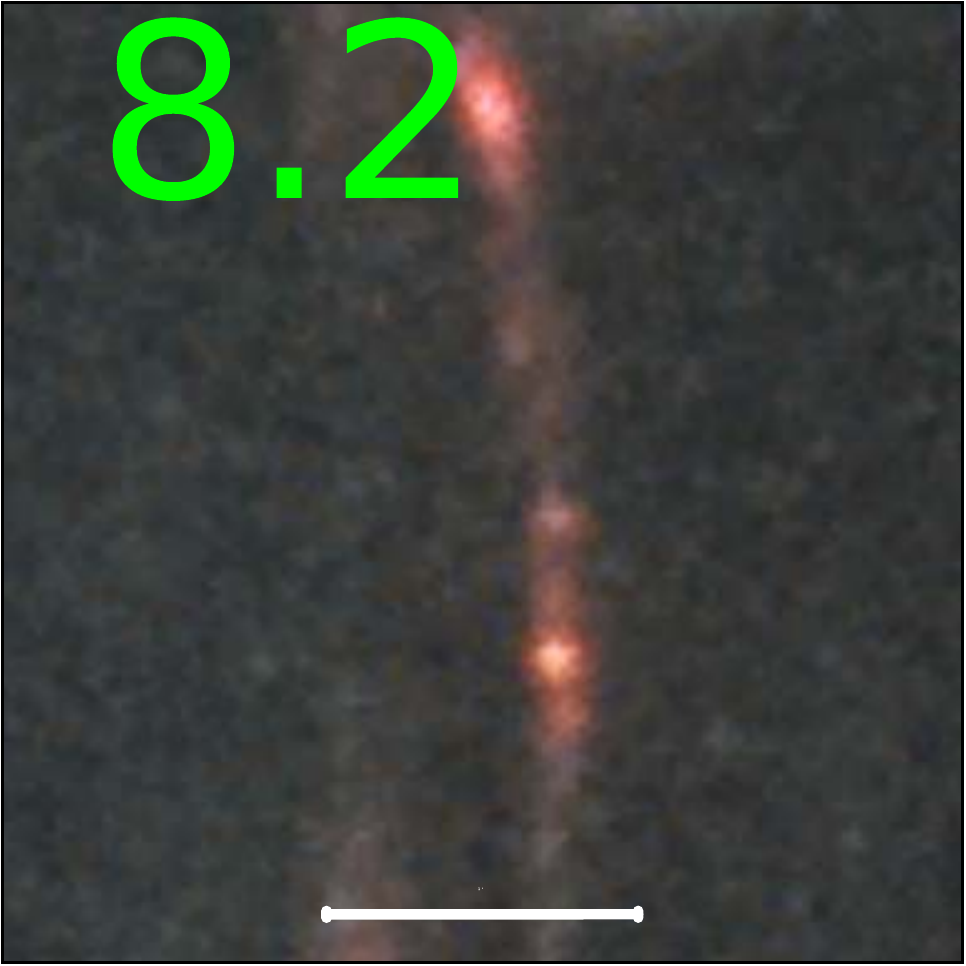}
    \includegraphics[width = 0.328\columnwidth]{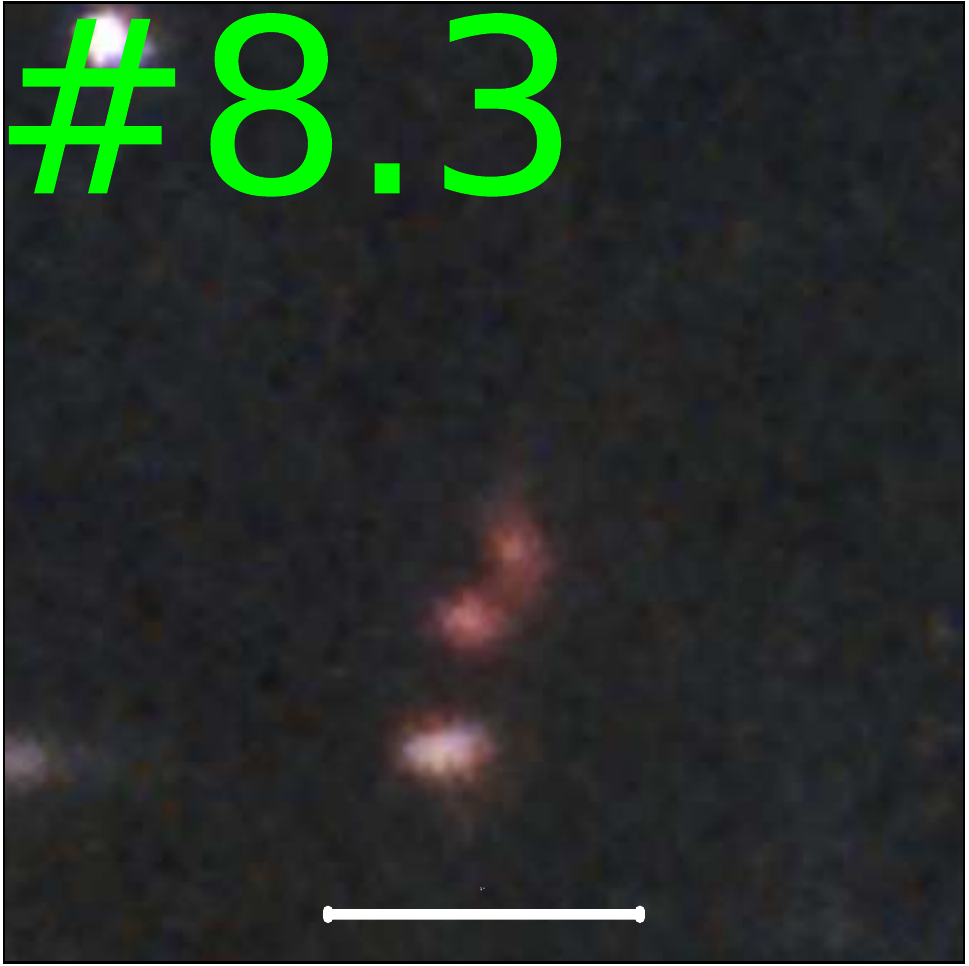}

    \includegraphics[width = 0.328\columnwidth]{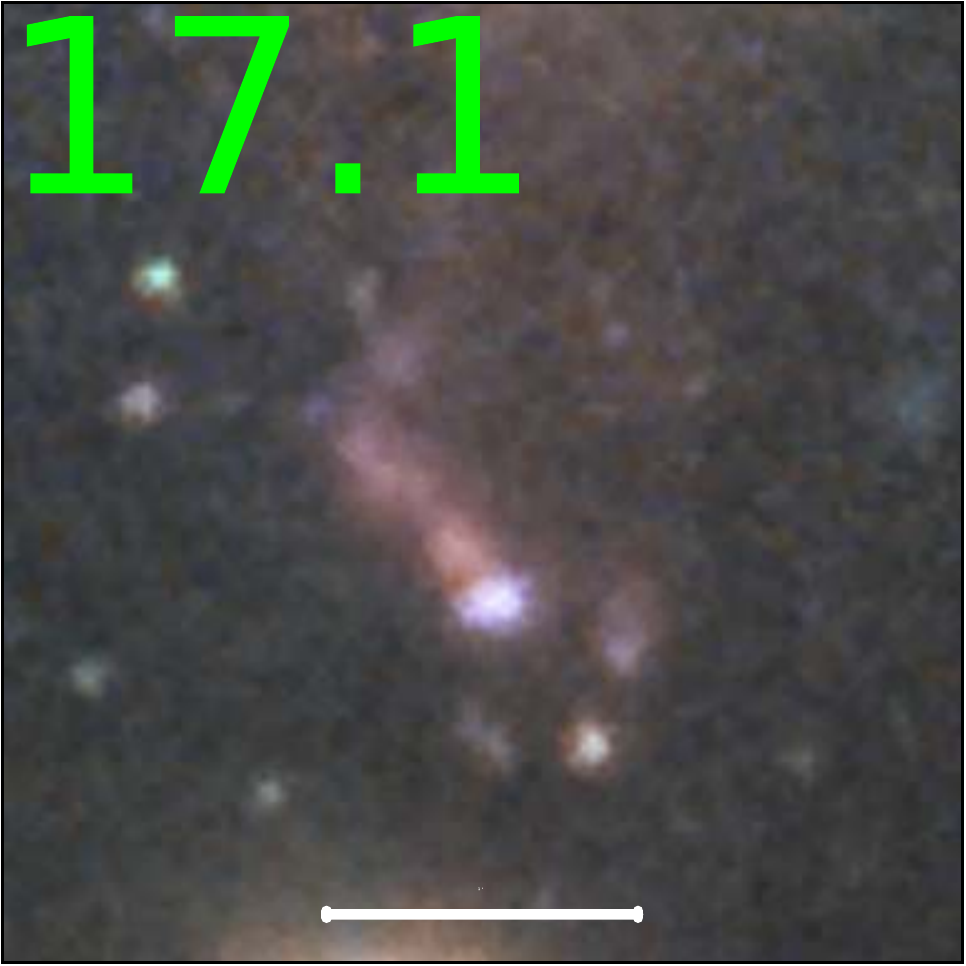}
    \includegraphics[width = 0.328\columnwidth]{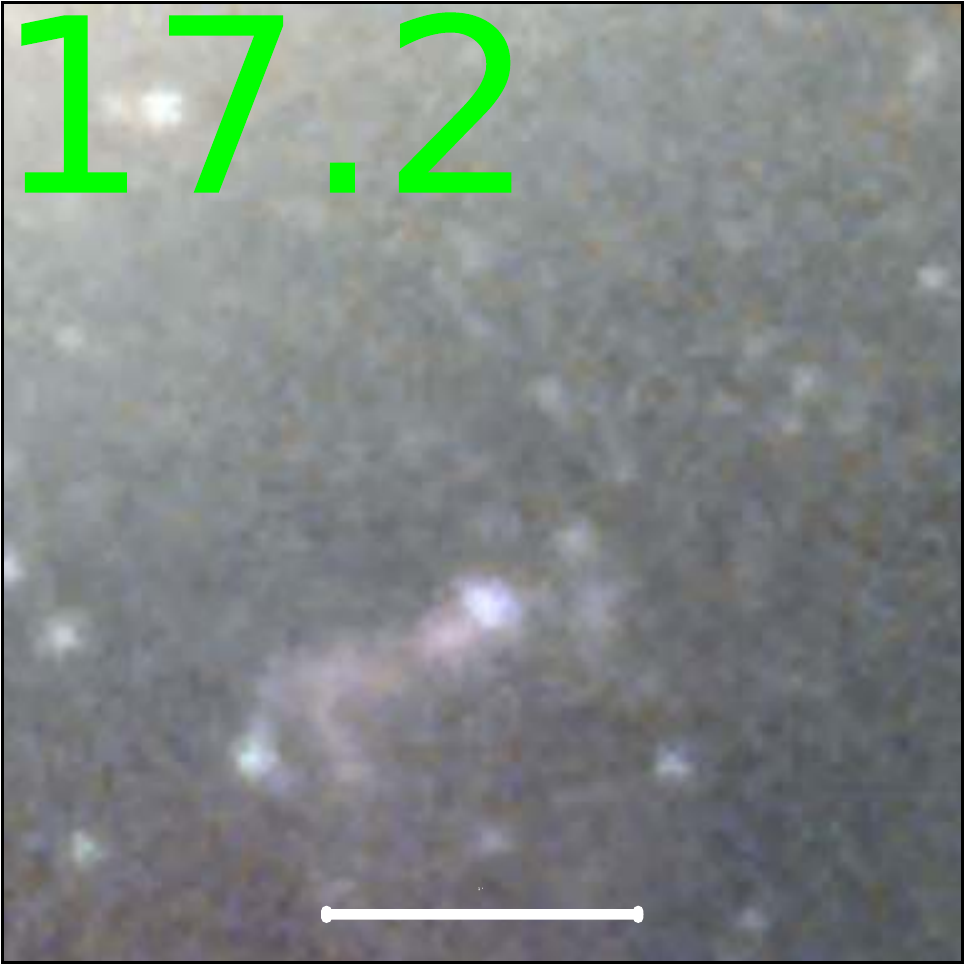}
    \includegraphics[width = 0.328\columnwidth]{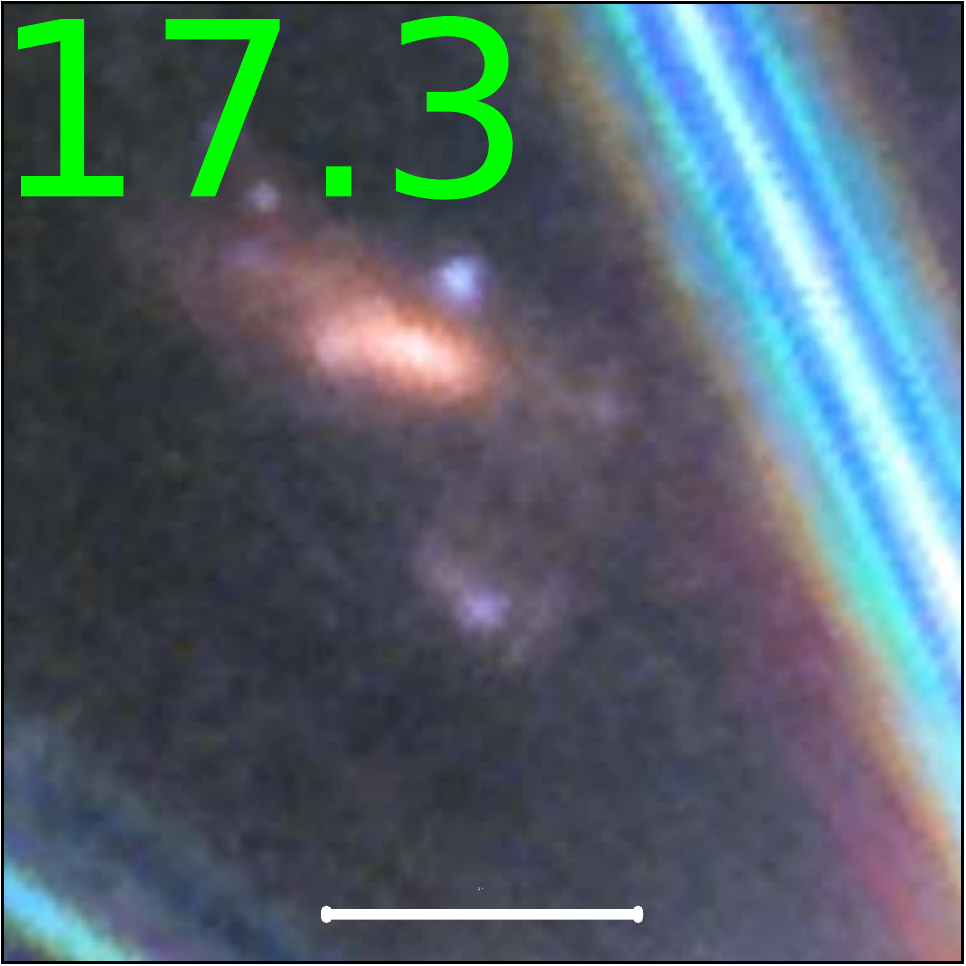}

    \includegraphics[width = 0.328\columnwidth]{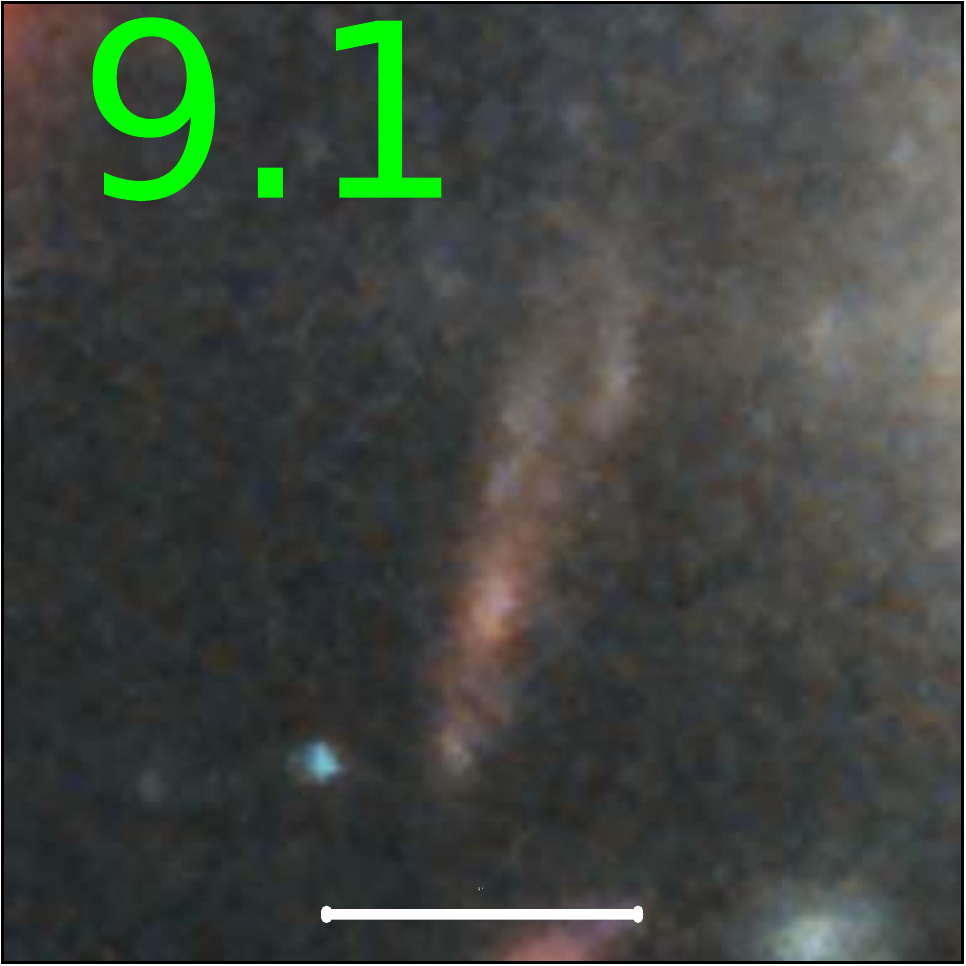}
    \includegraphics[width = 0.328\columnwidth]{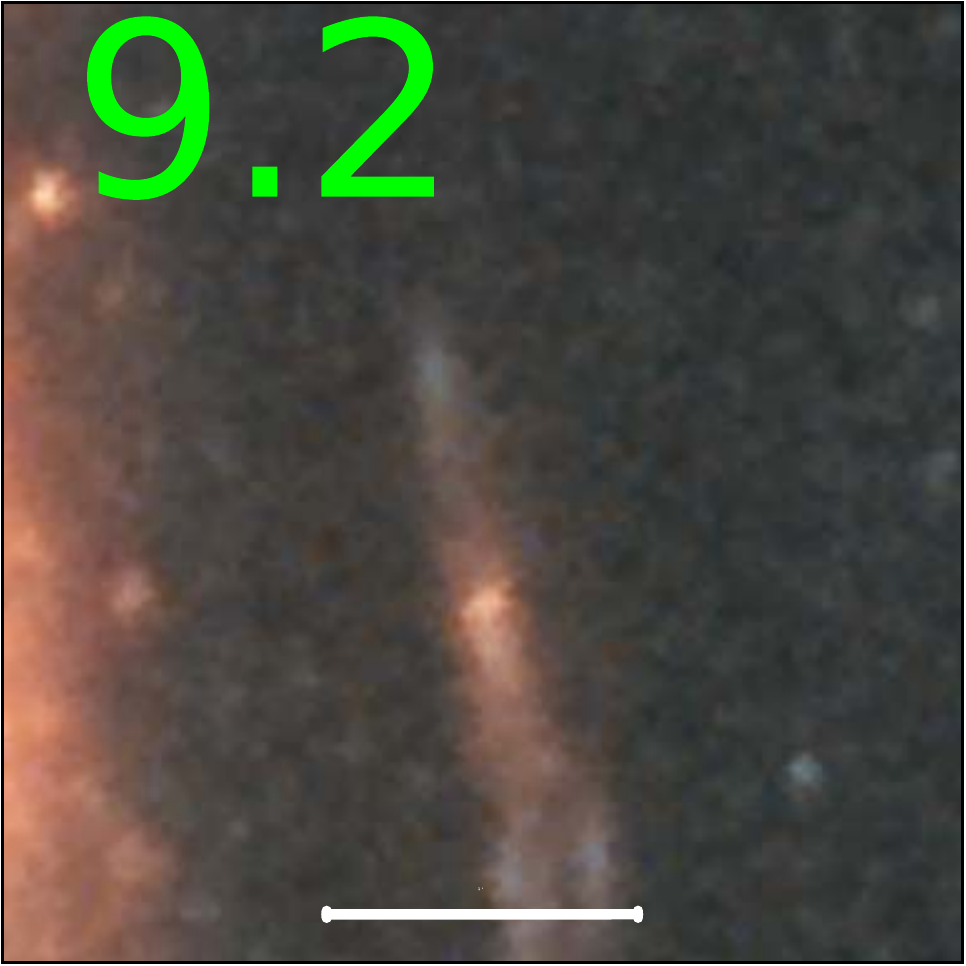}

    \includegraphics[width = 0.328\columnwidth]{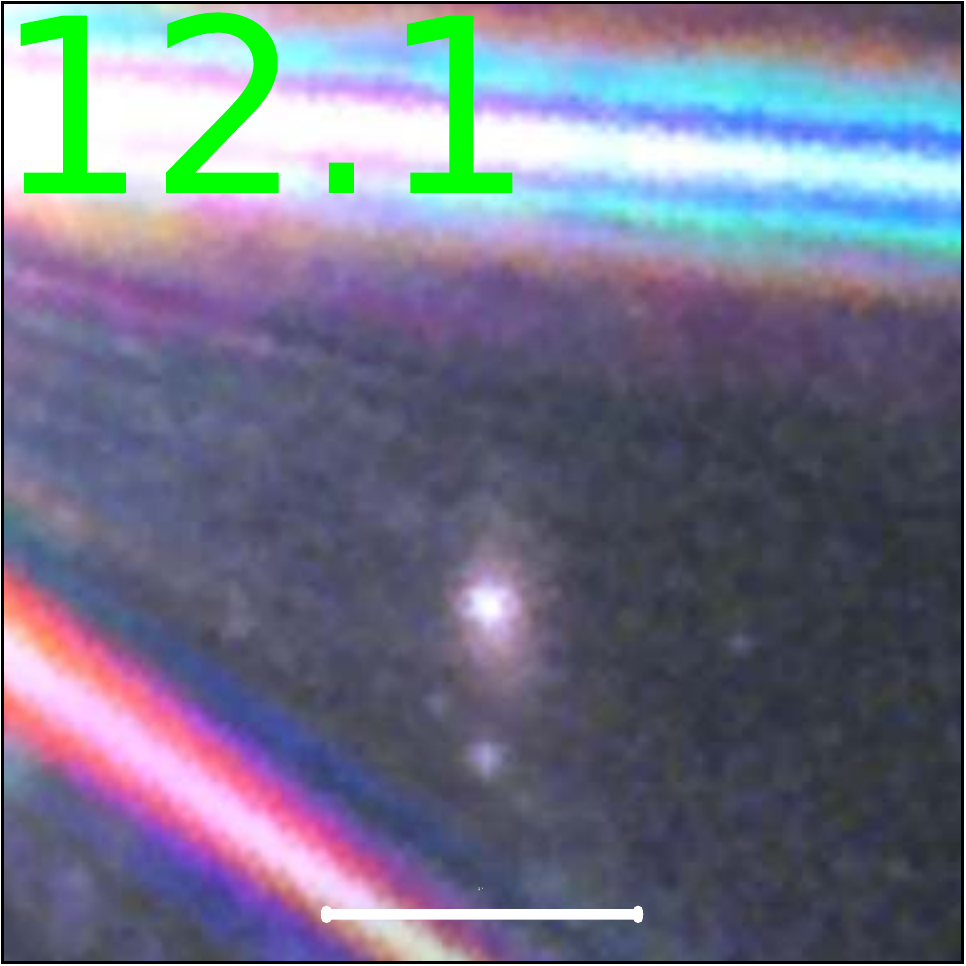}
    \includegraphics[width = 0.328\columnwidth]{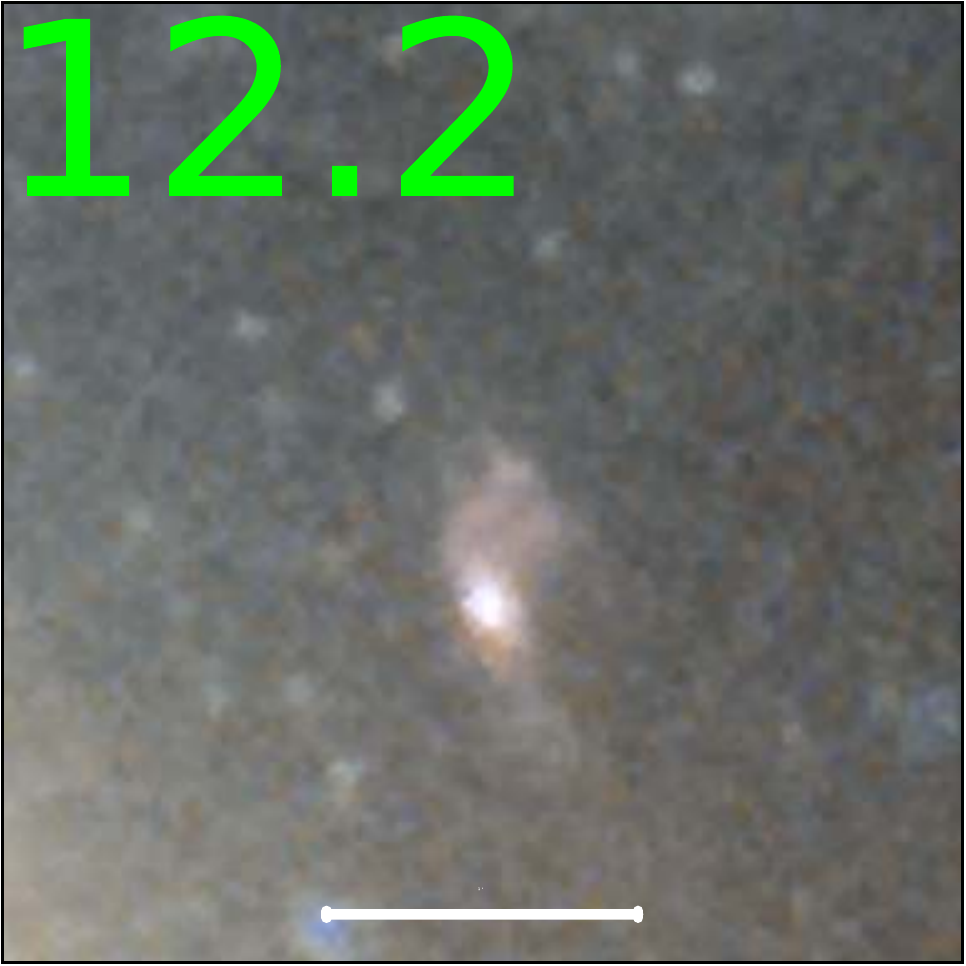}
    \includegraphics[width = 0.328\columnwidth]{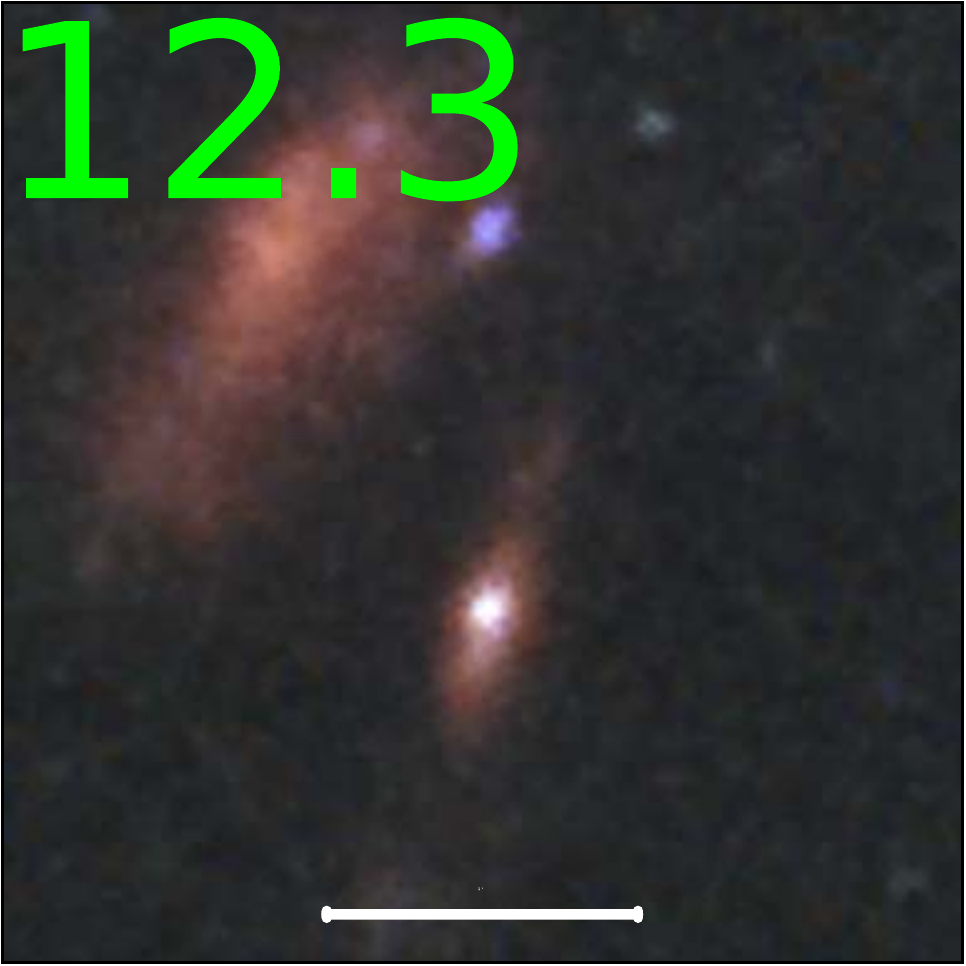}

    \includegraphics[width = 0.328\columnwidth]{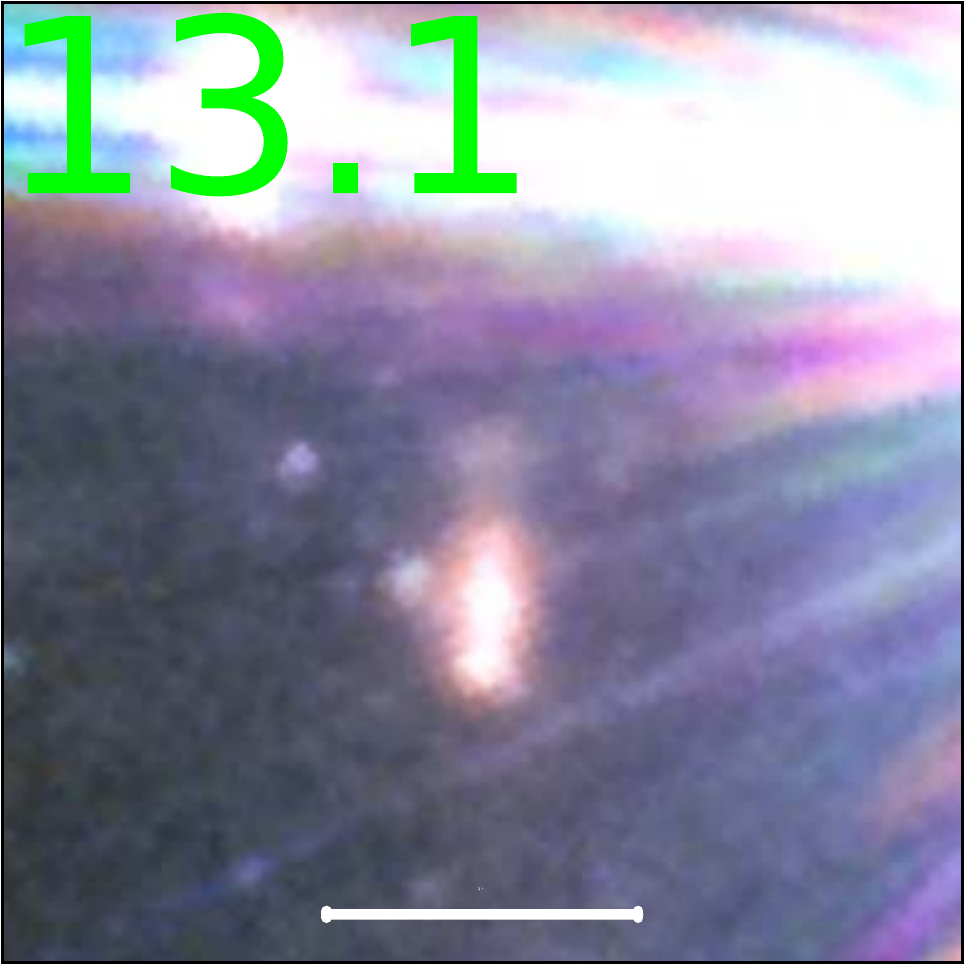}
    \includegraphics[width = 0.328\columnwidth]{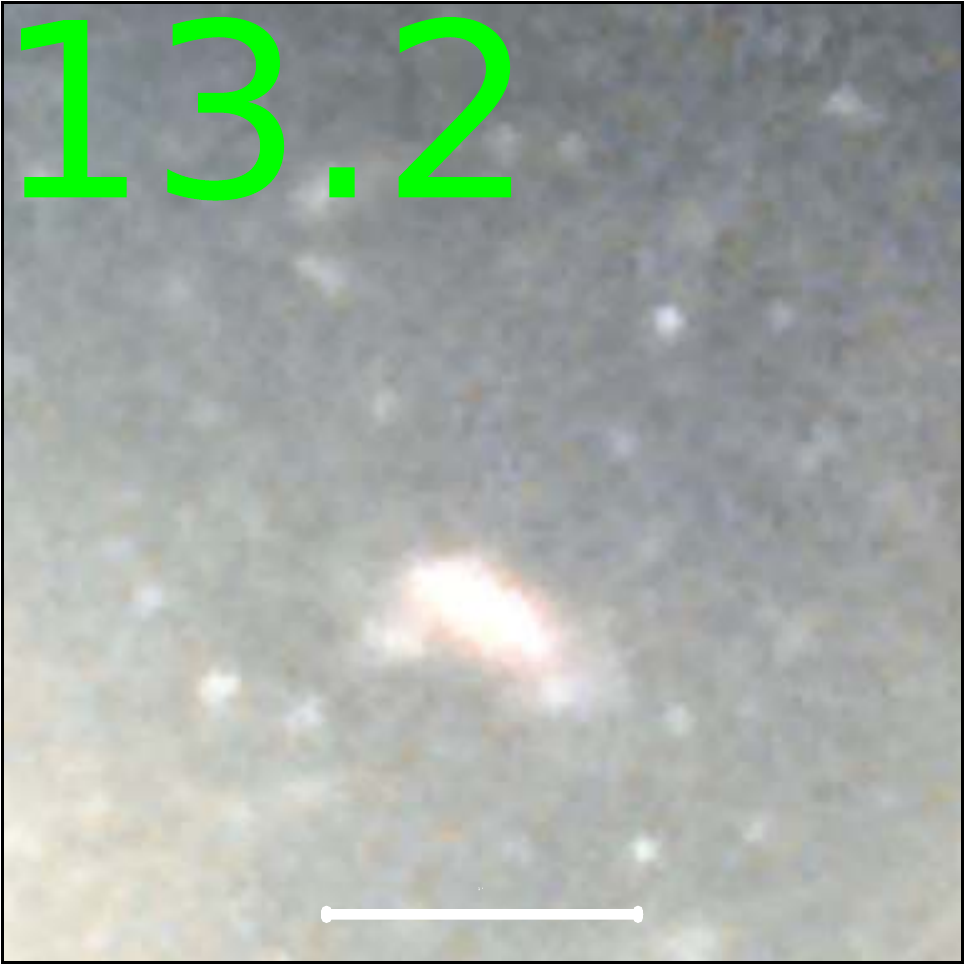}
    \includegraphics[width = 0.328\columnwidth]{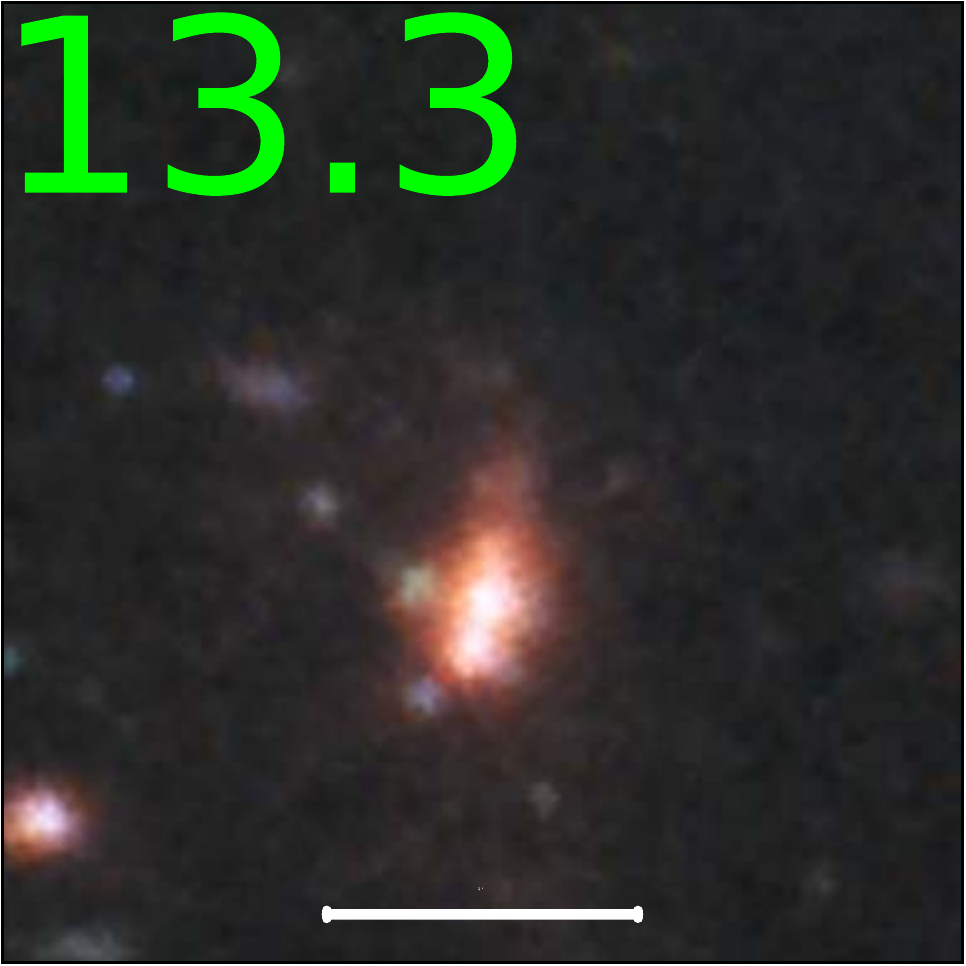}
    
    \includegraphics[width = 0.328\columnwidth]{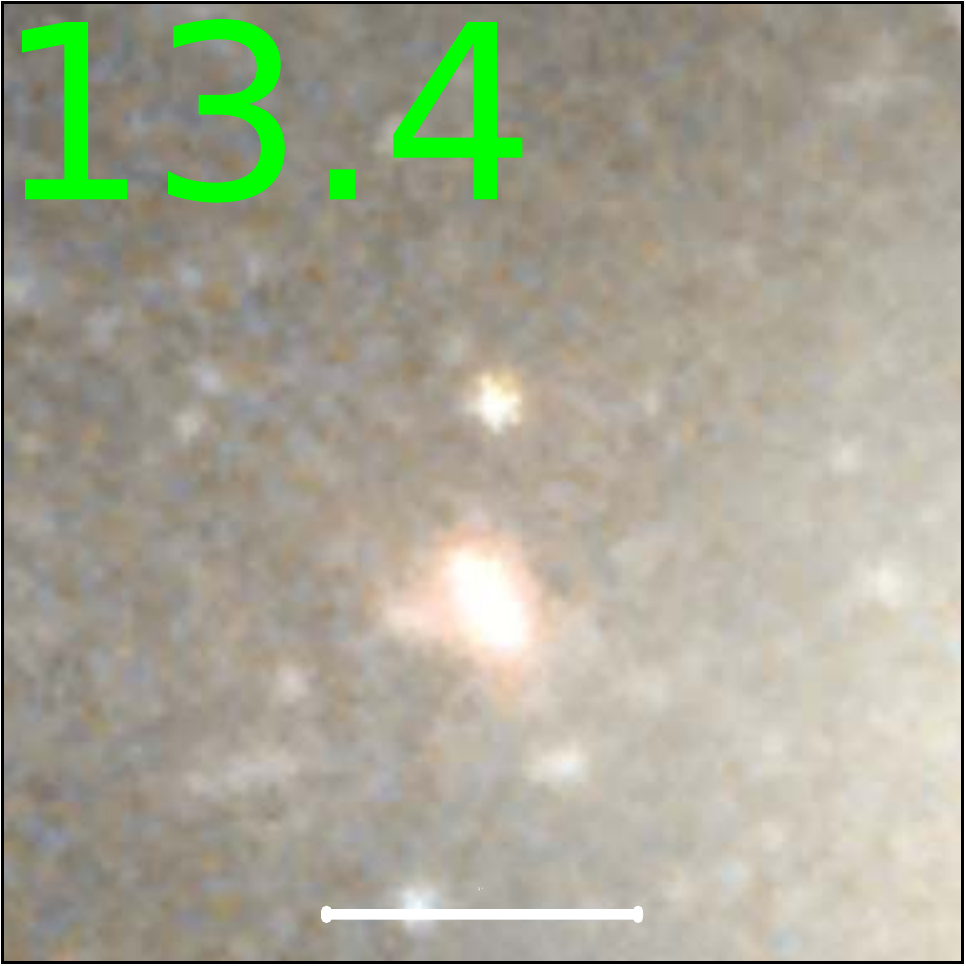}

  \caption{Continued.}
  \label{fig:cut_outs}
\end{figure}
 
\begin{figure}
    \setcounter{figure}{\value{figure}-1}

    \includegraphics[width = 0.328\columnwidth]{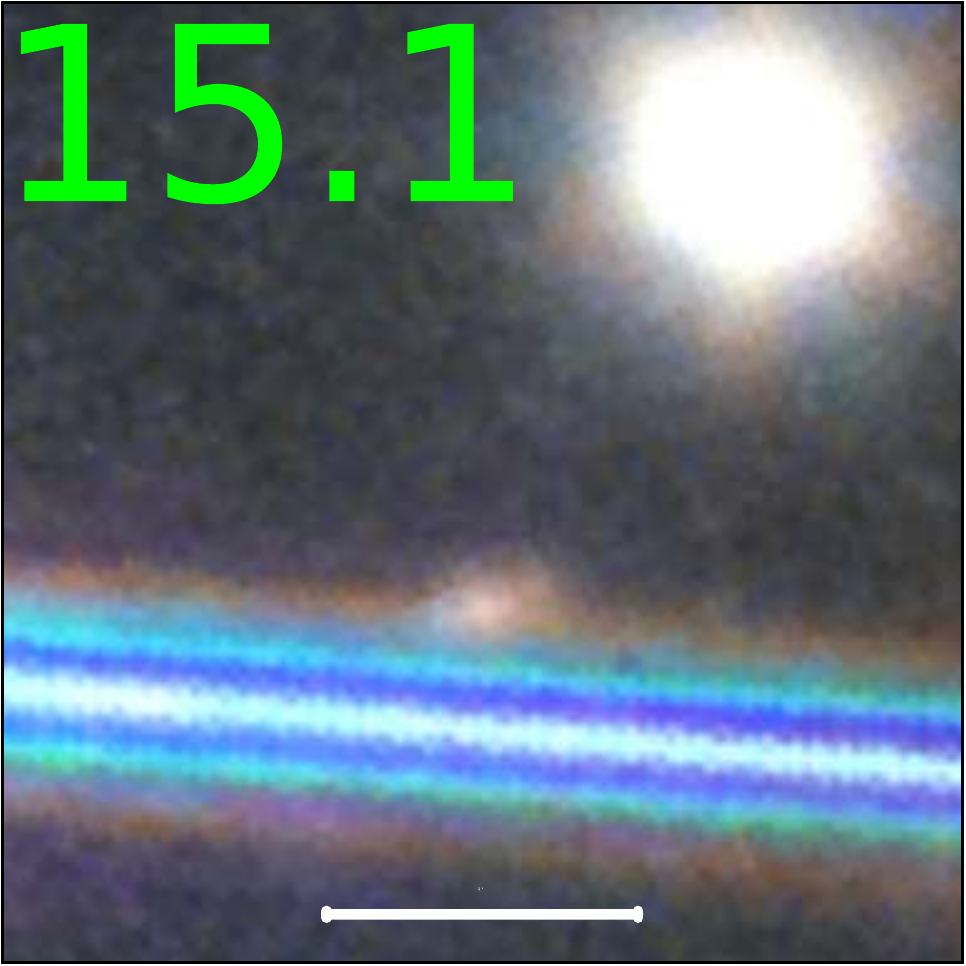}
    \includegraphics[width = 0.328\columnwidth]{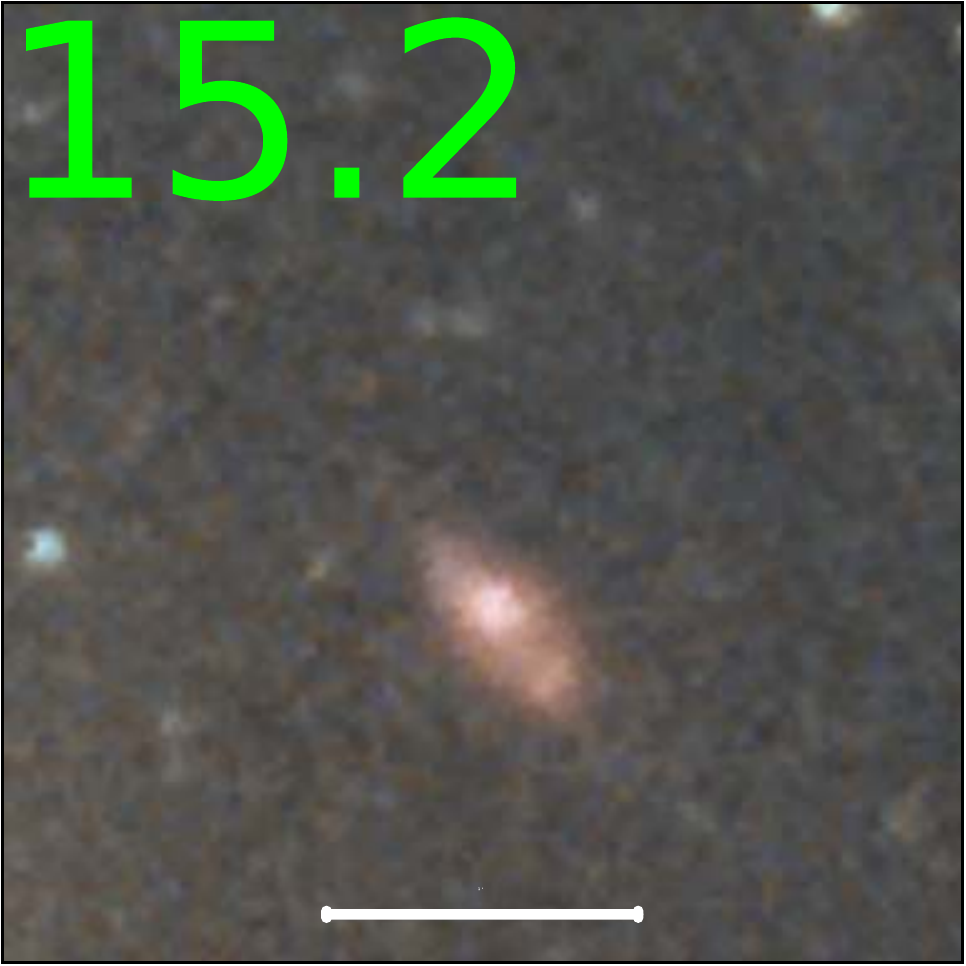}
    \includegraphics[width = 0.328\columnwidth]{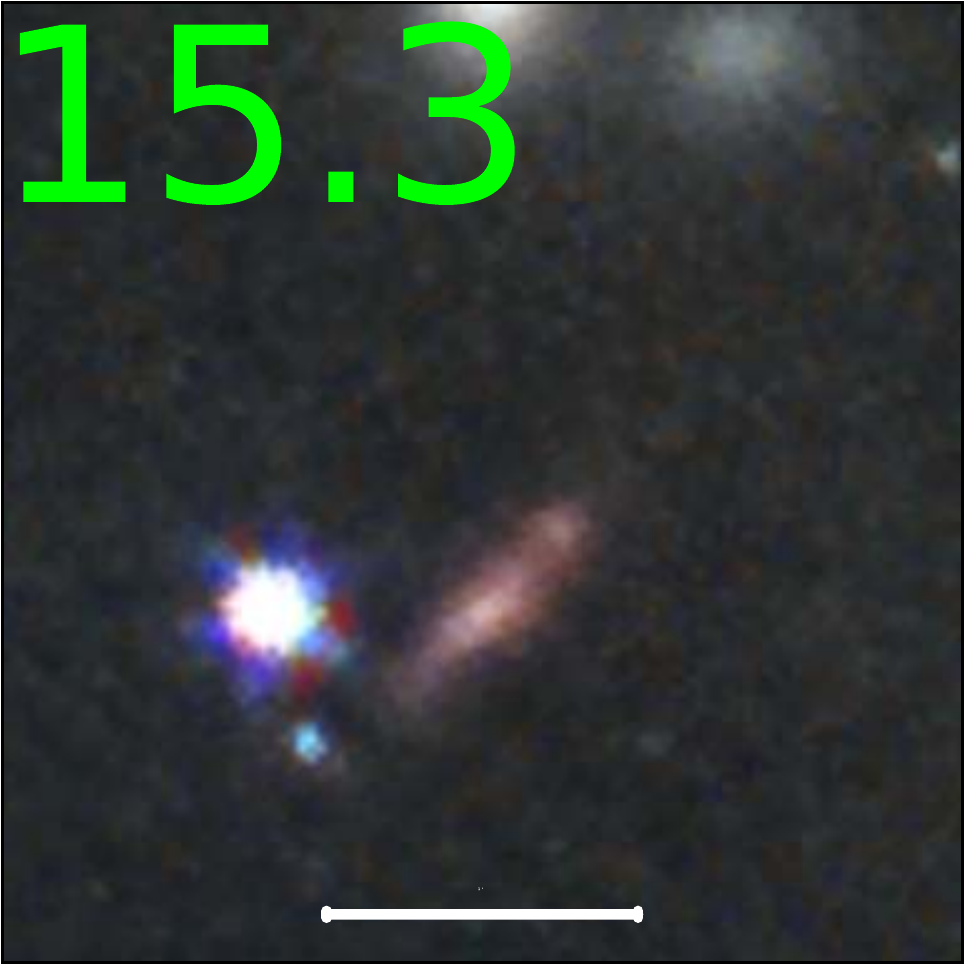}

    \includegraphics[width = 0.328\columnwidth]{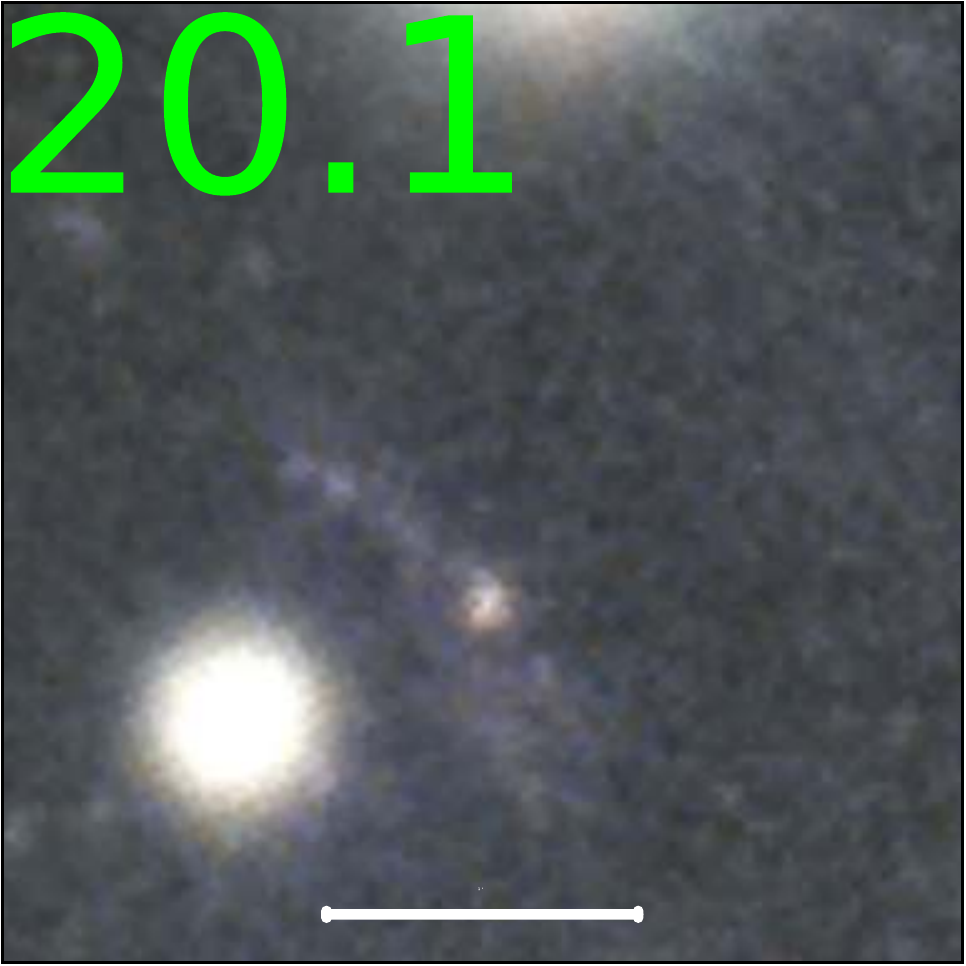}
    \includegraphics[width = 0.328\columnwidth]{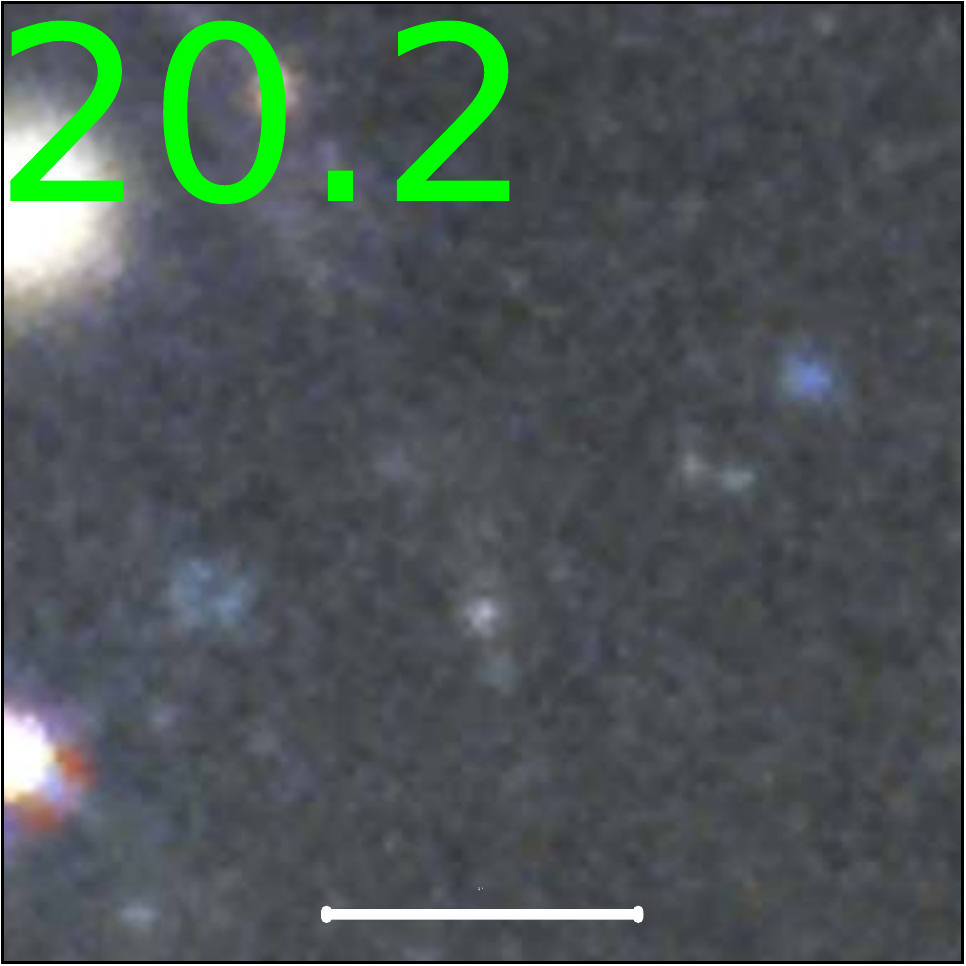}

    \includegraphics[width = 0.328\columnwidth]{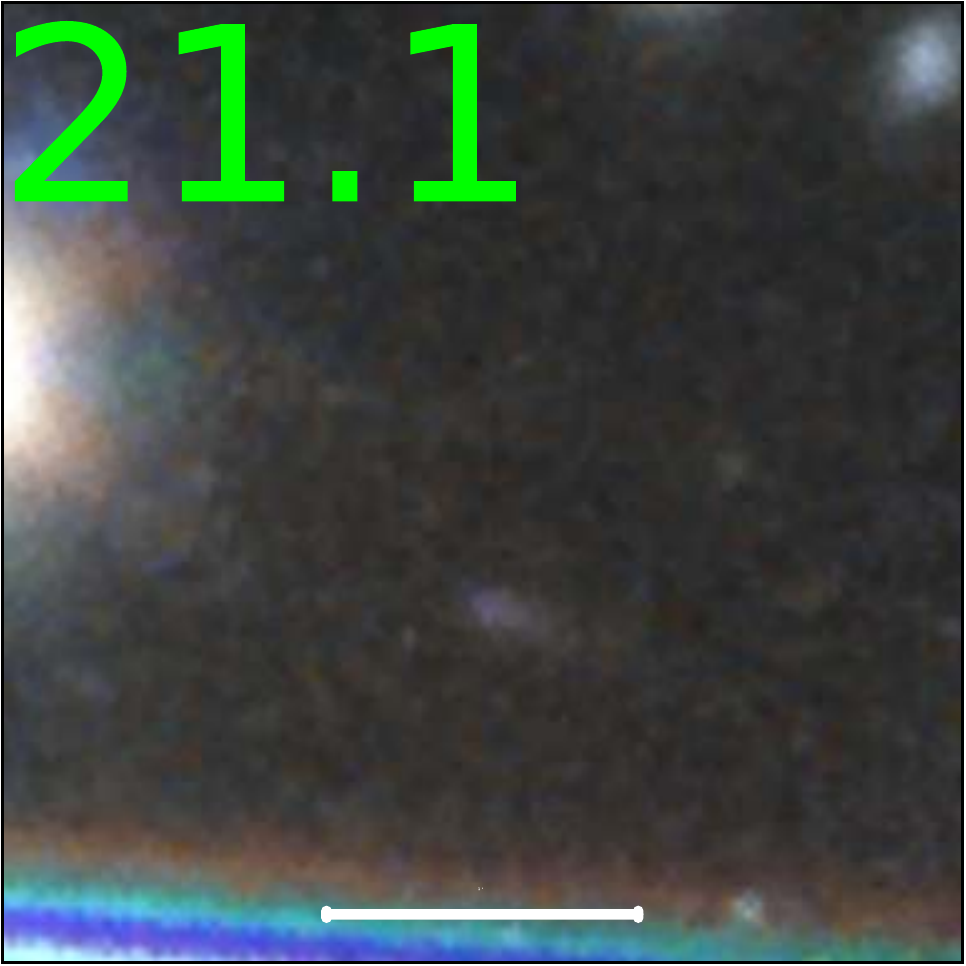}
    \includegraphics[width = 0.328\columnwidth]{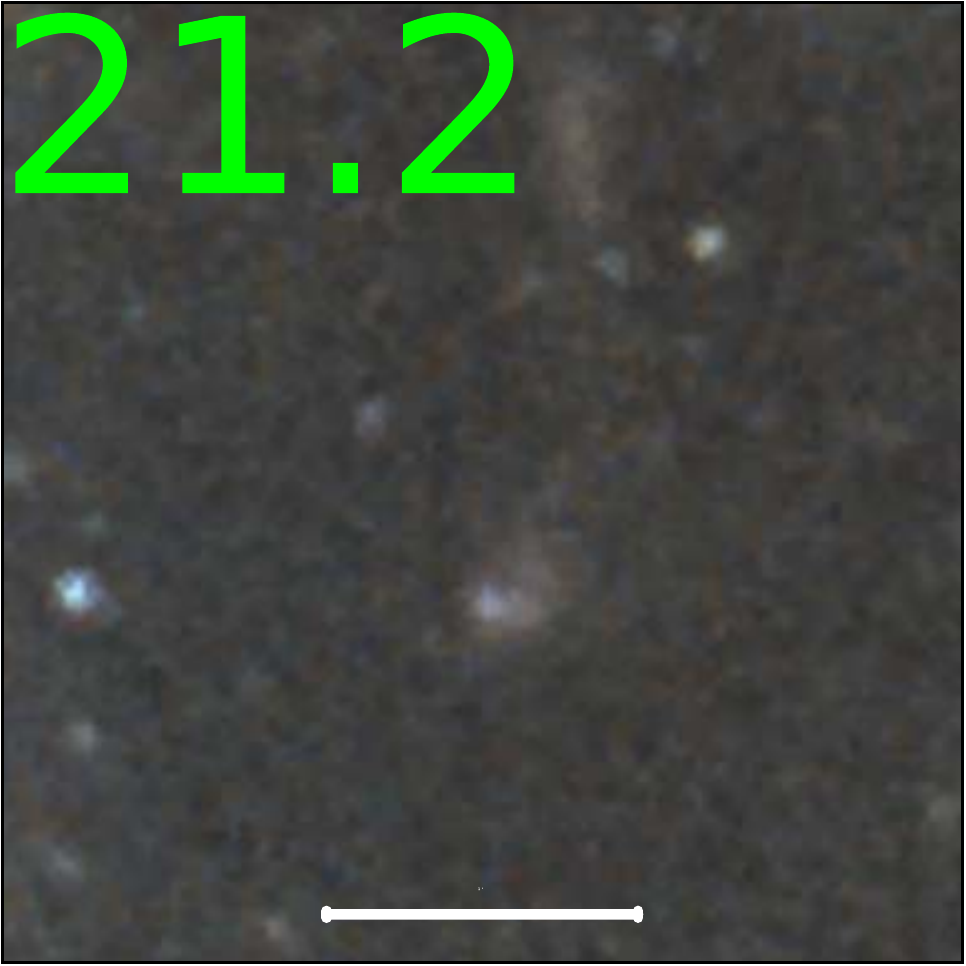}
    \includegraphics[width = 0.328\columnwidth]{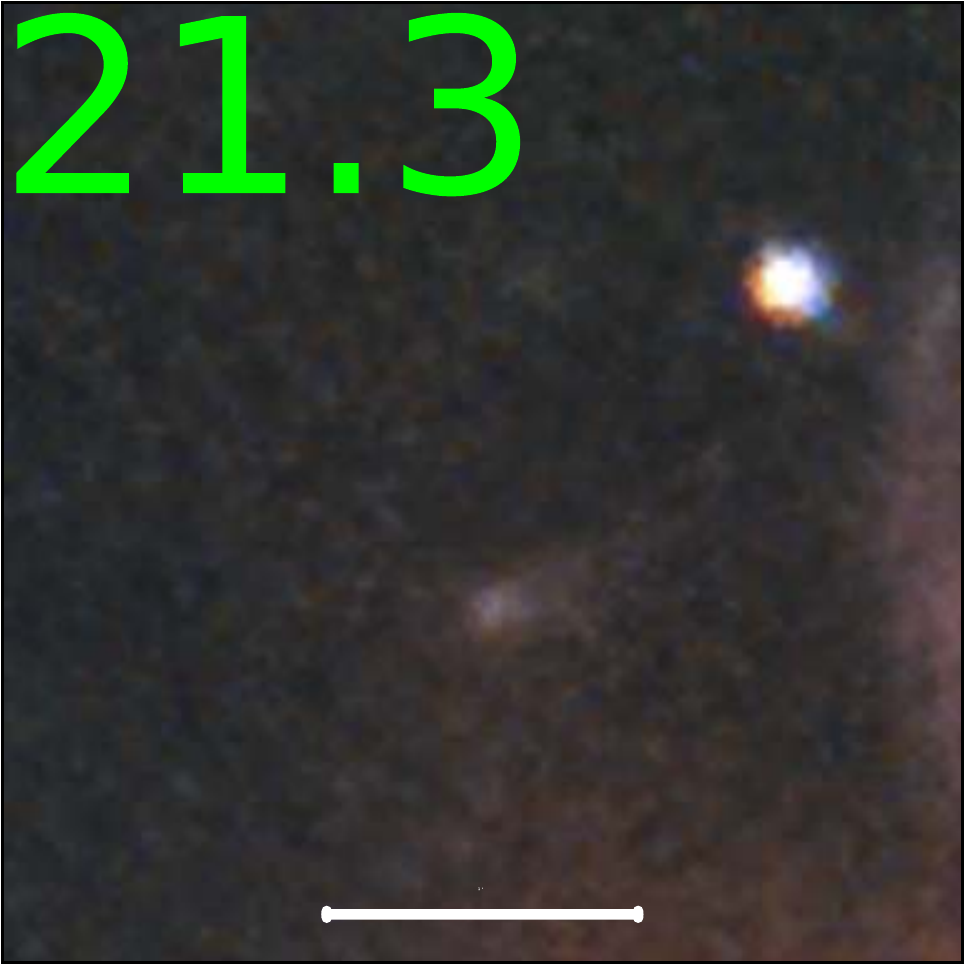}

    \includegraphics[width = 0.328\columnwidth]{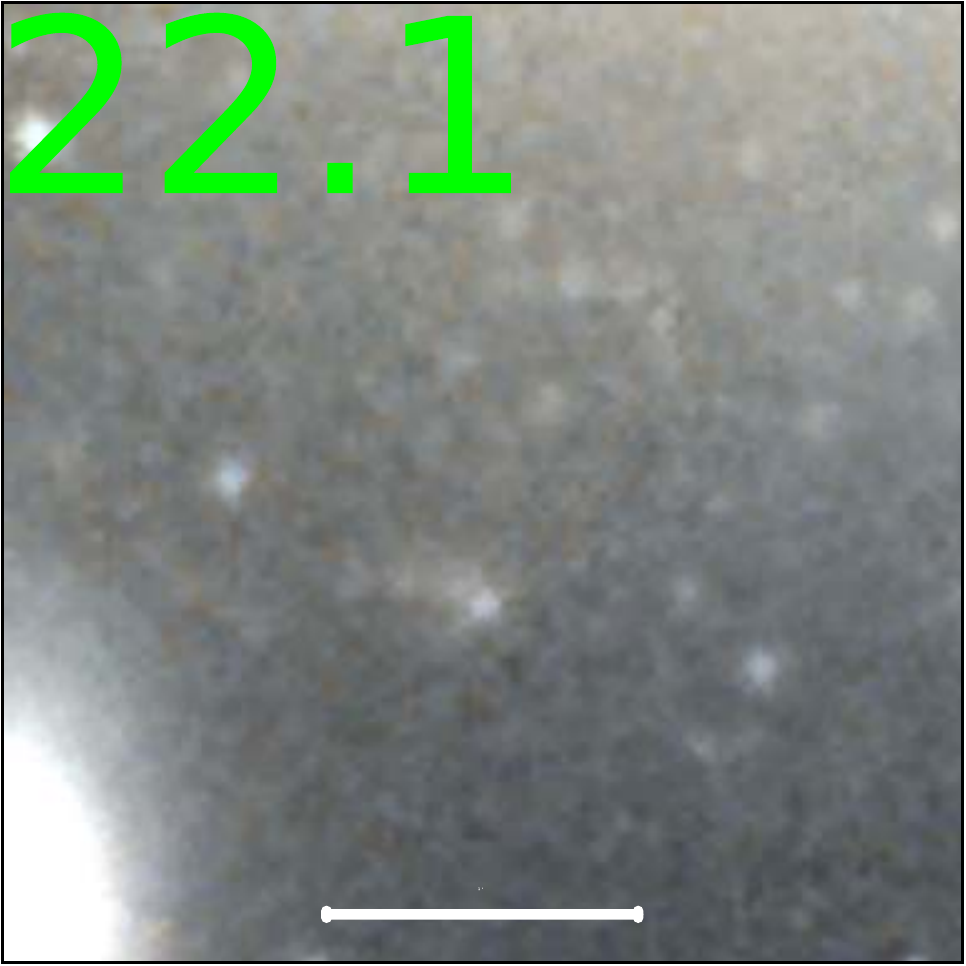}
    \includegraphics[width = 0.328\columnwidth]{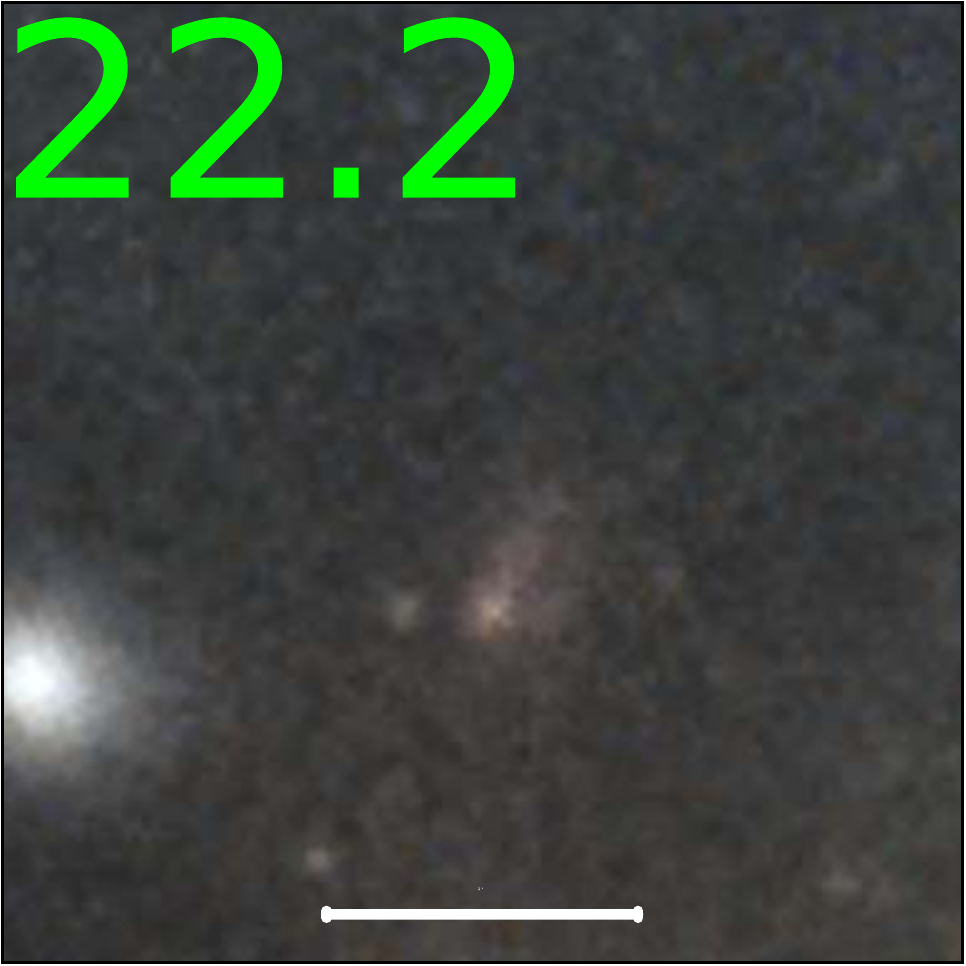}

  \caption{Continued.}
  \label{fig:cut_outs}
\end{figure}

\begin{table*}[!]
\centering
\scriptsize
\caption{MUSE spectroscopic catalogue.}
\begin{tabular}{l c c c c c c c c c } \hline \hline
ID & RA & Dec & $z_{\rm MUSE}$ & QF & mult. & $|\mu|$ & $\delta |\mu|$~min & $\delta |\mu|$~max\\
\hline
1	&	110.8284294	&	$-$73.4602574	&	0.0000	&	4	&	1	&	1.000	&	0.000	&	0.000	\\
2	&	110.8479302	&	$-$73.4616212	&	0.0000	&	4	&	1	&	1.000	&	0.000	&	0.000	\\
3	&	110.8334991	&	$-$73.4484003	&	0.0000	&	4	&	1	&	1.000	&	0.000	&	0.000	\\
4	&	110.8080566	&	$-$73.4466289	&	0.0000	&	4	&	1	&	1.000	&	0.000	&	0.000	\\
5	&	110.8075712	&	$-$73.4469594	&	0.0000	&	4	&	1	&	1.000	&	0.000	&	0.000	\\
6	&	110.8261351	&	$-$73.4486167	&	0.0000	&	4	&	1	&	1.000	&	0.000	&	0.000	\\
7	&	110.8235839	&	$-$73.4499063	&	0.0000	&	4	&	1	&	1.000	&	0.000	&	0.000	\\
8	&	110.8235958	&	$-$73.4474173	&	0.0000	&	4	&	1	&	1.000	&	0.000	&	0.000	\\
9	&	110.8066750	&	$-$73.4516338	&	0.0000	&	4	&	1	&	1.000	&	0.000	&	0.000	\\
10	&	110.8117026	&	$-$73.4532181	&	0.0000	&	4	&	1	&	1.000	&	0.000	&	0.000	\\
11	&	110.8354266	&	$-$73.4530012	&	0.0000	&	4	&	1	&	1.000	&	0.000	&	0.000	\\
12	&	110.8404555	&	$-$73.4533044	&	0.0000	&	4	&	1	&	1.000	&	0.000	&	0.000	\\
13	&	110.8166004	&	$-$73.4551109	&	0.0000	&	4	&	1	&	1.000	&	0.000	&	0.000	\\
14	&	110.8450243	&	$-$73.4553745	&	0.0000	&	4	&	1	&	1.000	&	0.000	&	0.000	\\
15	&	110.8354564	&	$-$73.4557020	&	0.0000	&	4	&	1	&	1.000	&	0.000	&	0.000	\\
16	&	110.8178433	&	$-$73.4544317	&	0.0000	&	4	&	1	&	1.000	&	0.000	&	0.000	\\
17	&	110.8040760	&	$-$73.4581786	&	0.0000	&	4	&	1	&	1.000	&	0.000	&	0.000	\\
18	&	110.8230292	&	$-$73.4593733	&	0.0000	&	4	&	1	&	1.000	&	0.000	&	0.000	\\
19	&	110.8367020	&	$-$73.4585449	&	0.0000	&	4	&	1	&	1.000	&	0.000	&	0.000	\\
20	&	110.8486872	&	$-$73.4593321	&	0.0000	&	4	&	1	&	1.000	&	0.000	&	0.000	\\
21	&	110.8495683	&	$-$73.4625124	&	0.3217	&	3	&	1	&	1.000	&	0.000	&	0.000	\\
22	&	110.8362871	&	$-$73.4568771	&	0.3241	&	3	&	1	&	1.000	&	0.000	&	0.000	\\
23	&	110.8252476	&	$-$73.4545586	&	0.3774	&	3	&	1	&	1.000	&	0.000	&	0.000	\\
24	&	110.8243781	&	$-$73.4599050	&	0.3806	&	3	&	1	&	1.000	&	0.000	&	0.000	\\
25	&	110.8557313	&	$-$73.4557427	&	0.3822	&	3	&	1	&	1.000	&	0.000	&	0.000	\\
26	&	110.8530957	&	$-$73.4566588	&	0.3836	&	3	&	1	&	1.000	&	0.000	&	0.000	\\
27	&	110.8161206	&	$-$73.4511937	&	0.3844	&	3	&	1	&	1.000	&	0.000	&	0.000	\\
28	&	110.8456507	&	$-$73.4513365	&	0.3847	&	3	&	1	&	1.000	&	0.000	&	0.000	\\
29	&	110.8538419	&	$-$73.4500630	&	0.3848	&	3	&	1	&	1.000	&	0.000	&	0.000	\\
30	&	110.8185163	&	$-$73.4552171	&	0.3849	&	3	&	1	&	1.000	&	0.000	&	0.000	\\
31	&	110.8045012	&	$-$73.4561504	&	0.3864	&	3	&	1	&	1.000	&	0.000	&	0.000	\\
32	&	110.8550113	&	$-$73.4501908	&	0.3865	&	3	&	1	&	1.000	&	0.000	&	0.000	\\
33	&	110.8377986	&	$-$73.4536001	&	0.3865	&	3	&	1	&	1.000	&	0.000	&	0.000	\\
34	&	110.8326866	&	$-$73.4569116	&	0.3869	&	3	&	1	&	1.000	&	0.000	&	0.000	\\
35	&	110.8562438	&	$-$73.4506971	&	0.3872	&	2	&	1	&	1.000	&	0.000	&	0.000	\\
36	&	110.8257093	&	$-$73.4586938	&	0.3888	&	3	&	1	&	1.000	&	0.000	&	0.000	\\
37	&	110.8376015	&	$-$73.4561834	&	0.3896	&	3	&	1	&	1.000	&	0.000	&	0.000	\\
38	&	110.8024677	&	$-$73.4586733	&	0.3905	&	3	&	1	&	1.000	&	0.000	&	0.000	\\
39	&	110.8182380	&	$-$73.4546131	&	0.3909	&	3	&	1	&	1.000	&	0.000	&	0.000	\\
40	&	110.8401314	&	$-$73.4558252	&	0.3911	&	3	&	1	&	1.000	&	0.000	&	0.000	\\
41	&	110.8268052	&	$-$73.4546385	&	0.3914	&	3	&	1	&	1.000	&	0.000	&	0.000	\\
42	&	110.8264520	&	$-$73.4549705	&	0.3915	&	3	&	1	&	1.000	&	0.000	&	0.000	\\
43	&	110.8184078	&	$-$73.4482682	&	0.3932	&	2	&	1	&	1.000	&	0.000	&	0.000	\\
44	&	110.8006165	&	$-$73.4485206	&	0.3940	&	3	&	1	&	1.000	&	0.000	&	0.000	\\
45	&	110.8487541	&	$-$73.4603147	&	0.3970	&	3	&	1	&	1.000	&	0.000	&	0.000	\\
46	&	110.8368347	&	$-$73.4565146	&	0.3980	&	3	&	1	&	1.000	&	0.000	&	0.000	\\
47	&	110.8049026	&	$-$73.4494579	&	0.4138	&	2	&	1	&	1.097	&	-0.003	&	0.003	\\
48	&	110.8418679	&	$-$73.4567000	&	0.4233	&	3	&	1	&	1.194	&	-0.005	&	0.005	\\
49	&	110.8521136	&	$-$73.4555097	&	0.5189	&	3	&	1	&	1.700	&	-0.035	&	0.039	\\
50	&	110.8327947	&	$-$73.4466290	&	0.6002	&	3	&	1	&	1.527	&	-0.032	&	0.036	\\
51	&	110.8302846	&	$-$73.4604916	&	0.7455	&	3	&	1	&	2.144	&	-0.081	&	0.083	\\
52	&	110.8089116	&	$-$73.4609025	&	0.7460	&	3	&	1	&	1.936	&	-0.060	&	0.061	\\
53	&	110.8384789	&	$-$73.4492352	&	0.7469	&	3	&	1	&	2.315	&	-0.088	&	0.098	\\
54	&	110.8035266	&	$-$73.4573758	&	0.9366	&	3	&	1	&	3.876	&	-0.167	&	0.195	\\
55	&	110.8523022	&	$-$73.4594938	&	1.0815	&	3	&	1	&	3.119	&	-0.135	&	0.149	\\
56	&	110.8533363	&	$-$73.4595550	&	1.0815	&	3	&	1	&	3.195	&	-0.143	&	0.157	\\
57	&	110.8535225	&	$-$73.4591817	&	1.0823	&	3	&	1	&	3.314	&	-0.147	&	0.171	\\
58	&	110.8101344	&	$-$73.4605069	&	1.0826	&	3	&	1	&	2.733	&	-0.137	&	0.144	\\
59	&	110.8566353	&	$-$73.4587921	&	1.0833	&	3	&	1	&	3.378	&	-0.150	&	0.189	\\
60	&	110.8557579	&	$-$73.4588674	&	1.0833	&	3	&	1	&	3.371	&	-0.148	&	0.190	\\
61	&	110.8066813	&	$-$73.4523137	&	1.2696	&	9	&	1	&	22.076	&	-3.915	&	5.736	\\
62	&	110.8363455	&	$-$73.4587211	&	1.3782	&	3	&	3	&	63.360	&	-17.166	&	34.317	\\
63	&	110.8385802	&	$-$73.4509557	&	1.3782	&	3	&	3	&	8.717	&	-0.878	&	1.073	\\
64	&	110.8405311	&	$-$73.4551527	&	1.3782	&	3	&	3	&	6.234	&	-0.581	&	0.625	\\
65	&	110.8156178	&	$-$73.4602436	&	1.3802	&	9	&	1	&	3.540	&	-0.242	&	0.265	\\
66	&	110.8207008	&	$-$73.4506760	&	1.3825	&	9	&	3	&	5.193	&	-0.439	&	0.482	\\
67	&	110.8162336	&	$-$73.4534926	&	1.3825	&	9	&	3	&	12.655	&	-1.484	&	1.603	\\
68	&	110.8171040	&	$-$73.4589089	&	1.3825	&	9	&	3	&	8.959	&	-1.002	&	1.160	\\
69	&	110.8222252	&	$-$73.4526676	&	1.4254	&	3	&	1	&	28.789	&	-5.146	&	6.543	\\
70	&	110.8236935	&	$-$73.4518167	&	1.4254	&	3	&	2	&	54.813	&	-6.289	&	8.633	\\
71	&	110.8405974	&	$-$73.4510474	&	1.4503	&	3	&	3	&	6.682	&	-0.635	&	0.694	\\
72	&	110.8427896	&	$-$73.4546490	&	1.4503	&	3	&	3	&	22.004	&	-3.654	&	4.245	\\
73	&	110.8388169	&	$-$73.4587230	&	1.4503	&	3	&	3	&	10.271	&	-1.064	&	1.336	\\
74	&	110.8485993	&	$-$73.4606870	&	1.4792	&	9	&	1	&	3.450	&	-0.201	&	0.228	\\
75	&	110.8520623	&	$-$73.4525378	&	1.4816	&	3	&	1	&	11.676	&	-1.151	&	1.523	\\
76	&	110.8519575	&	$-$73.4523356	&	1.4817	&	3	&	1	&	10.683	&	-0.969	&	1.290	\\
77	&	110.8053717	&	$-$73.4499939	&	3.3136	&	9	&	1	&	143.596	&	-77.257	&	357.679	\\
78	&	110.7998624	&	$-$73.4549695	&	3.4287	&	9	&	1	&	14.849	&	-2.354	&	3.257	\\
\hline
\hline
\end{tabular}
\label{tab:muse_cat}
\tablefoot{\tiny Spectroscopic confirmations from MUSE. The columns QF is the quality flag of the MUSE confirmation, where 3 corresponds to a secure measurement, 2 to a probable secure redshift, 9 to a measurement based on a single emission line, and 4 to a secure star confirmation. Multiple images are indicated in the mult.~column, and magnification factors with the error bars are also indicated in columns $\mu$ and $\delta_m$min/max (68\% confidence interval). Entries with $|\mu|=1$ are cluster members and foreground objects not subject to the cluster lensing magnification. Coordinates in this table are obtained from \texttt{SExtractor} \citep{1996A&AS..117..393B}, and they are slightly different from Table \ref{tab:multiple_images} for which we fine tuned the positions of corresponding clumps of multiple images.}
\end{table*}

\begin{table}[h]
\centering
    \caption{Recovered mass parameters from the \jwst lens model.}
    \begin{tabular}{c c c c c} \hline \hline
    ~ & Median & 68\% CL & 95\% CL & 99.7\% CL \\
\hline
$x~[\arcsec]$ & $0.42$ & $_{-0.46}^{+0.51}$ & $_{-0.88}^{+1.07}$ &$_{-1.29}^{+1.68}$ \\
$y~[\arcsec]$ & $1.04$ & $_{-0.08}^{+0.08}$ & $_{-0.15}^{+0.17}$ &$_{-0.23}^{+0.27}$ \\
$\varepsilon$ & $0.59$ & $_{-0.07}^{+0.07}$ & $_{-0.15}^{+0.13}$ &$_{-0.22}^{+0.19}$ \\
$\theta$~[deg] & $0.95$ & $_{-0.76}^{+0.65}$ & $_{-1.75}^{+1.25}$ &$_{-2.95}^{+1.81}$ \\
$r_{\rm core}~[\arcsec]$ & $18.7$ & $_{-1.2}^{+1.4}$ & $_{-2.2}^{+3.0}$ &$_{-3.2}^{+5.1}$ \\
$\sigma$~[km/s] & $1132$ & $_{-31}^{+36}$ & $_{-61}^{+78}$ &$_{-91}^{+136}$ \\
$\gamma_{\rm shear}$ & $0.08$ & $_{-0.01}^{+0.03}$ & $_{-0.02}^{+0.08}$ &$_{-0.03}^{+0.14}$ \\
$\theta_{\rm shear}$~[deg] & $-27.2$ & $_{-19.3}^{+12.9}$ & $_{-34.4}^{+19.6}$ &$_{-42.7}^{+23.8}$ \\
$r^{\rm gal}_{\rm cut}~[\arcsec]$ & $0.96$ & $_{-0.35}^{+0.74}$ & $_{-0.45}^{+2.02}$ &$_{-0.46}^{+4.40}$ \\
$\sigma^{\rm gal}$~[km/s] (km/s) & $453$ & $_{-104}^{+108}$ & $_{-180}^{+171}$ &$_{-233}^{+218}$ \\

\hline\hline
    \end{tabular}
    \tablefoot{The first 6 rows are the cluster halo parameters, the next 2 are the external shear parameters, and the last 2 are the cluster member normalisation parameters.  The angles are defined as zero along the x-axis and increase counterclockwise. The parameter $\varepsilon$ is given by $(a^2 - b^2)/(a^2+b^2)$, where $a$ and $b$ are the semi-major and minor axis, respectively. The positions $x$ and $y$ are relative distances to the central galaxy (RA=$110.8268104$, Dec=$-73.4546454$). The reference magnitude value used for the members scaling relation is $mag_{\rm F160W} = 17.649$.}
\label{tab:model_params}
\end{table}

\end{appendix}

\end{document}